\documentclass[traditabstract]{aa} 
\usepackage{graphicx}
\usepackage{epstopdf}
\usepackage{deluxetable}
\usepackage{mathrsfs}
\usepackage{natbib}
\usepackage{amsmath}
\usepackage{amssymb}
\bibpunct{(}{)}{;}{a}{}{,} 
\usepackage[english]{babel}
\usepackage{longtable}
\usepackage{url}
\usepackage{hyperref}

\authorrunning{Taddia et al.}
\titlerunning{SNe~Ic-BL from (i)PTF}

\begin{document}

\title{Analysis of broad-lined Type Ic supernovae from the (intermediate) Palomar Transient Factory}

\author{F. Taddia\inst{1},
J. Sollerman\inst{1},
C. Fremling\inst{2},
C. Barbarino\inst{1},
E. Karamehmetoglu\inst{1},
I. Arcavi\inst{3},
S.~B. Cenko\inst{4,5} 
A.~V. Filippenko\inst{6,7},
A. Gal-Yam\inst{8},
D. Hiramatsu\inst{9,10},
G. Hosseinzadeh\inst{11}, 
D.~A. Howell\inst{9,10}, 
S.~R. Kulkarni\inst{2},
R. Laher\inst{12},
R. Lunnan\inst{1,2},
F. Masci\inst{12},
P.~E. Nugent\inst{6,13},
A. Nyholm\inst{1},
D.~A. Perley\inst{14},
R. Quimby\inst{15,16}, and
J.~M. Silverman\inst{6,17}.}

\institute{The Oskar Klein Centre, Department of Astronomy, Stockholm University, AlbaNova, 106 91 Stockholm, Sweden.\\ \email{francesco.taddia@astro.su.se}
\and Cahill Center for Astrophysics, California Institute of Technology, Pasadena, CA 91125, USA.
\and The School of Physics and Astronomy, Tel Aviv University, Tel Aviv 69978, Israel 
\and Astrophysics Science Division, NASA Goddard Space Flight Center, Mail Code 661, Greenbelt, MD 20771, USA. 
\and Joint Space-Science Institute, University of Maryland, College Park, MD 20742, USA 
\and Department of Astronomy, University of California, Berkeley, CA 94720- 3411, USA. 
\and Miller Senior Fellow, Miller Institute for Basic Research in Science, University of California, Berkeley, CA 94720, USA. 
\and Department of Particle Physics and Astrophysics, Weizmann Institute of Science, Rehovot 76100, Israel. 
\and Las Cumbres Observatory, Goleta, CA 93117, USA. 
\and Department of Physics, University of California, Santa Barbara, CA 93106, USA. 
\and Harvard-Smithsonian Center for Astrophysics, 60 Garden Street, Cambridge, MA 02138, USA. 
\and Infrared Processing and Analysis Center, California Institute of Technology, Pasadena, CA 91125, USA. 
\and Lawrence Berkeley National Laboratory, 1 Cyclotron Road, MS 50B-4206, Berkeley, CA 94720, USA. 
\and Astrophysics Research Institute, Liverpool John Moores University, IC2, Liverpool Science Park, 146 Brownlow Hill, Liverpool L3 5RF, UK. 
\and Department of Astronomy/Mount Laguna Observatory, San Diego State University, 5500 Campanile Drive, San Diego, CA 92812-1221, USA. 
\and Kavli IPMU (WPI), The University of Tokyo Institutes for Advanced Study, The University of Tokyo, Kashiwa, Chiba 277-8583, Japan 
\and Samba TV, San Francisco, CA 94107, USA.} 

\date{Received  XX / Accepted XX}

\abstract{We study 34 Type Ic supernovae that have broad spectral features 
(SNe~Ic-BL). This is the only SN type found in
  association with long-duration gamma-ray bursts (GRBs). We obtained our
  photometric data with the Palomar Transient Factory (PTF) and its continuation, the  intermediate Palomar Transient Factory (iPTF). This is the first large, homogeneous sample of SNe~Ic-BL from an untargeted
  survey. Furthermore, given the high observational cadence of (i)PTF,
  most of these SNe~Ic-BL were discovered soon after explosion. 
We present K-corrected $Bgriz$ light curves of these SNe, obtained
through photometry on template-subtracted images. We analyzed the
shape of the $r$-band light curves, finding a correlation between the
decline parameter $\Delta m_{15}$ and the rise parameter $\Delta
m_{-10}$.  We studied the SN colors and, based on $g-r$, we estimated the
host-galaxy extinction for each event. 
Peak $r$-band absolute magnitudes have an average of 
$-18.6\pm0.5$ mag. We fit each $r$-band light curve with that of
SN~1998bw (scaled and stretched) to derive the explosion
epochs. We computed the bolometric light curves using bolometric
corrections, $r$-band data, and $g-r$ colors. Expansion velocities from
\ion{Fe}{ii} were obtained by fitting spectral templates of SNe~Ic. 
Bolometric
light curves and velocities at peak were fitted using the semianalytic
Arnett model to estimate ejecta mass $M_{\rm ej}$, explosion
energy $E_{K}$ and $^{56}$Ni mass $M(^{56}$Ni) for each SN. 
We find average values of $M_{\rm ej} = 4\pm3~{\rm M}_{\odot}$, 
$E_{K} = (7\pm6) \times 10^{51}$~erg, 
and $M(^{56}$Ni) $= 0.31\pm0.16~{\rm M}_{\odot}$. 
The parameter distributions were
compared to those presented in the literature and are overall in agreement with them. We also estimated the degree of $^{56}$Ni mixing using scaling relations derived from hydrodynamical models and we find that all the SNe are strongly mixed.
The derived explosion parameters imply that at least 21\% of the progenitors of SNe~Ic-BL are compatible with massive ($>28$~${\rm M}_{\odot}$), possibly single stars, whereas at least 64\% might come from less massive stars in close binary systems.}


\maketitle

\section{Introduction}
\label{sec:intro}

Core-collapse supernovae (CC~SNe) are explosions of massive 
($M_{\rm ZAMS}$~$\gtrsim$~8~$\rm M_{\odot}$) 
stars. So-called
stripped-envelope (SE) SNe -- which show deficiency
or lack
of H, or even He, and are hence classified as SNe~IIb/Ib or Ic -- can be
found among CC~SNe 
\citep[e.g.,][for a review]{filippenko97}. 
These SE~SNe could arise from single massive
($M_{\rm ZAMS}$~$\gtrsim$~30~$\rm M_{\odot}$) stars stripped of their H/He
envelopes by strong stellar winds \citep{conti76}, or they
could be the explosions of somewhat less massive stars
($M_{\rm ZAMS}$~$\lesssim$~20~$\rm M_{\odot}$) stripped by their companions in
binary systems \citep[e.g.,][]{yoon10,lyman16,taddia18}.

Among He-poor SNe~Ic, there are events
characterized by broad lines in their spectra, indicating 
particularly fast expansion velocities. 
These SNe are called broad-lined SNe~Ic, or SNe~Ic-BL. As shown by
\citet{modjaz16}, normal SNe~Ic have \ion{Fe}{ii} velocities of
$\sim$10$^4$~km~s$^{-1}$ at light-curve peak, whereas SNe~Ic-BL show
\ion{Fe}{ii} velocities of 
$\sim$ (1.5--3) $\times 10^4$~km~s$^{-1}$. 
In some cases, however, it is not straightforward to distinguish between normal SNe~Ic and SNe~Ic-BL, as there are SNe, such as SN~2004aw \citep{taubenberger06}, showing intermediate properties. 

SNe~Ic-BL is
the only\footnote{With the exceptions of the spectroscopically normal SNe~Ic 2002lt and 2013ez \citep{cano14_slowgrbsne} and the SLSN~Ic 2011kl associated with the UL$-$GRB 111209A \citep{greiner15}.} SN class found to occur in correspondence with long-duration
gamma-ray bursts (GRBs; \citealp{woosley06}), the first clear example being
SN~1998bw \citep{galama98}. These GRB-SNe are believed to be
accompanied by the launch of collimated relativistic outflows, as shown in the case of SN~1998bw, whose strong radio emission revealed relativistic ejecta \citep{kulkarni98}. 
On the other hand, many SNe~Ic-BL seem not to be associated with any GRB 
\cite[see, e.g., SN~2002ap;][]{berger02}. 
Even SN~2012ap \citep{mili15_2012ap} and SN~2009bb \citep{pignata11}, which show evidence for a central engine and ejecta with relativistic velocities, are not coincident with a GRB. 
The reason for the
difference between SNe~Ic-BL with and without a GRB 
is still debated. 
It has been suggested that some SNe Ic-BL are accompanied by an off-axis
GRB, which may be initially 
invisible but later should emerge in radio
emission. However, this has never been observed \citep{soderberg10}. 
\citet{corsi16}
estimated that $<85$\% of their 15 SNe~Ic-BL 
from the (intermediate) Palomar Transient Factory 
with radio observations could hide off-axis GRBs expanding in media with relatively high densities. 

The light curves of SNe~Ic-BL rise faster than those of SNe~Ib or IIb,
and their peak magnitudes are brighter than those of the other SE~SNe
\citep[e.g.,][]{cano13,taddia15,lyman16,prentice16,taddia18}.
The modeling of the light curves has suggested that SNe~Ic-BL eject larger
$^{56}$Ni masses and have higher explosion energies ($\sim 10^{52}$
erg) than the other SE~SN types
\citep[e.g.,][]{cano13,taddia15,lyman16,prentice16,taddia18}, but that
the ejecta masses are not very different. Given the high explosion
energies of these SNe, they are sometimes also called ``hypernovae.'' 
SNe~Ic-BL and
GRB-SNe have been suggested (also by host galaxy studies) to have progenitors that are younger and
more massive than those of normal SNe~Ic
\citep[e.g.,][]{sanders12,cano13}. 

With the exception of the small SN~Ic-BL sample from SDSS-II studied by
\citet{taddia15}, the other analyzed samples
\citep[e.g.,][]{cano13,lyman16,prentice16} come from literature collections.
Most of those SNe~Ic-BL were discovered by targeted surveys, and the
nonuniform data collections were obtained with many different telescopes.
In addition, 
light curves of 10 SNe~Ic-BL from the Center for Astrophysics (CfA) survey 
were released by
\citet{bianco14}, and \citet{modjaz16} performed a detailed analysis
of the spectra of 10 SNe~Ic-BL without GRBs and 11 SNe~Ic-BL with GRBs. 

Thanks to the Palomar Transient Factory
(PTF; \citealp{rau09,law09}) and its continuation the intermediate
Palomar Transient Factory (iPTF; \citealp{kulkarni13}), we can
present a large (34 objects) sample of 
SNe~Ic-BL 
with optical observations 
from an untargeted survey.
Almost all of the photometry has been obtained with two telescopes and reduced in the same way.
This allows for a systematic study of the properties of this SN population
and its progenitor stars.

This paper is structured as follows. 
In Sect.~\ref{sec:sample}, we describe the basic information about
our sample of SNe~Ic-BL from the PTF+iPTF surveys. 
The SN observations and data reduction methods are
presented in Sect.~\ref{sec:data}. The light curves in 
different bands are presented and
analyzed (in particular, the $r$ band) in Sect.~\ref{sec:analysis}, as
are the SN colors (especially $g-r$). These colors are also used to derive the host galaxy extinction.
In Sect.~\ref{sec:spectra} we present the SN spectra, whereas in 
Sect.~\ref{sec:boloprop} we produce and discuss the bolometric properties of our SNe. By modeling the derived bolometric properties, in Sect.~\ref{sec:model} we obtain progenitor and explosion parameters for our SNe. A discussion of the results (which are compared with those in the literature) and final conclusions are given in Sect.~\ref{sec:discussion} and \ref{sec:conclusions}, respectively. 

\section{The SN sample}
\label{sec:sample}

Our study includes 34 events discovered and monitored by PTF and iPTF
in the years 2009--2017 that we classified as SNe~Ic-BL.
Their names and coordinates are reported in Table~\ref{tab:sample}. 
These 34 events are a significant part of the total SE~SN sample from PTF and iPTF, which counts 204 events, including SNe~IIb (58), SN~Ib (38), spectroscopically normal SNe~Ic (60), and  SNe~Ibc (14). Supernovae Ibc are objects with spectra showing possible traces of \ion{He}{} in their spectra, making them intermediate cases between spectroscopically normal SNe~Ic and Ib. The SNe~Ic and SNe~Ib light curves will be studied by Barbarino et al. in prep., and the spectra have been presented and analyzed by \citet{fremling18}. We do not consider superluminous Type I SNe (SLSNe~I) in our study; these are typically characterized by absolute peak magnitudes $\lesssim -21$ mag, long rise times, and peculiar spectral features at early times \citep[][]{quimby11,galyam12,decia18,lunnan18}.

\begin{figure}
 \centering
\includegraphics[width=18cm,angle=0]{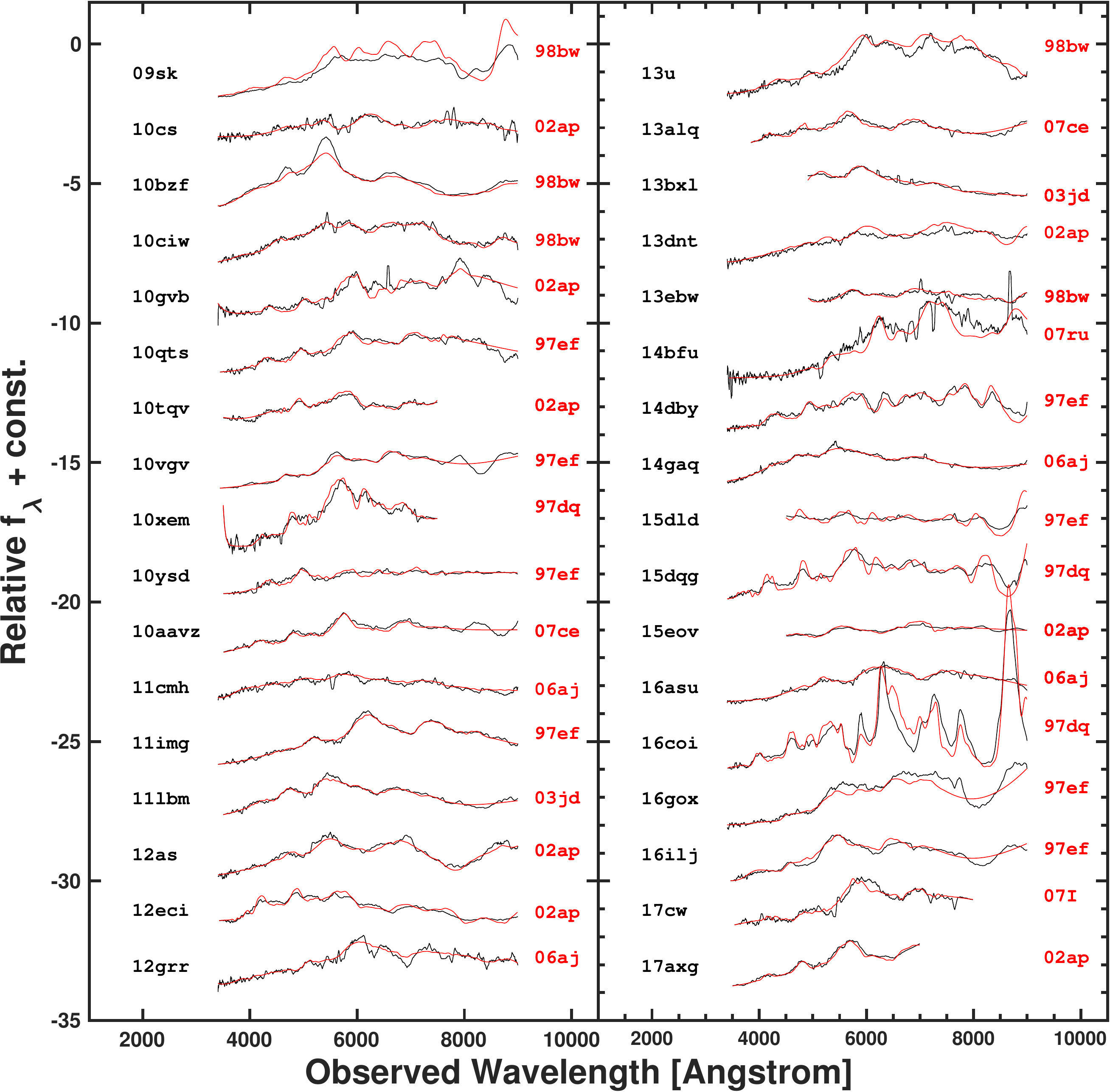}
  \caption{ \label{snid} SNID fits to the spectra of our 34 SNe~Ic-BL. 
   Next to each observed spectrum (in black) we report the SN
    name. The SN~Ic-BL template spectrum that fits best is presented in red and the corresponding SN name is reported in red next to each spectrum. We selected a single spectrum for each SN, in order to show that they belong to the SN~Ic-BL class.}
 \end{figure}

The inclusion of SNe in our sample was based on spectroscopic
classification, comparing our SN spectra 
with those of known SNe~Ic-BL.
We inspected the spectra of all SNe classified as possible SNe~Ic and
SNe~Ic-BL in the PTF archive, and defined our final sample of SNe~Ic-BL
by cross-correlating our SN spectra with those included as templates
in the Supernova Identification \citep[SNID;][]{snid} program. We also used 
as templates
those by \citet{liu14}, specific for SE~SNe. In Fig.~\ref{snid}
we present for each SN
an example of SNID matches to SN~Ic-BL spectral templates.
As in \citet{modjaz16}, 
PTF10bzf (SN 2010ah), PTF10qts, and PTF10vgv are included in our SN~Ic-BL sample. 
If we use the method proposed by \citet{prentice17} to classify helium-poor SNe based on the number of absorption features in the spectra at
and before the light-curve peak, our spectra show between two, three, and four features in most
of the cases. This corresponds to SNe~Ic with larger line blending caused by
their fast velocities. Spectroscopically normal SNe~Ic typically show six or seven
features. 
We report the Ic-$<N>$  classification \citep{prentice17}
for each SN
observed before peak in Table~\ref{tab:speclog}, after visual
inspection of all the spectra. 
When only past-peak spectra are available, upper limits
for $<N>$ are provided. 

Given their spectra, we include iPTF15eov and iPTF16asu in our sample,
even though these SNe are somewhat peculiar objects: the former is a
slow riser, slow decliner, superluminous SN, whereas the latter is
a very luminous fast-rising event. We describe their peculiarities with
respect to the average properties of our sample throughout the paper,
and discuss their underlying powering mechanism in Sect.~\ref{sec:prog}. We also notice that if we use the 
classification code presented by \citet{quimby18}, the spectra of iPTF15eov are better matched by SLSN-I than SN~Ic spectra, whereas iPTF16asu is better matched by SN~Ic spectra. 

\begin{figure}
 \centering
\includegraphics[width=12cm,angle=0]{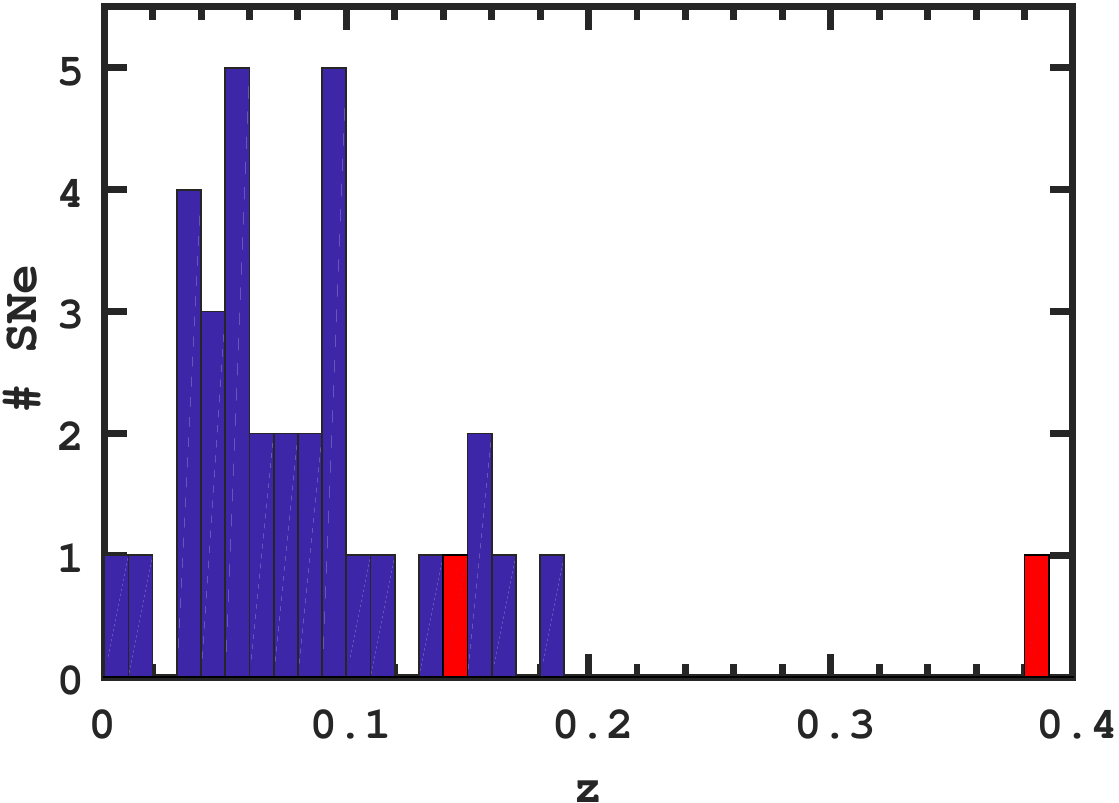}
  \caption{Redshift distribution for our sample of SNe~Ic-BL. All events are at $z<0.2$, except iPTF14bfu at $z=0.384$, which was targeted by iPTF in order to look for the optical counterpart of GRB 140606B \citep{cano15}. Also, iPTF13bxl ($z=0.145$) was discovered while trying to find the optical counterpart of GRB 130702A \citep{singer13}. The two GRB SNe are marked in red.\label{redshift}}
 \end{figure}

We measured the redshift of our SNe from their host galaxy spectral lines,
often detected in the SN spectra or in archival host galaxy
spectra. When the galaxy lines were not detected, as in the case of
PTF12grr, the best SNID fit was used to estimate the redshift. 
Our SNe have the redshift distribution shown in
Fig.~\ref{redshift}. The average redshift is $0.090\pm0.069$;
the most distant object are located at $z=0.384$ and the most nearby at
$z=0.0036$. The SN redshifts and host galaxy names
are reported in Table~\ref{tab:sample}.
We did not include in our sample the transient at $z=1.9733$ named iPTF14yb, which was presented by \citet{cenko14aft} and associated with GRB 140226A; given its large distance, for this transient we obviously have only afterglow data available.

We computed the luminosity distance of each SN based on its redshift
and assuming cosmological parameters from the five-year Wilkinson Microwave Anisotropy Probe (WMAP) observations \citep[H$_0 = 70.5$~km~s$^{-1}$~Mpc$^{-1}$, 
$\Omega_{m}=0.27$, $\Omega_{\Lambda}=0.73$;][]{komatsu09}, including corrections for peculiar motions (Virgo+GA+Shapley, as in \citealp{mould00}).
The Milky Way (MW) color excess was obtained from \citet{schlafly11} via the NASA/IPAC Extragalactic Database
(NED). Both the distances and $E(B-V)^{\rm MW}$ are listed 
in Table~\ref{tab:sample}.

We report $g$-band absolute magnitudes ($M_g^{\rm gal}$) for each 
host galaxy, as obtained through the apparent magnitude listed in the Sloan Digital Sky Survey
(SDSS)
catalog\footnote{\href{http://skyserver.sdss3.org}{http://skyserver.sdss3.org}}
or, when not observed by SDSS, in the Pan-STARRS
catalog\footnote{\href{http://archive.stsci.edu/panstarrs/search.php}{http://archive.stsci.edu/panstarrs/search.php}}. 
These galaxy magnitudes were K-corrected using their $g-r$ 
colors\footnote{\href{http://kcor.sai.msu.ru/about/}{http://kcor.sai.msu.ru/about/}} \citep{kc1,kc2}. 
For the host galaxies of iPTF14bfu and iPTF13bxl, we obtained the $g$ and $r$ magnitudes from \citet{cano15} and \citet{delia15}, respectively. 
The absolute galaxy magnitudes span a relatively large range, from $M_{g}^{\rm gal} = -14.6$~mag to $M_{g}^{\rm gal} =-21.2$~mag, with average $<M_{g}^{\rm gal}> = -18.3\pm0.3$~mag. 

From the post-peak $g-r$ colors of our SNe, we inferred the host galaxy extinction using a method similar to that illustrated by \citet{stritzinger17}. The details of the method are given in Sect.~\ref{sec:hostextinction}. The values of $E(B-V)^{\rm host}$ are listed in Table~\ref{tab:sample}. 

We also attempted to measure 
the equivalent width (EW) of the narrow \ion{Na}{i}~D line in the SN spectra, and derive the host galaxy color excess using the formula provided by \citet{taubenberger06}. These $E(B-V)^{\rm host}$ values from \ion{Na}{i}~D are also reported in Table~\ref{tab:sample}. However, we detected the \ion{Na}{I}~D narrow absorption lines in just five SNe, likely because our spectra of most SNe, which are rather distant,
do not have the necessary resolution and signal-to-noise ratio. 
Since we could not measure the \ion{Na}{i}~D EW for most of our SNe, and given that it is known
that the \ion{Na}{i}~D EW is a poor proxy for host extinction
\citep[see, e.g.,][]{phillips13}, we resorted to use the SN colors to estimate the host extinction.

\begin{figure}
 \centering
\includegraphics[width=18cm,angle=0]{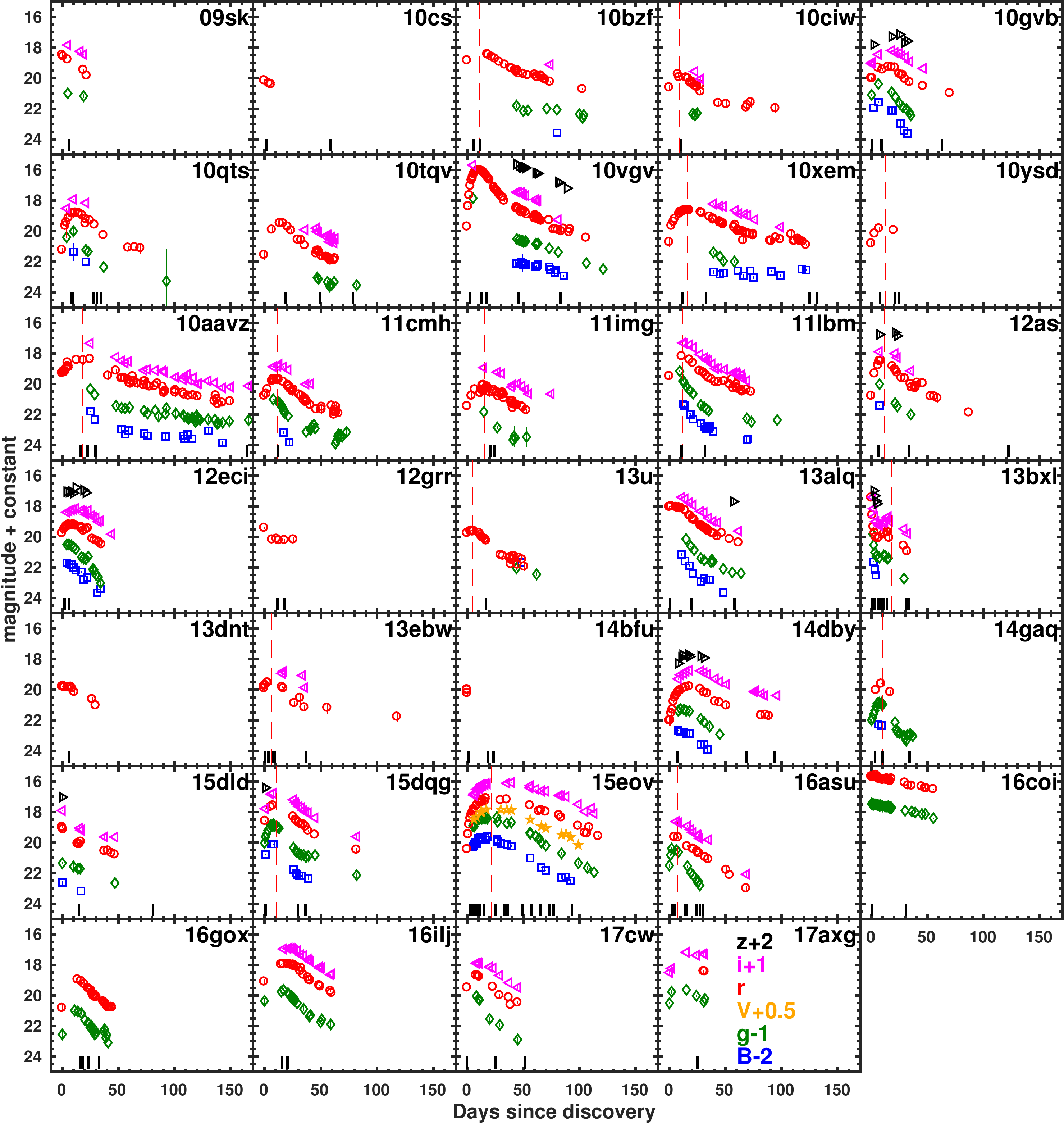}
  \caption{$B$ (blue squares), $g$ (green diamonds), $r$ (red
    circles), $i$ (magenta triangles), and $z$ (black triangles)
    photometry for our sample of SNe~Ic-BL. In the case of iPTF15eov we also have a $V$-band light curve (yellow stars). The light curves as a
    function of discovery epoch have been shifted for clarity (see the
    legend in the last subpanel). The epochs of the observed $r$-band
    maxima are marked by vertical red dashed lines. The spectral
    epochs are shown by black segments at the bottom of each subpanel.\label{lc}}
 \end{figure}

Light curves of each SN are shown in 
Fig.~\ref{lc}. Our sample was mainly observed in the $r$ and $g$
bands (Sect.~\ref{sec:data}) and some photometric data were also obtained in the $B$, $i$, and $z$ bands.
The light curves are presented, in
the observer frame, in  
apparent magnitudes as a function of days since PTF/iPTF discovery. The different bands are shifted by different
amounts for clarity, and
these shifts are indicated in the last subpanel of Fig.~\ref{lc}.
Of our 34 events, we have 28 SNe observed before $r$-band maximum brightness (or
$g$-band maximum in the case of iPTF16asu and iPTF17axg). In the
following sections, we only analyze those
28 SNe with at least one band observed before maximum brightness. 

Our SNe were observed in the $r$ band with a median cadence of
3~d and for a median duration of 61~d and with a
minimum duration of 6~d (PTF10cs) and a maximum duration of 150~d 
(PTF10aavz). We thus observed our SNe~Ic-BL mainly during
their photospheric phases. 

For each SN, the collaboration also obtained spectra, mainly thanks to Caltech astronomers using the Keck (1 and 2) and Palomar (P200) telescopes.
The epochs of the spectra are indicated by vertical black
segments in Fig.~\ref{lc}. The first spectral observation occurred
before $r$-band maximum for 20 out of the 34 SNe. Many of our SNe were,
in fact, imaged right after explosion, as demonstrated by the tight
pre-explosion limits discussed in Sect.~\ref{sec:explo}. 

In summary, this PTF+iPTF photometric dataset of SNe~Ic-BL is 
characterized by its untargeted nature, its large size,
early coverage, high cadence, multiband coverage,
and the availability of pre-explosion images. In addition, we present
accompanying low-resolution spectroscopy to aid the analysis.

Several of the SNe in our sample have already been presented elsewhere.
We report a list of references in the last column of Table~\ref{tab:sample}. 
Nine of our SNe were described in single-object papers (marked in
boldface in Table~\ref{tab:sample}). 
All data from PTF+iPTF were already published for these SNe in the abovementioned papers, except in the cases of  iPTF14bfu, iPTF15dld, and iPTF16coi, for which we provide additional data. 
The other previously reported SNe in our sample were included in sample papers. In particular, \citet{corsi16} presented radio observations and also the PTF/iPTF $r$-band light curves and classification spectra of 15 SNe.
 For all these 21 previously studied SNe --- as well as for the 
 13 SNe presented for the first time in this paper --- we provide a new analysis in the context of our large, untargeted, and homogeneous SN~Ic-BL sample. 

\section{Photometry acquisition and reduction}
\label{sec:data}

Photometric observations were performed with the 48 inch Samuel
Oschin Telescope at Palomar Observatory (P48), equipped with the
92~Mpx mosaic camera CFH12K \citep{rahmer08}, a Mould $r$-band filter
\citep{ofek12}, and an SDSS-like $g$-band filter (and in one case also
an $i$-band filter). The P48 observations allowed the discovery of our
SNe. Twenty-four of our SNe were also observed with the automated Palomar 60 inch telescope (P60; \citealp{cenko06}) in the $Bgriz$ bands.

 Point spread function (PSF) photometry was obtained on template-subtracted 
 P48 and P60 images using the Palomar Transient Factory
 Image Differencing and Extraction (PTFIDE) pipeline \citep{masci17}
 for P48 and the {\tt FPipe} pipeline presented by \citet{fremling16} for
 P60. The P48 templates were obtained from the stack of pre-explosion
 images. As P60 templates, we used  SDSS
 images in the corresponding filters (and Pan-STARRS images in the
 cases of PTF10cs, iPTF14bfu, and iPTF15eov, which did not have SDSS coverage). The photometry was calibrated against SDSS (or Pan-STARRS) stars \citep{ahn14} in the SN field and is presented in AB magnitudes. When P48 and P60 photometry in $g$ and $r$ did not match perfectly owing to slightly different filters, we scaled the P48 photometry to match the P60 data points by adding to the P48 light curve the mean magnitude difference between P48 and P60 in the time range where P48 and P60 data overlap.
For example, P48 mounts a Mould $r$ filter, while P60 is equipped with an
$r$ filter closer to the standard SDSS. 
In the case of iPTF15eov, $BgVri$ data were also obtained by the Las Cumbres Observatory (LCO) and reduced using the LCO pipeline {\tt lcogtsnpipe} \citep{valenti16}, a PyRAF-based photometric reduction pipeline. We performed point spread function photometry. Reference images were obtained with SBIG and Sinistro after the SN faded and image subtraction was performed using PyZOGY \citep{guevel17}, an implementation in Python of the subtraction algorithm described by Zackay et al. \citep{zackay16}. The $BV$ data and $gri$ data were calibrated to Vega magnitudes and AB magnitudes, respectively. 
 iPTF16asu was also observed with Telescopio Nazionale Galileo (TNG) in $gri$ at late epochs (see \citealp{whitesides17}).  
 The final SN magnitudes in each band will be made available on WISeREP \citep{yaron12}.
        
\section{Supernova light curves and colors}
\label{sec:analysis}

Figure~\ref{lc} provides all of the photometric observations obtained in the optical bands for our SN sample. We now proceed to the analysis of the 
28 SNe observed before peak brightness. The first step is to correct the observed light curves for time dilation and K-corrections, given their small but not negligible effects on the light-curve shape.

\subsection{K-corrections} 
        \label{sec:Kcorr}

        We first determined the observed peak epoch in the $r$ band ($t_{r}^{\rm max}$) , given that
most of our SNe were observed at peak in that filter. This was done by
fitting a polynomial to the $r$-band light curves. The peak epochs are
indicated with a dashed red line in Fig.~\ref{lc}. In the cases of iPTF16asu and iPTF17axg, the $r$-band peak epoch was estimated from the $g$-band peak epoch using the formula presented by \citet{taddia15}.

Armed with measured $t_{r}^{\rm max}$, we determined a phase $t$ for
each spectrum of the SNe in our sample. All phases are given in the
rest frame, after correcting for time dilation --- that is, correcting the
observed phase by a factor $(1+z)$. 
A log of the spectra is provided in Table~\ref{tab:speclog}, and all of the spectra are presented and analyzed in Sect.~\ref{sec:spectra}.  
With these spectra, we computed average K-corrections for the $Bgriz$
bands as functions of redshift and time since $t_{r}^{\rm max}$. 

\begin{figure}
 \centering
\includegraphics[width=16cm,angle=0]{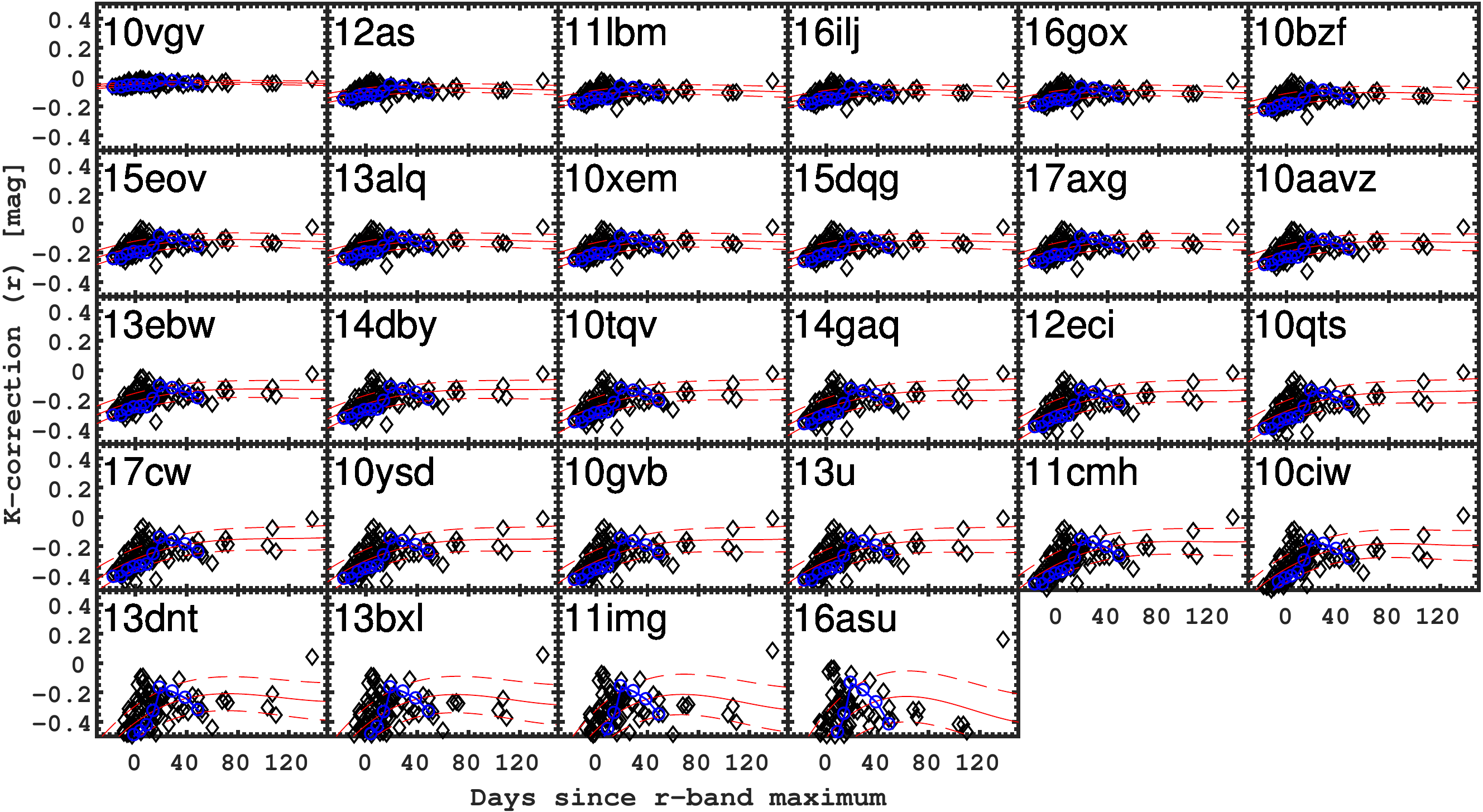}
  \caption{ \label{kcorr_r}K-corrections in the $r$ band for 
  our SN sample. For each SN redshift we determined the $K_{\rm corr}^{r}(z,t)$ $=$  $r_{\rm observed}$ $-$ $r_{\rm K-corrected}$ 
    from all the SN spectra in our sample, and we fit them with a
    second-order polynomial, shown as a solid red line. The 1$\sigma$ uncertainties are shown as red dashed lines. The SNe are ordered based on redshift, with the nearest ones on top. For the most distant events, K-corrections in $r$ have values up to 0.4 mag around peak. For comparison, in blue we show the K-corrections from the SN~Ic-BL spectral templates available at \href{https://c3.lbl.gov/nugent/nugent_templates.html}{Peter Nugent's Spectral Templates webpage}. The results are similar and compatible within the uncertainties.}
 \end{figure}

Ideally, K-corrections for each individual SN should be
calculated using a comprehensive spectral sequence for the given
object. Unfortunately, we have very few spectra per SN, but
instead have a large sample of SE~SN spectra in general.
We therefore proceeded in the following way.
For each given SN observed in a filter centered on $\lambda$,
(i) we considered the redshift $z$ of the SN, and shifted all the
rest-frame spectra in our entire sample to that redshift using
$f_{\lambda}^{\rm rest}=f_{\lambda}^{\rm observed}(1+z)$ and 
$\lambda_{\rm rest}~=~\lambda_{\rm observed}/(1+z)$.
(ii) Next, we computed for each spectrum the synthetic magnitude in the
given filter $\lambda$ at that redshift ($m_{z,t}$) and in the rest
frame ($m_{z=0,t}$). From each spectrum, we first removed the host galaxy
emission lines and we discarded those spectra that were strongly
contaminated by the host continuum. We also first corrected all of the spectra for MW extinction. 
 (iii) Then, we defined the K-correction from each spectrum in the sample as $K_{\rm corr}^{\lambda}(z,t)~=~m_{z,t}-m_{z=0,t}$, with $t$ being the phase of the spectrum. 
 (iv) We plotted all of the obtained $K_{\rm corr}^{\lambda}(z,t)$ as a
 function of phase and fit them with a second-order polynomial. The
 fits for the  $r$ band are shown in Fig.~\ref{kcorr_r}.
 The K-corrections are larger at early epochs. 
 (v) Once we obtained the K-correction polynomials in all of the bands and
 for each SN, we K-corrected all of the $Bgriz$ light curves by the
 values obtained with the interpolation of the polynomials at the
 epochs of the different light-curve observations.  The average
 1$\sigma$ uncertainty in the K-corrections between $-20$ and +40 days for
 the $r$ band is 0.07~mag. 

 In the following analysis, we always refer to our K-corrected and
 time-dilation-corrected light curves. K-corrections can be as large
 as $\sim0.5$ mag at peak brightness in the $r$ band (see Fig.~\ref{kcorr_r}),
 but for most of the events these corrections are smaller than 0.3 mag at the same
 phase and subsequently even lower. 
For comparison, we also obtained the $r$-band K-corrections from
Peter Nugent's spectral template of SNe~Ic-BL\footnote{available at this webpage:
\href{https://c3.lbl.gov/nugent/nugent_templates.html}{https://c3.lbl.gov/nugent/nugent\_templates.html}}, and we plot
these as well in Fig.~\ref{kcorr_r}. The results are compatible within the
uncertainties.  
 
\subsection{Light-curve shape} 
        \label{sec:lcshape}
        
Armed with the K-corrected light curves, we proceeded to fit the
$r$-band light curves with the function provided by \citet{contardo00}
to characterize their shape. This function includes an exponential
rise, a Gaussian peak, and a linear late decline, as also discussed in
\citet{taddia18}. Only the 27 SNe observed both before and after $r$-band peak were included; iPTF17axg was observed before peak but only in $g$ band. 
From the functional fit, it is easy to derive the peak epoch and peak magnitude, the decline parameter $\Delta m_{15}$, and the rise parameter $\Delta m_{-10}$, as well as the late linear-decline slope.

The results of our best fits to the $r$-band light curves are shown
in Fig.~\ref{contardo_r}. In the top panel, each SN is reported in
individual subpanels, whereas in the bottom panel we overplot all of our SNe to
better show the general shape of their light curves. The latter panel
highlights the large variety of rise and decline rates; broad and narrow light curves are part of our sample. iPTF15eov emerges as the broadest event, clearly separated from the rest of the SNe. 

We estimate uncertainties in each of the light-curve-fit parameters by
Monte Carlo simulating 1000 light curves according to their
photometric uncertainties and refitting them with the same function. The
standard deviations of the best-fit parameters were taken as the
uncertainty on each parameter. The parameters and their estimated
uncertainties are reported in Table~\ref{tab:contardo_r}.

 Figure~\ref{dm15dmm10_r} suggests a correlation between $\Delta
 m_{15}$  and $\Delta m_{-10}$, where  fast-rising objects are also
 fast decliners.  The correlation is characterized by a $p$-value of
 0.006 if we perform a Spearman correlation test. We also tested if
 $\Delta m_{15}$ correlates with the late-time linear slope, but given
 the small number of SNe observed at late epochs (seven SNe with a
 measured late-time slope and $\Delta  m_{15}$) we could not
 statistically confirm the correlation found by
 \citet{taddia18}. Our test gives a $p$-value of 0.55.

\begin{figure}
 \centering
\includegraphics[width=12cm,angle=0]{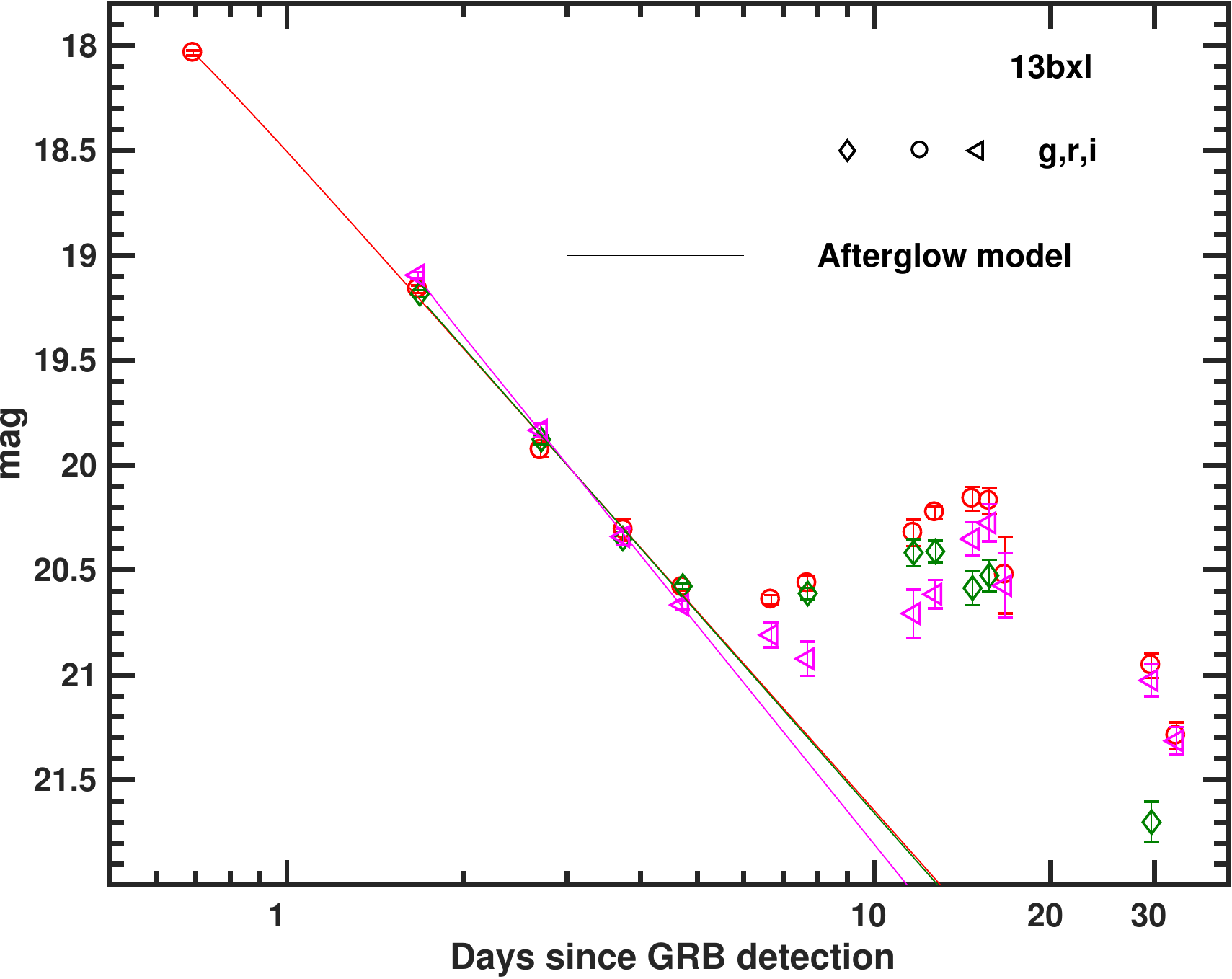}
  \caption{ \label{afterglow}Fit of the afterglow emission from the $gri$ light curves of iPTF13bxl.}
 \end{figure}

\begin{figure}
 \centering
\includegraphics[width=16cm,angle=0]{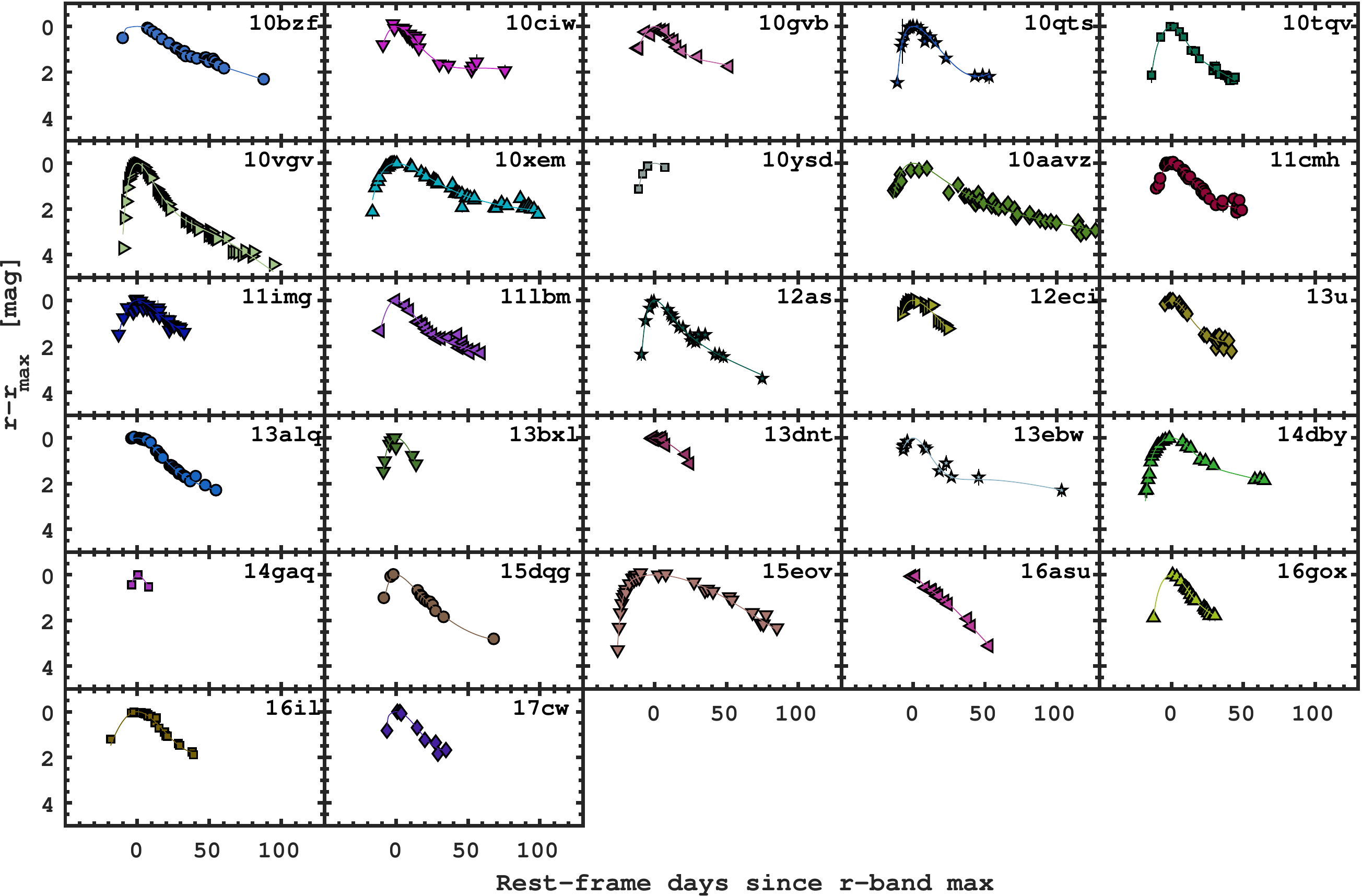}\\
\includegraphics[width=16cm,angle=0]{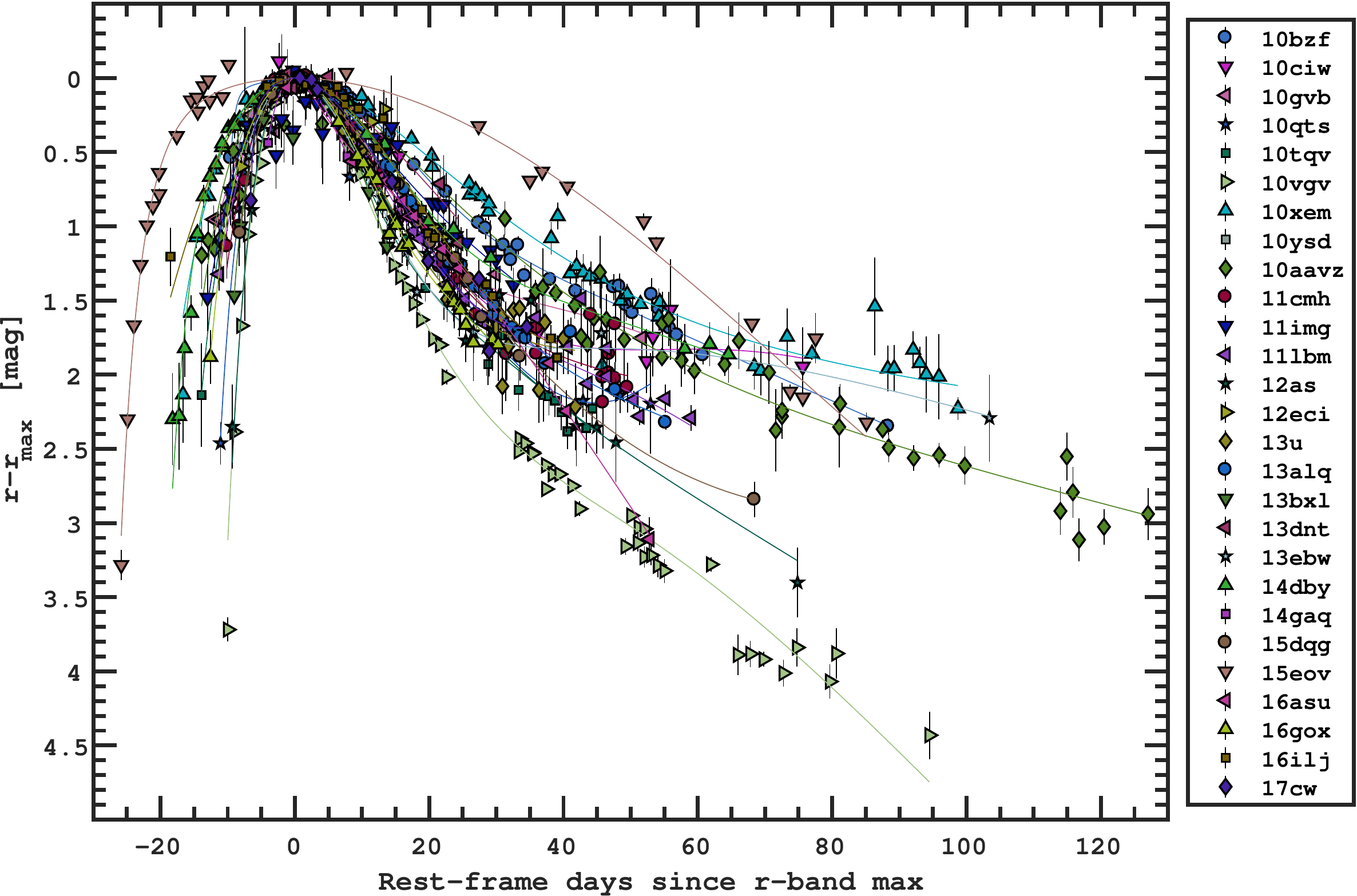}
  \caption{ \label{contardo_r}(Top panel) Rest-frame $r$-band light curves fit by the \citet{contardo00} equation. The best fits are shown as solid lines, and the light curves have been normalized to peak brightness. (Bottom panel) Same as in the top panel, but now all SNe were overplotted to highlight the variety of rise and decay timescales. }
 \end{figure}

\begin{figure}
 \centering
\includegraphics[width=12cm,angle=0]{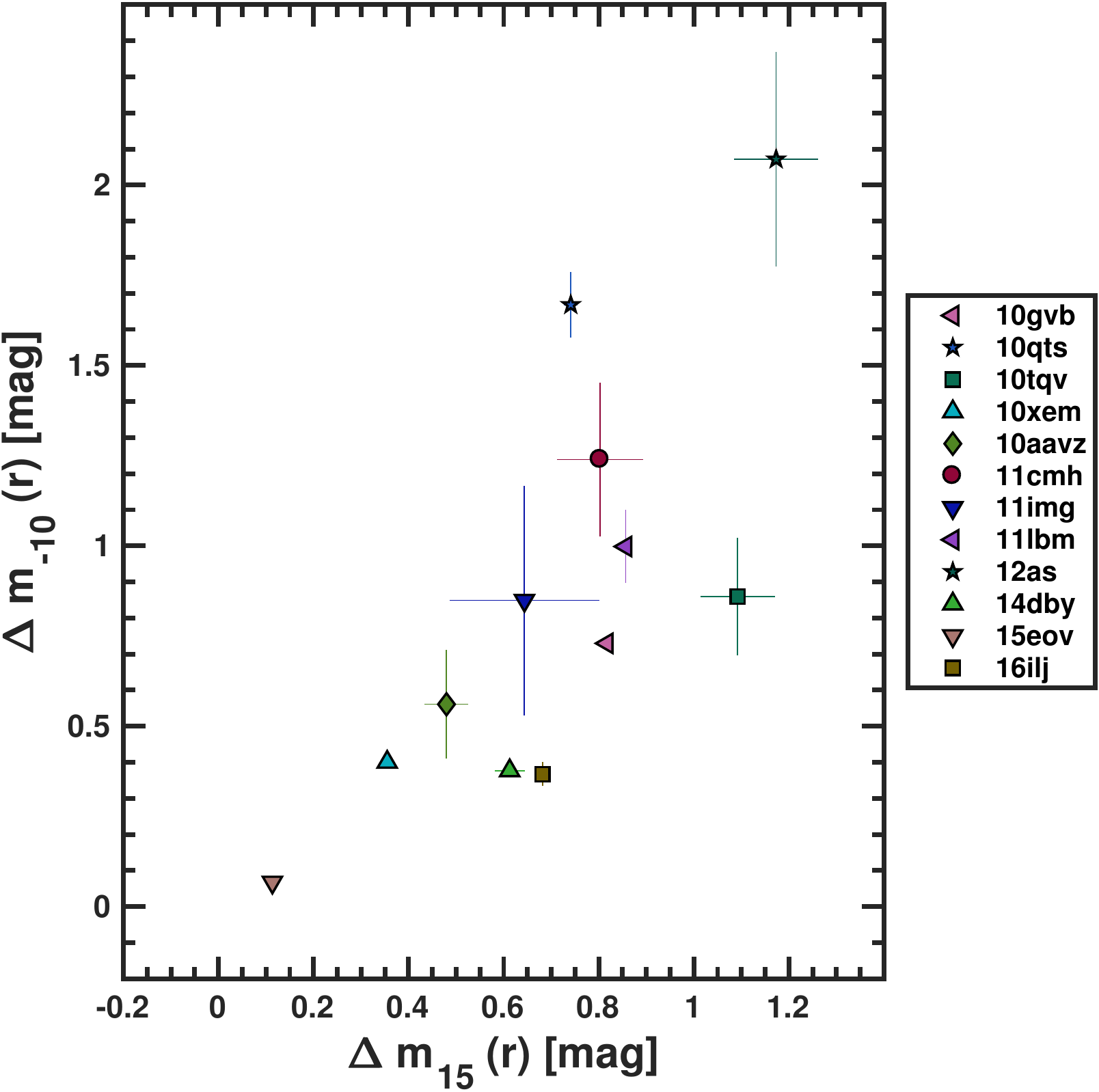}
  \caption{ \label{dm15dmm10_r}$\Delta m_{-10}$ vs $\Delta m_{15}$ in $r$ for the sample of SNe~Ic-BL. The correlation is significant ($p$-value = 0.006 using a Spearman test).}
 \end{figure}

 We note that in the case of iPTF13bxl, the light curves are
 characterized by a strong GRB afterglow in the optical that precedes the SN emission (see Fig.~\ref{afterglow}). This characteristic of the GRB SN iPTF13bxl was studied in detail by \citet{singer13}, \citet{delia15}, and \citet{toy16}, where the association of iPTF13bxl to GRB 130702A was investigated.
 Therefore, before performing the light-curve fit with the
 \citet{contardo00} function, we fit and removed the afterglow
 emission using the method presented by \citet[][see their
 Eq. 2]{cano11}. The afterglow fits to the epochs $< 5$~d since
 the GRB are indicated by solid lines in Fig.~\ref{afterglow} for the $gri$ light curves.

\subsection{Observed g-r colors and host extinction}
\label{sec:hostextinction}

We proceeded with the colors of our SNe, in particular $g-r$. We made use of the $g-r$ colors to estimate the host extinction, following the approach of \citet{stritzinger17}. 

For each SN with both $g$ and $r$
observations both before and after $r$-band maximum,
we interpolated the $r$ observations to the $g$-band epochs, and plotted
the obtained $g-r$, after having corrected both bands for 
MW extinction. This was done adopting the MW $E(B-V)$ given in Table~\ref{tab:sample}, assuming $R_V=3.1$ and the \citet{fitzpatrick99} reddening law. 

\begin{figure}
 \centering
\includegraphics[width=16cm,angle=0]{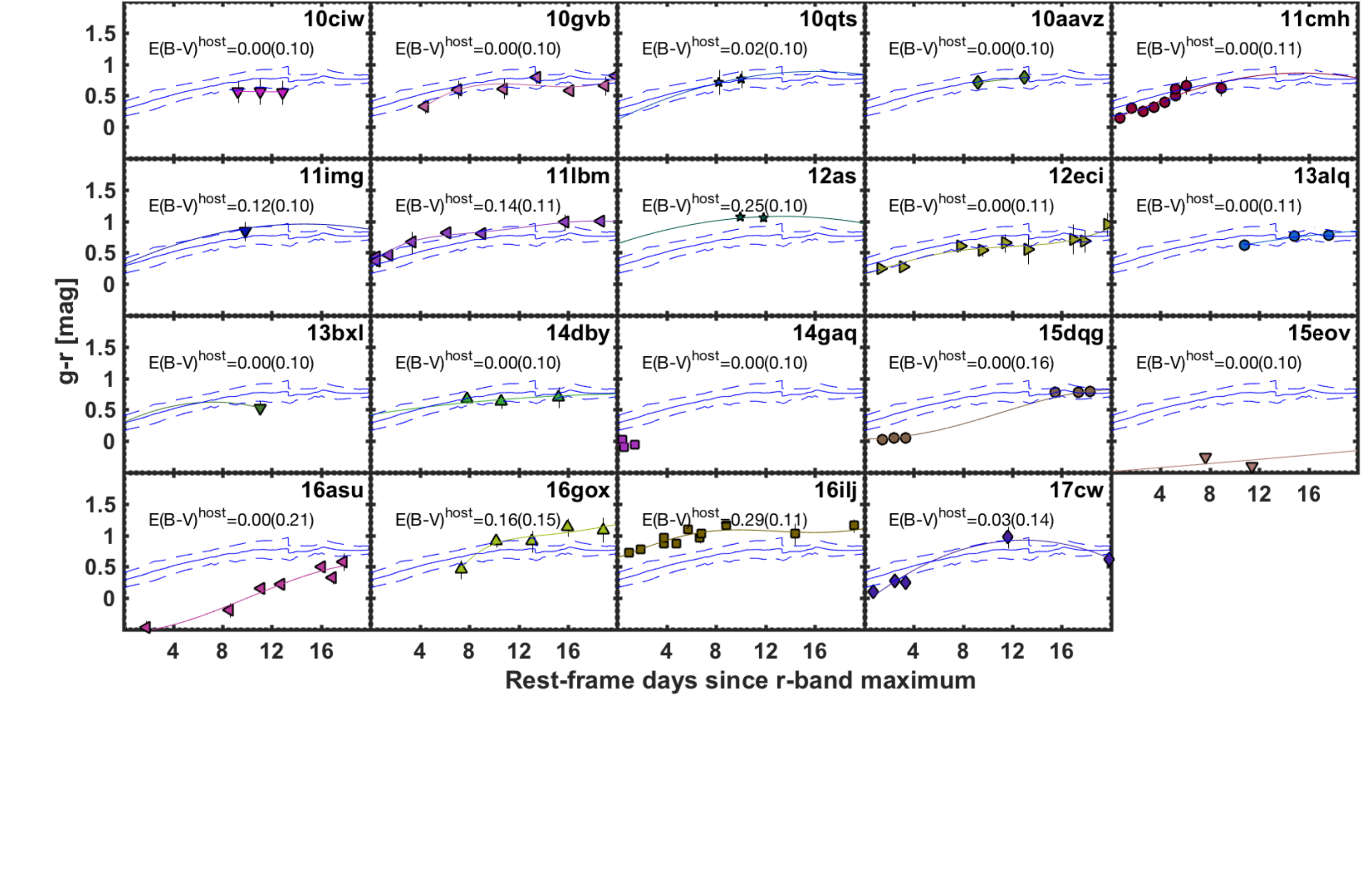}
  \caption{ $E(B-V)^{\rm host}$ from the $g-r$ colors measured between 0 and 20~d after $r$-band peak brightness and corrected only for the MW extinction. The blue curves represent the intrinsic color and its uncertainty (obtained from the dereddened $g-r$ colors of six SNe~Ic-BL from the literature \citep[see][their Fig. 10]{prentice16}. Each SN color is fit with a low-order polynomial (colored curves with color matching the symbol color) in order to compare it to the intrinsic color and compute the average $g-r$ color excess.\label{gmr020}}
 \end{figure}

The $g-r$ colors in the time interval between 0 and 20~d after
$r$-band maximum are shown in Fig.~\ref{gmr020}. We also illustrate a
$g-r$ color template between 0 and 20~d after $r$ maximum,
which we obtained from the extinction-corrected $g-r$ colors presented 
for six relatively nearby ($z<0.05$) SNe~Ic-BL
\cite[][see their Fig.~10]{prentice16}. 

The $g-r$ colors of the SNe presented by
\citet{prentice16} were corrected for the host extinction mainly based
on the measurement of narrow \ion{Na}{i}~D absorption in the SN spectra. 
We took the average and standard deviation of these colors at each epoch between 0 and 20~d as the intrinsic color template. The standard deviation of our $g-r$ template is $\sim 0.1$ mag. 

An alternative approach to obtaining the intrinsic color for SNe~Ic-BL
would be to identify apparently unreddened events in our sample and
choose their colors as intrinsic.
We could identify, among
the objects with the bluest colors 
that also had no detected \ion{Na}{i}~D at
the host galaxy rest wavelength, those with absolute magnitude
(as computed assuming no host extinction) not too far from the middle of the
absolute-magnitude distribution in our sample. 
Similar assumptions were made in
the selection of the intrinsic colors of SE~SNe by
\citet{stritzinger17}, but they only had two SNe~Ic-BL. 
For our sample,
PTF12eci would 
match these criteria, 
and we could therefore assume its $g-r$ color as the
intrinsic $g-r$ color for SNe~Ic-BL.
In fact, the $g-r$ color of PTF12eci (see Fig.~\ref{gmr020}) turns out to be consistent, within the errors, with the $g-r$ template that we obtained from the dereddened $g-r$ color of the best-observed SNe~Ic-BL from the literature 
\citep[from][]{prentice17}.
Therefore, we proceed assuming the $g-r$ template as the intrinsic $g-r$ color.  

We fit the $g-r$ colors of all the SNe with 
low-order polynomials, shown as solid, colored lines in Fig.~\ref{gmr020}. We then
computed the average $E(g-r)$ for each of our SNe in the range between
0 and 20~d since peak brightness by computing the average difference between
the fit to the observed $g-r$ color and the assumed intrinsic $g-r$ color.
Next we converted
$E(g-r)$ into $E(B-V)~{\rm host}$ assuming $R_V^{\rm host}=3.1$.
Using the canonical $R_V^{\rm host}=3.1$ allows for easy comparisons
with other work in the literature. We also notice that
for the only SN~Ic-BL (SN~2009bb) for which 
\citet{stritzinger17} 
could estimate the $R_V^{\rm host}$, the value was consistent with the
usual 3.1 (it was $3.3^{+0.4}_{-0.3}$). 

The computed $E(B-V)^{\rm host}$ are reported in Table~\ref{tab:sample} and shown in Fig.~\ref{gmr020}. The uncertainties in the host galaxy extinction include the uncertainty of the $g-r$ template, and they also account for the standard deviation of the difference between each epoch of the intrinsic and  measured $g-r$ color. For the SNe without observed $g-r$ color in the range between 0 to 20~d past maximum, we adopted the host extinction obtained from the measured equivalent width of \ion{Na}{i}~D, reported in Table~\ref{tab:sample}, and then used Eq.~1 of \citet{taubenberger06}. For those cases, we adopted an uncertainty of $E(B-V)=0.2$~mag. If no \ion{Na}{i}~D was detected and no $g-r$ color excess could be computed, we assumed the host extinction to be negligible.  This was done for 11 SNe, which are those in Table~\ref{tab:sample} without indications for $E(B-V)^{\rm host}$. 
We notice that the colors of iPTF15eov are particularly blue, as are
those of iPTF16asu. These two SNe are also 
peculiar in other
respects within our sample --- iPTF15eov because of its unusually broad
light curve and its extraordinary large peak brightness (see
Sect.~\ref{sec:absmag}), and iPTF16asu for its nonstandard light curve
with an unprecedented fast rise (as discussed in detail by \citealp{whitesides17}).

\subsection{Absolute magnitudes}
\label{sec:absmag}

\begin{figure}
 \centering
 $\begin{array}{cc}
\includegraphics[width=11cm,height=7cm,angle=0]{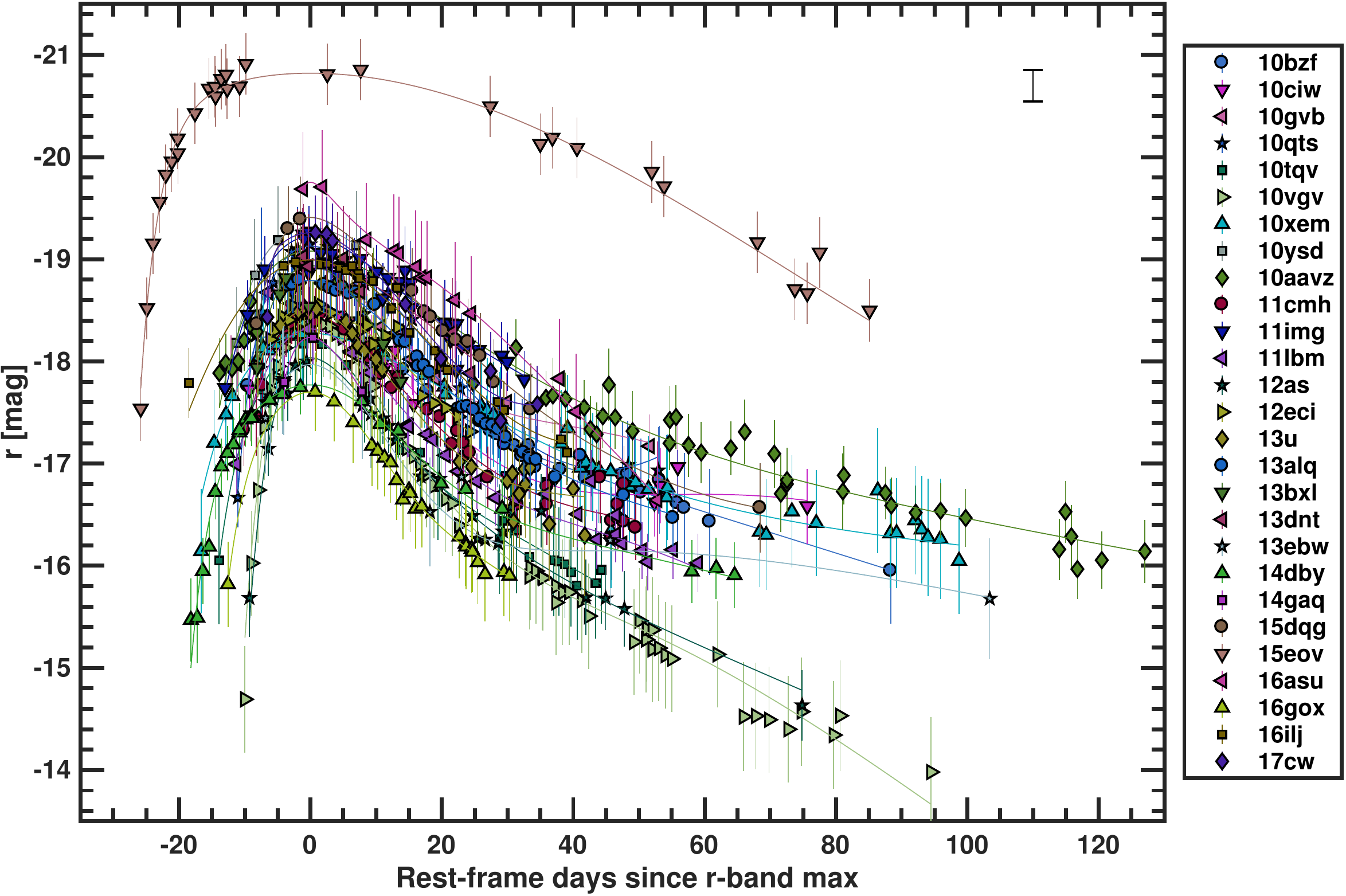}&
\includegraphics[width=7cm,height=7cm,angle=0]{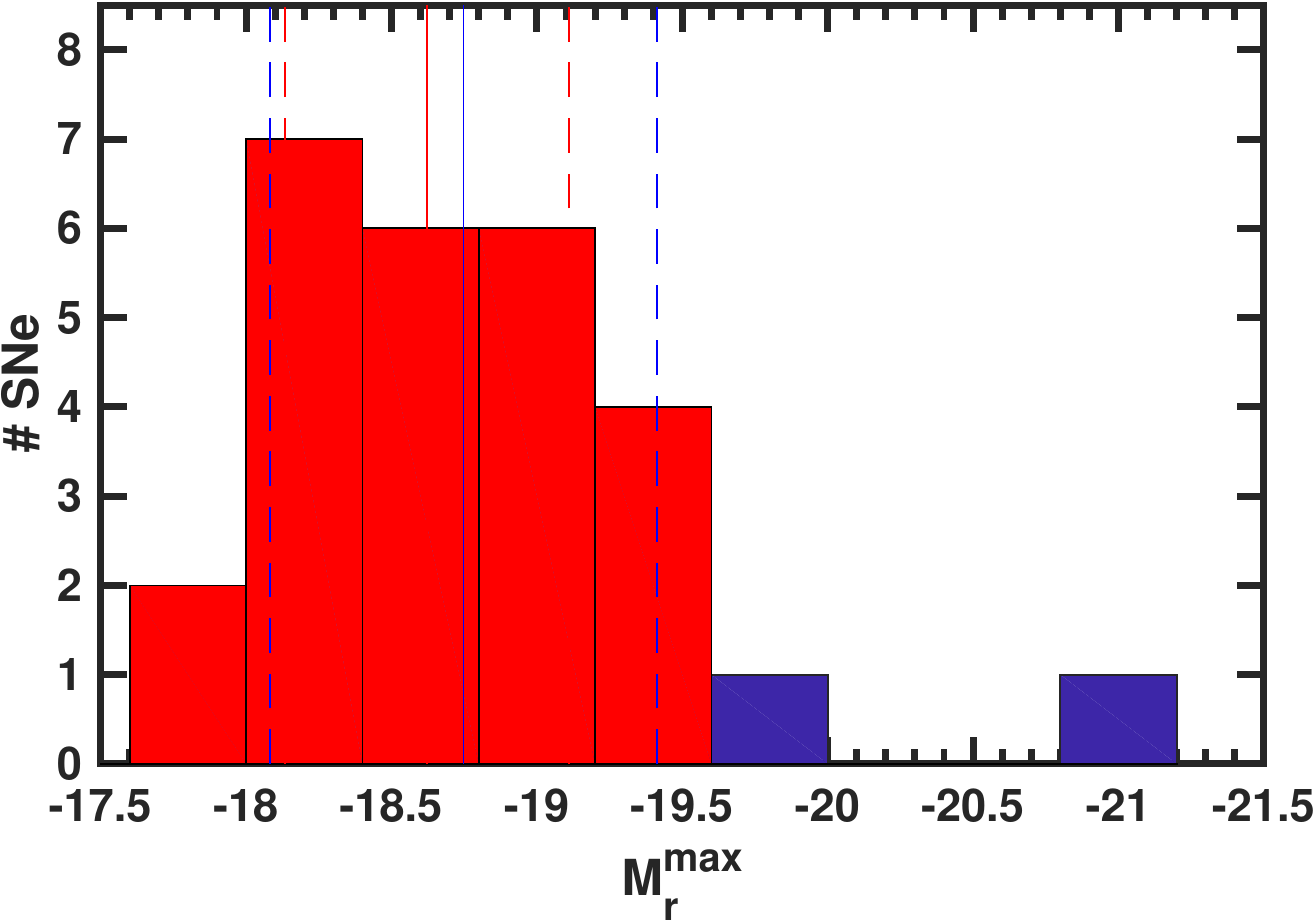}
\end{array}$ 
  \caption{(Left panel) Absolute $r$-band magnitude light curves,
    after (host + MW) extinction correction and K-correction. The best
    fits are the same as those presented in Fig.~\ref{contardo_r}. The systematic uncertainty in the distance is shown in the upper-right corner. (Right panel)
    Distribution of the absolute $r$ peak magnitudes for our sample. iPTF15eov and iPTF16asu are marked in blue. Blue vertical lines indicate the average and the standard deviation of the distribution when including all of the objects. 
    Red vertical lines mark the average and the standard deviation of the distribution when excluding iPTF15eov and iPTF16asu. \label{absmag_r}}
 \end{figure}

With the computed host galaxy extinctions, distances, and K-corrected
light curves, we proceeded to compute the absolute-magnitude light
curves, which for the $r$ band are shown in the left-hand panel of
Fig.~\ref{absmag_r}. The uncertainty in the absolute magnitudes
accounts for that in the host extinction and in the
photometry. We further report the systematic uncertainty from the distance ($\pm0.15$~mag) in the upper-right corner of the left panel in Fig.~\ref{absmag_r}. In the right-hand panel of Fig.~\ref{absmag_r},
the distribution of the peak $r$-band absolute magnitudes is
shown. Our sample ranges over 3~mag
in {$M_{r}^{\rm max}$}, 
between $-17.7$ and $-20.8$ mag, with an average of $-18.7\pm0.7$
mag. iPTF15eov is clearly brighter than all of the other SNe, $\sim1$ mag
brighter than the second brightest (iPTF16asu). Its $r$-band peak is $4\sigma$ brighter than the average of the other SNe in our sample.

If we exclude iPTF15eov and iPTF16asu from the average, as they might
not belong to the sample of normal SNe~Ic-BL (see
Sect.~\ref{sec:prog}),
the average peak is $-18.6\pm0.5$ mag. 
If we compute the average peak absolute magnitude by weighting for the inverse effective search volume for each SN, we obtain $-18.4\pm0.4$ mag. This value should be closer to the intrinsic absolute magnitude average since it takes into account the volume effect.

\begin{figure}
 \centering
$ \begin{array}{cc}
\includegraphics[width=8cm,angle=0]{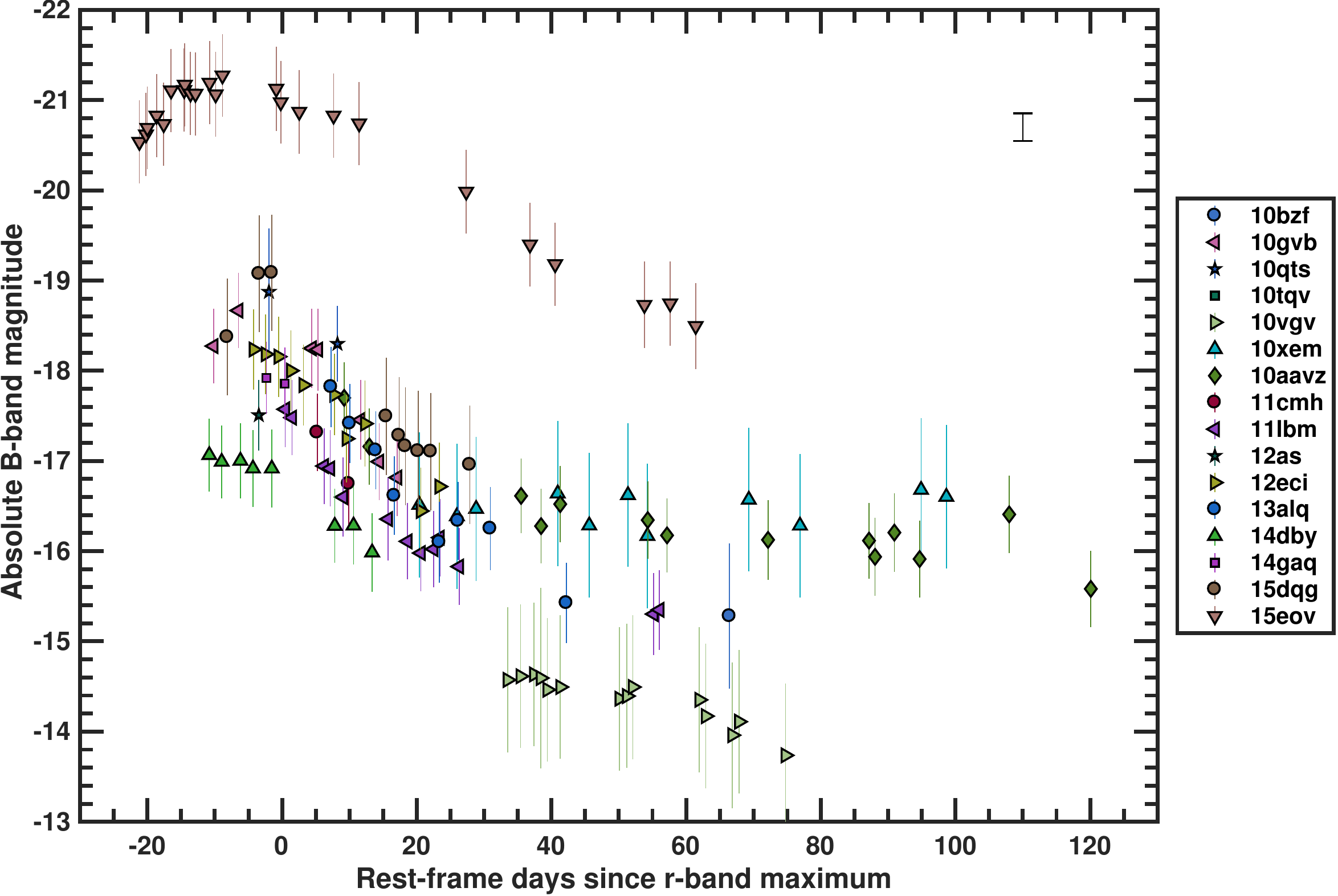}&
\includegraphics[width=8cm,angle=0]{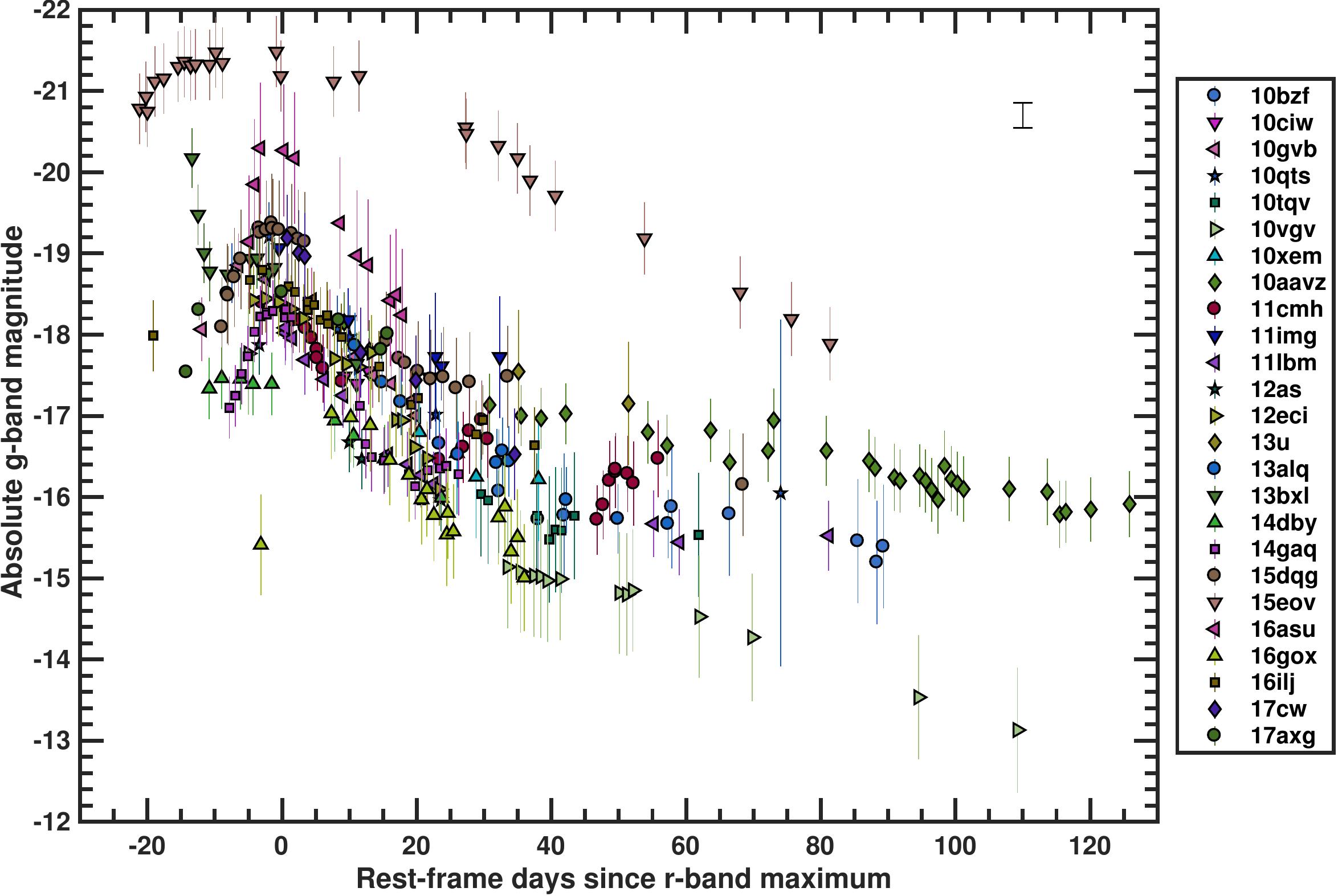}\\
\includegraphics[width=8cm,angle=0]{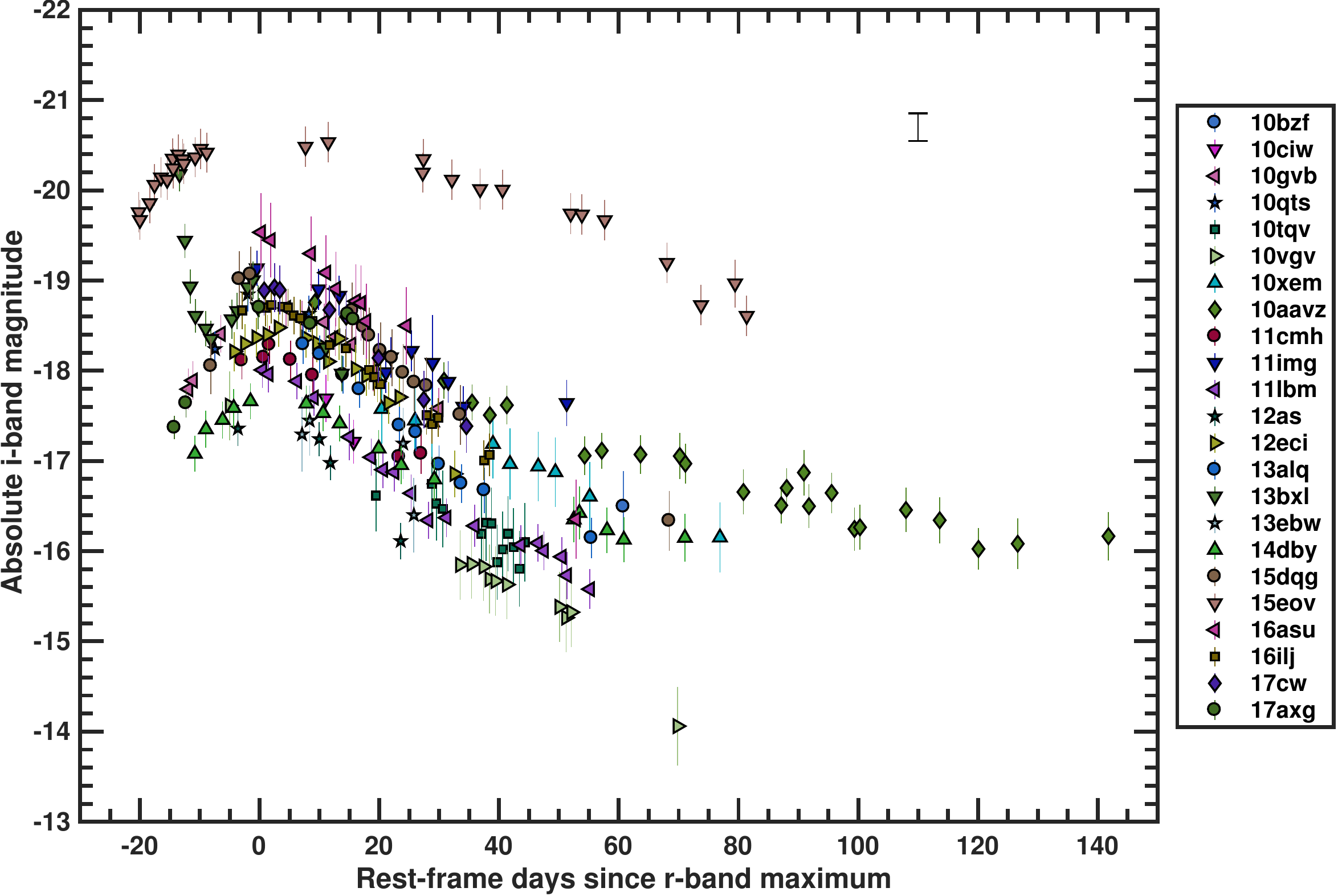}&
\includegraphics[width=8cm,angle=0]{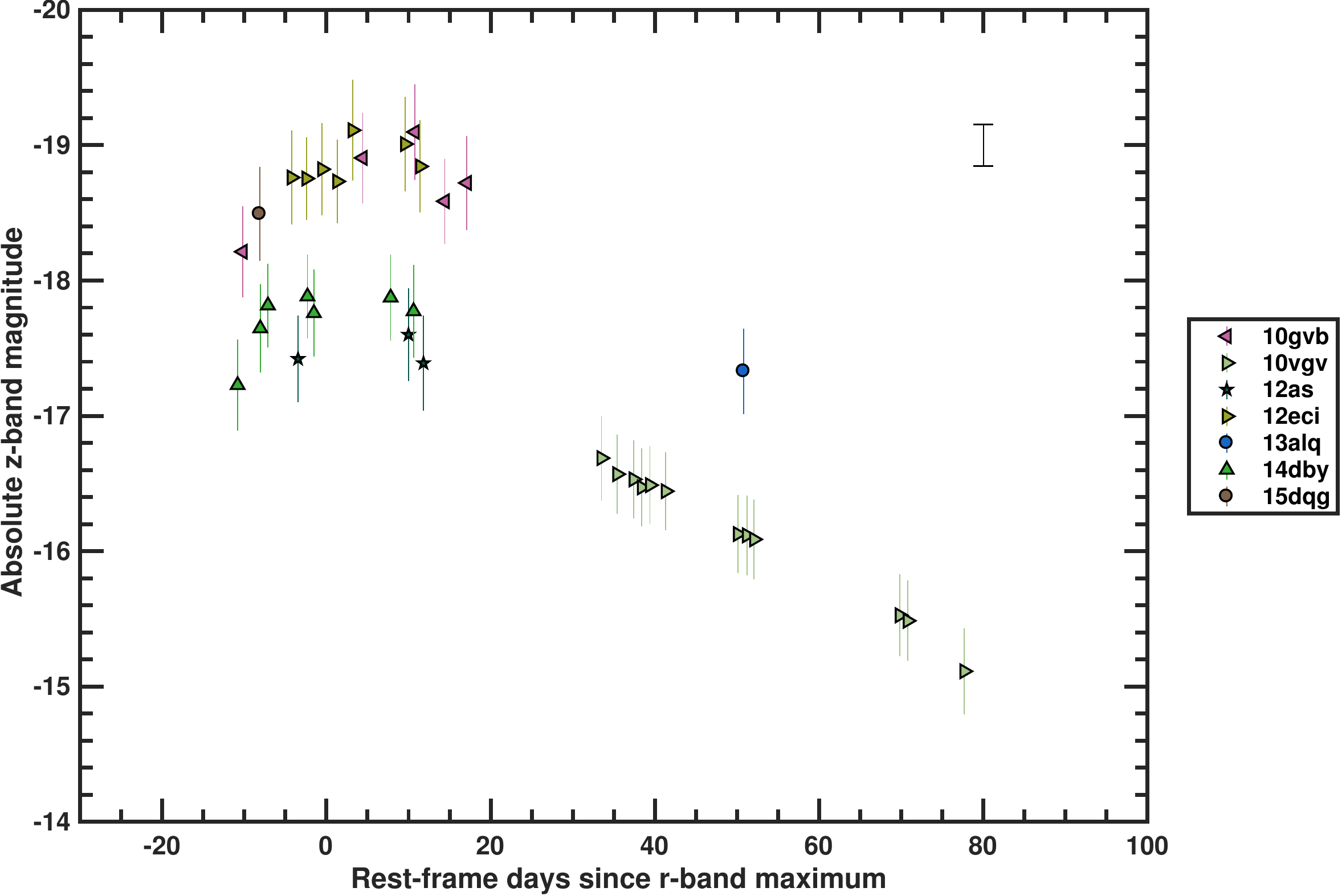}
\end{array}$
  \caption{ Absolute $Bgiz$ magnitudes, after (host + MW) extinction correction and K-correction.\label{absmag_other}}
 \end{figure}

All of the $r$-band peak magnitudes are reported in Table~\ref{tab:contardo_bolo}, where the uncertainty accounts for the error on the host extinction and on the fit to the light curve. An additional systematic uncertainty of 0.15 mag due to the systematic uncertainty in the distance should also be considered.
The spread of 
absolute peak magnitudes is larger than the total uncertainty in the measured peaks, implying that there is an actual range of peak luminosity in this family of transients.
The distribution of the peak $r$-band absolute magnitudes is shown in the left panel of Fig.~\ref{absmag_r}. 
In the other bands ($Bgiz$), we have fewer observations that include the
peak of the light curves. For completeness, we plot their absolute
magnitudes in Fig.~\ref{absmag_other} as a function of time since
$r$-band maximum.

\begin{figure}
 \centering
\includegraphics[width=12cm,angle=0]{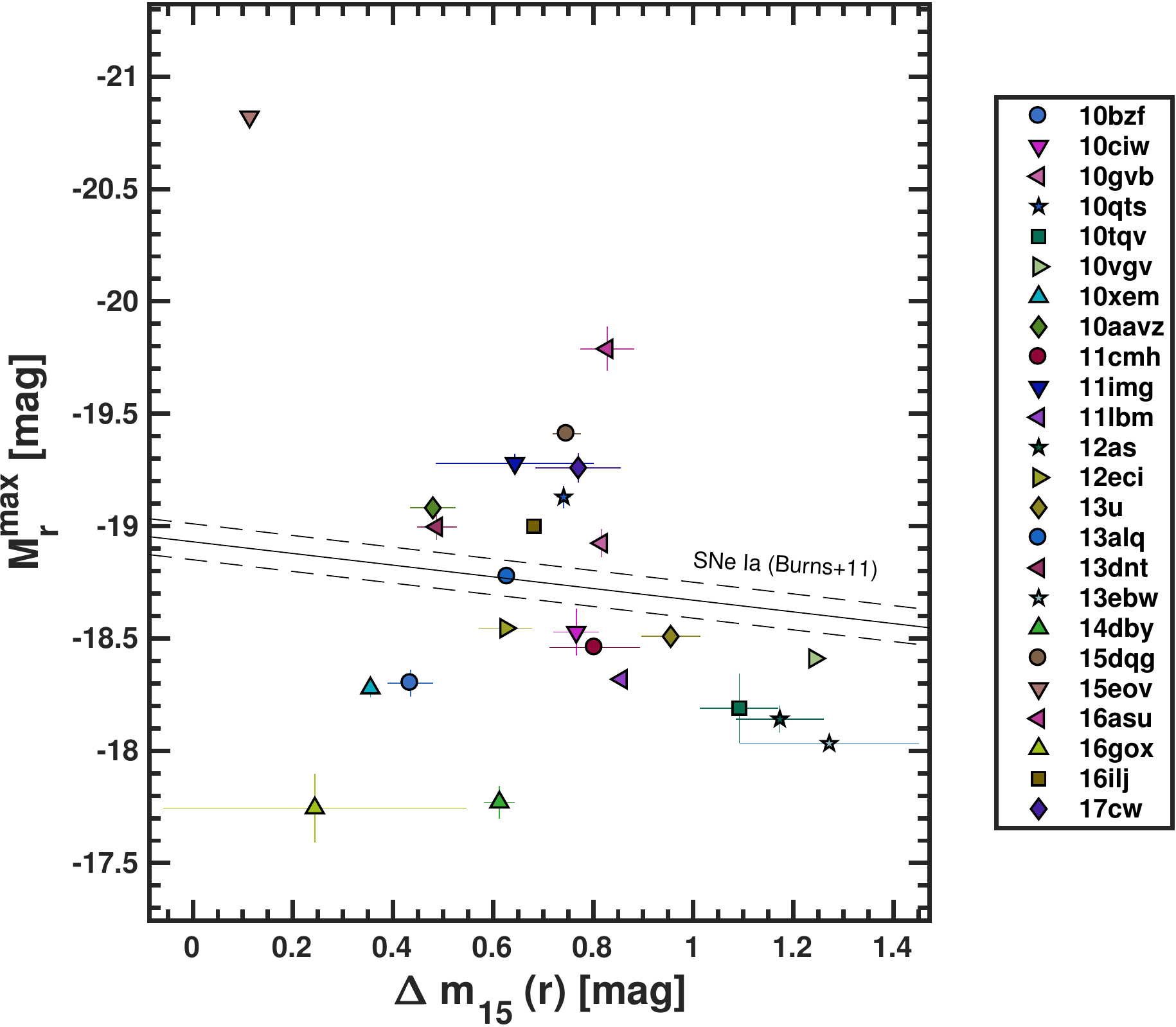}
  \caption{ \label{phillips_r}Absolute peak
    magnitude in the $r$ band vs. $\Delta m_{15}(r)$ for the sample of
    23 SNe~Ic-BL. The Phillips relation from \citet{burns11} for the
    $r$ band is shown as a solid line, with the dashed lines
    representing the error bars. Our data show no evidence for such a relation.
  }
 \end{figure}

In Fig.~\ref{phillips_r} we plot the absolute $r$ magnitude peak
{$M_{r}^{\rm max}$}
versus $\Delta m_{15}(r)$ to test if there is a Phillips relation \citep{phillips93} 
as in SNe~Ia. Such a relation was also
seen for GRB~SNe by \citet{cano14} and
\citet{li14}.
The SNe~Ic-BL in our sample do not follow such a relation (shown by a
black line from \citealp{burns11} in the figure). It is not always the
case that the less luminous SNe are also those that decline faster.  This
was also seen by \citet{cano14} for SNe~Ic-BL that are not
associated with GRBs and for other SE~SNe. In addition, we tested if there is
any correlation between {$M_{r}^{\rm max}$}
and $\Delta m_{40}(r)$, but did not find any such correlation.
Furthermore, the peak $r$-band magnitudes do not correlate with the slopes at late epochs.

\subsection{Colors}
\label{sec:color}

\begin{figure}
 \centering
 $\begin{array}{cc}
 \includegraphics[width=9cm,angle=0]{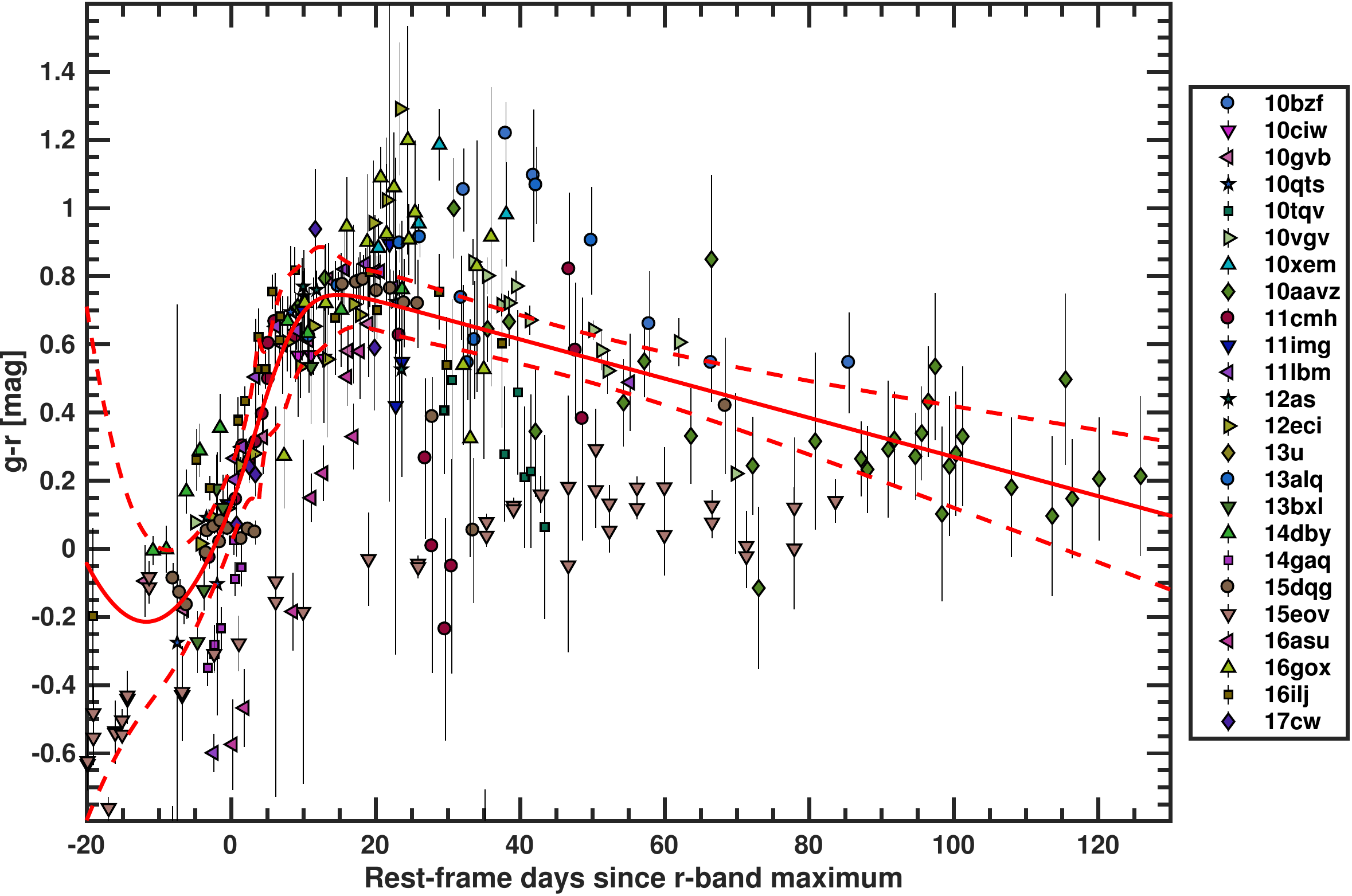}&
\includegraphics[width=9cm,angle=0]{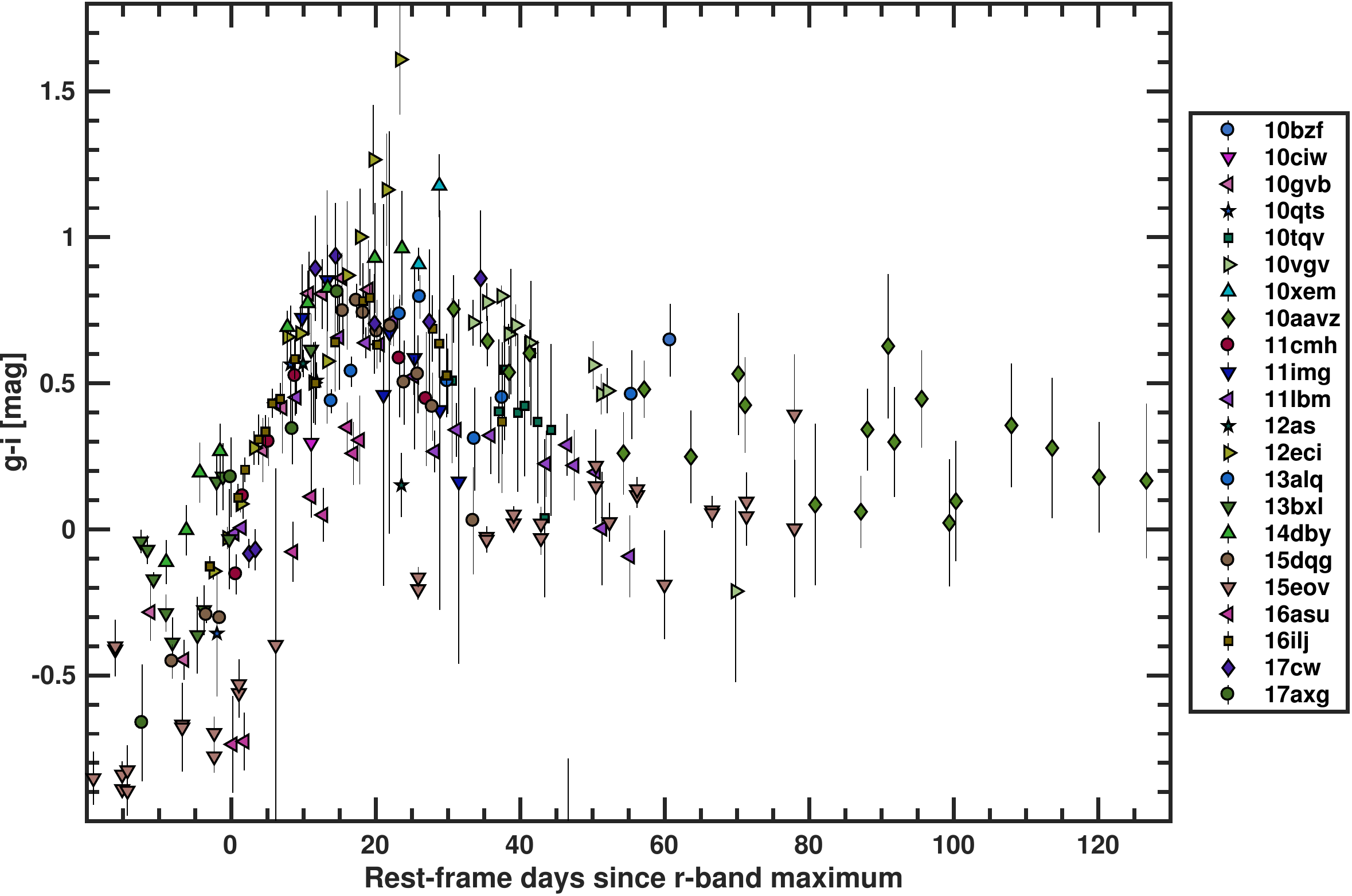}\\
\includegraphics[width=9cm,angle=0]{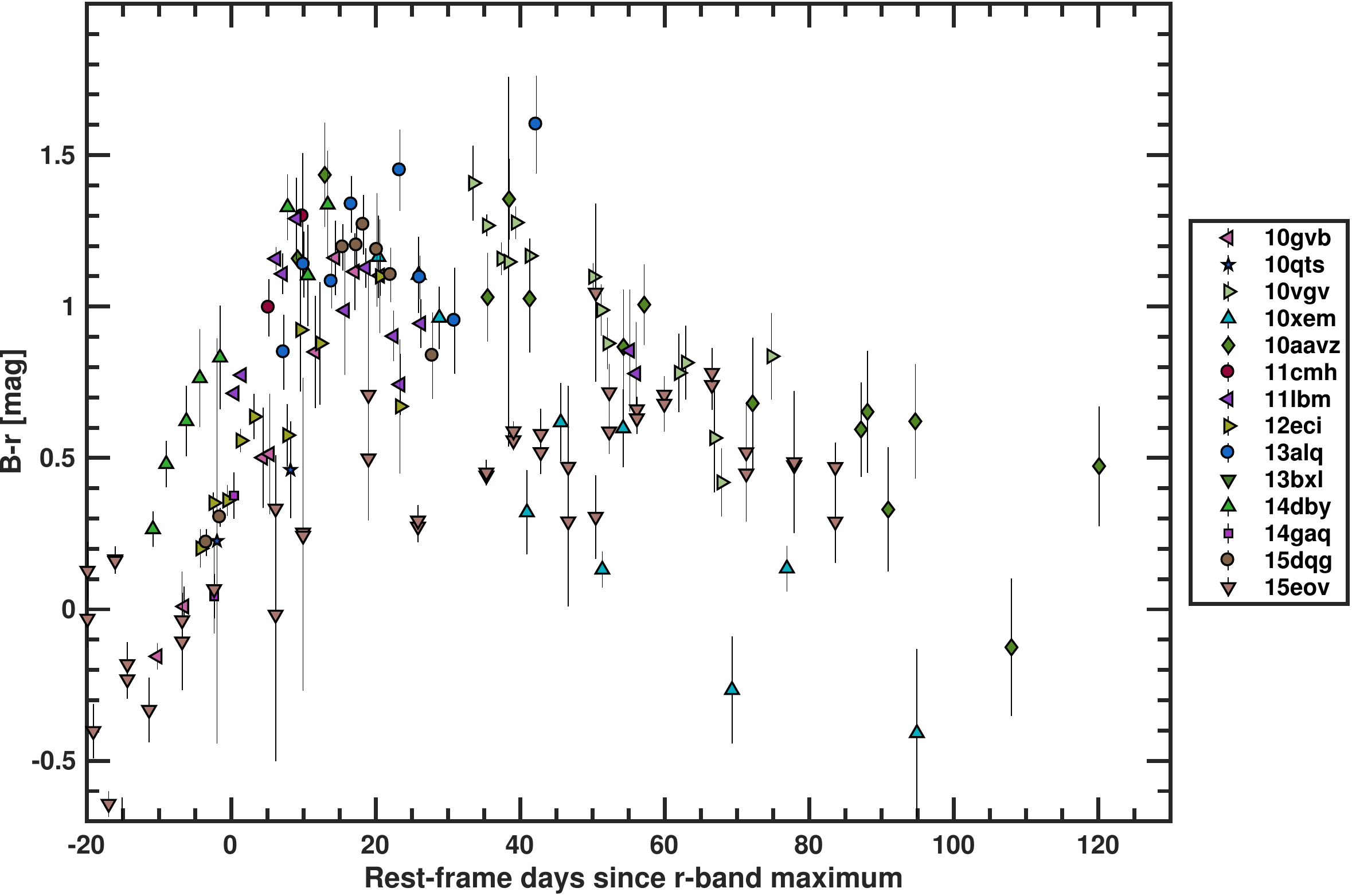}&
\includegraphics[width=9cm,angle=0]{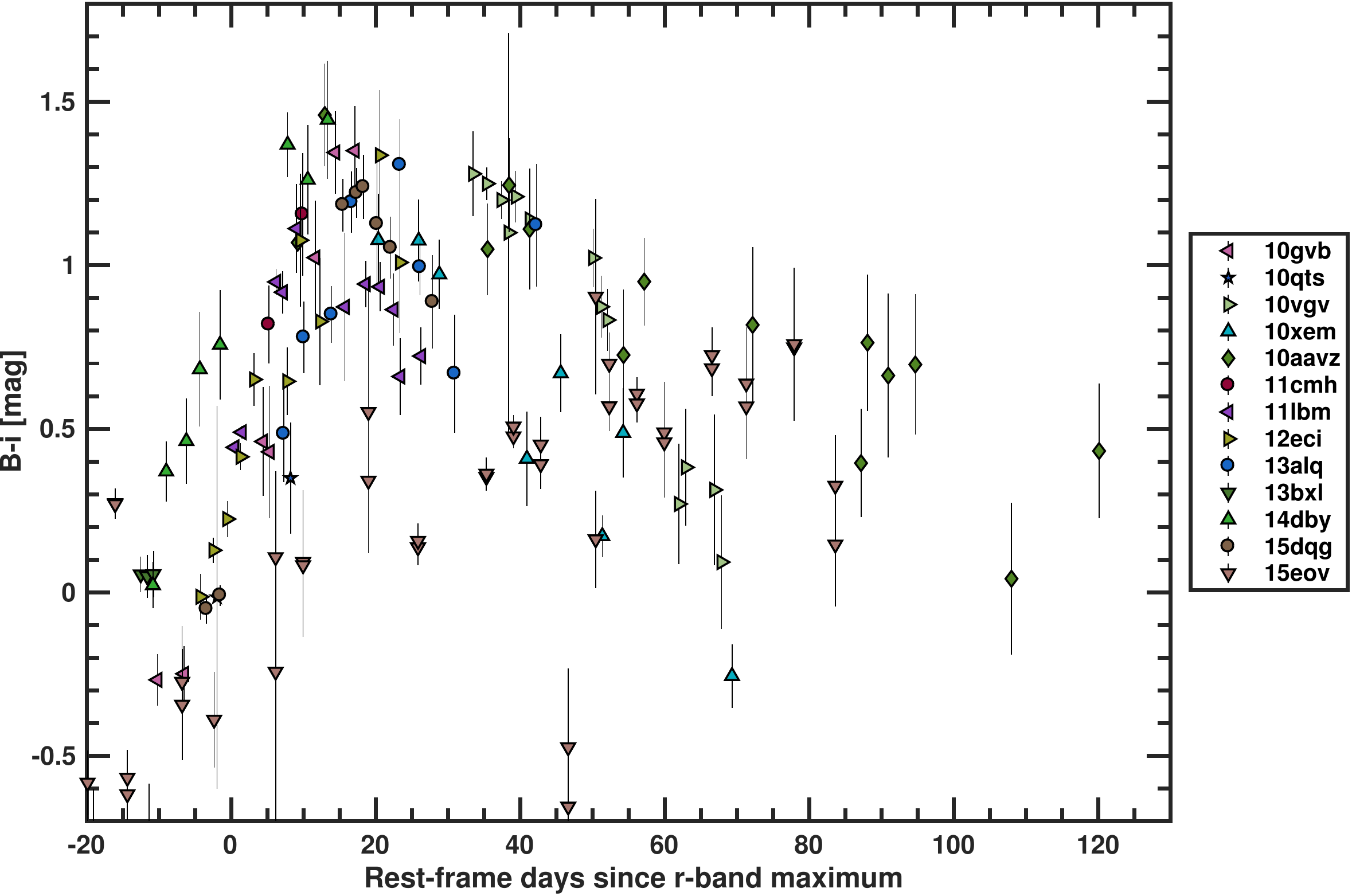}\\
\end{array}$
  \caption{ $g-r$, $g-i$, $B-r$, and $B-i$ color evolution for our SNe~Ic-BL. The $g-r$ color is fit with the expression used by \citet{burns14} and \citet{stritzinger17}, and the best fit with its 3$\sigma$ uncertainties is shown by red lines. All of the colors are MW and host-galaxy extinction corrected.\label{color}}
 \end{figure}

We compute the intrinsic $g-r$ colors of our SNe including the
corrections for host extinction. The intrinsic colors are shown in the
upper right panel of
Fig.~\ref{color}. In the other panels, we provide some additional
colors: $g-i$, $B-r$, and $B-i$. All colors tend to exhibit a
sharp rise from a few days before the epoch of  $r$-band peak to $\sim 20$~d after peak,
after which they start a slow decline to lower (bluer) values. 
This behavior becomes apparent when we fit $g-r$ with the functional
formula used by \citet{burns14} and \citet{stritzinger17}, and display
the best fit and 3$\sigma$ uncertainties with thick red lines. The
epochs before $r$-band peak are fit with a second-order polynomial. 
We (again) excluded iPTF15eov and iPTF16asu from the fit, as they are
remarkably bluer and also display a different evolution.

\begin{figure}
 \centering
 \includegraphics[width=16cm,angle=0]{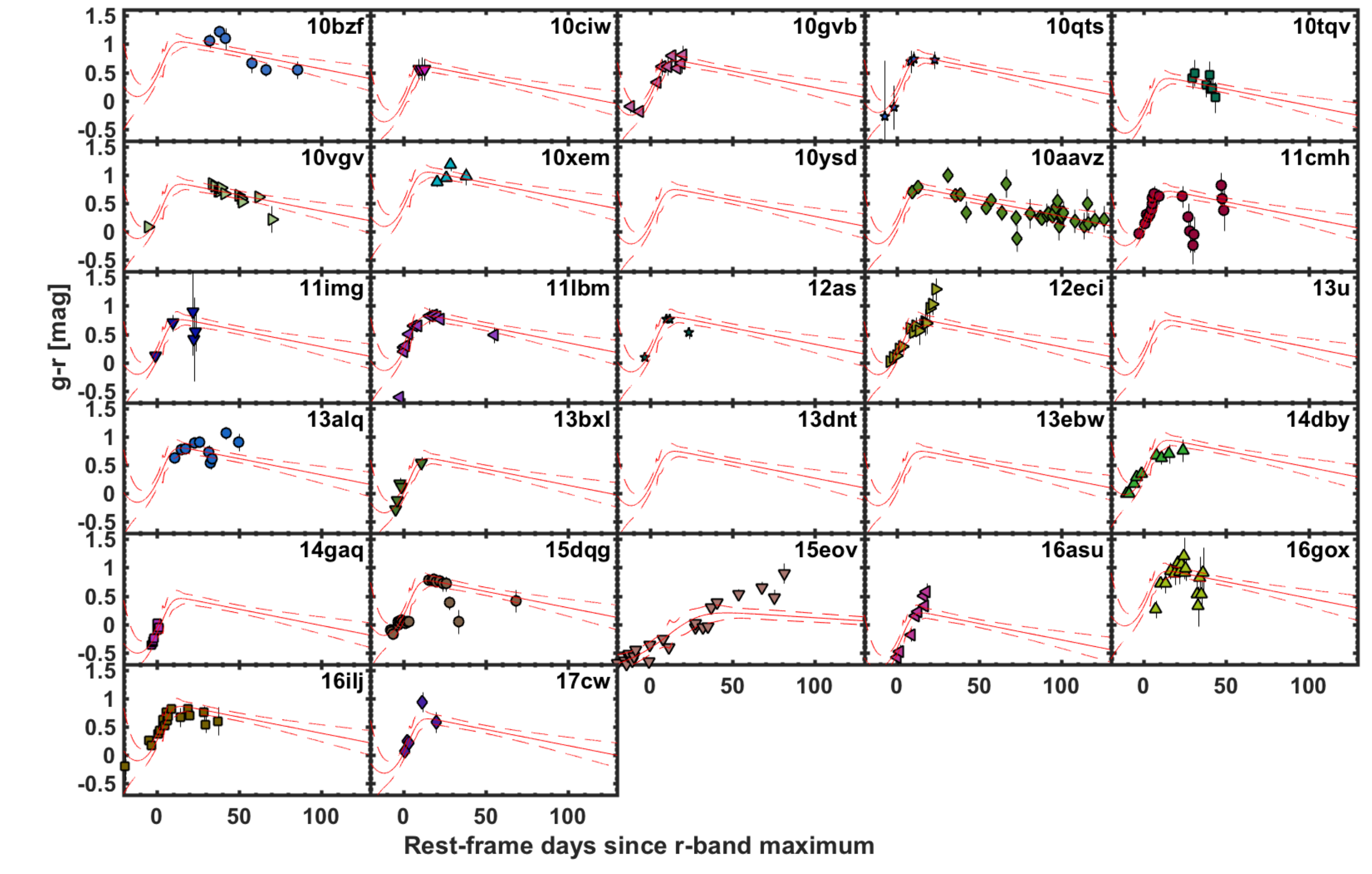}
  \caption{ Individual $g-r$ color evolution for our SNe, fit with the $g-r$ template shown in Fig.~\ref{color}. In the case of PTF10ysd, iPTF13u, iPTF13dnt, and iPTF13ebw, where we have no significant $g$-band observations, we assume the template of Fig.~\ref{color} to be their $g-r$ color.\label{color_fit}}
 \end{figure}

To provide a continuous description of the $g-r$ color of all our SNe,
we fit their individual $g-r$ colors with the function presented as
a red curve in Fig.~\ref{color}, using it as a template and allowing
it to shift up and down. In the case of iPTF15eov, where the color
evolution is clearly much slower than for the rest of the SNe, we also stretched the color template in time by a factor of 3.125 to fit the $g-r$ color of this event.
The best fits are presented in Fig.~\ref{color_fit}, and they 
are used to build the bolometric light curve in Sect.~\ref{sec:bololightcuvre}.

\section{Supernova spectra}
\label{sec:spectra}

The spectra of our SNe are the basis for their classification as SNe~Ic-BL, as discussed in Sect.~\ref{sec:sample}. 
As shown in Table~\ref{tab:speclog}, most of these 121 spectra were obtained using Keck~I and P200, equipped with LRIS \citep{oke95} and DBSP \citep{oke82}, respectively. Other telescopes were also used to obtain spectra: Keck~II equipped with DEIMOS \citep{faber03}, Gemini North with GMOS \citep{hook04}, 
Lick 3~m Shane with the Kast spectrograph \citep{miller93}, Kitt Peak National Observatory (KPNO) 4~m telescope with RCSpec, Hobby-Eberly Telescope (HET) with LRS \citep{hill98}, William Herschel Telescope (WHT) with ACAM \citep{benn08}, Apache Point Observatory (APO) telescope with DIS, TNG with DOLoRes, Magellan~I with IMACS \citep{dressler11}, University of Hawaii 2.2-meter telescope (UH88) with SNIFS \citep{lantz04}, Faulkes Telescope North (FTN) with FLOYDS, Nordic Optical Telescope (NOT) with ALFOSC, P60 with SEDM \citep{nadia17}, and Discovery Channel Telescope (DCT) with DeVeny/LMI. 
When possible, the spectra were obtained with an atmospheric dispersion compensator
(e.g., Keck LRIS) or with the slit along the parallactic angle to help ensure
accurate relative spectrophotometry \citep{filippenko82}.
The spectra were reduced in a standard manner, including wavelength calibration using lamp exposures and flux calibration with spectral standards observed the same night. 

In the following, we present the SN spectral sequences and their properties. 
From the spectra we want in particular to estimate the photospheric
velocities, which are later used for the modeling. 

\subsection{Spectral sequences}
\label{sec:specseq}

We present all spectra of our SNe in Appendix~\ref{appendix}, in
Figs.~\ref{spec_seq1} to \ref{spec_seq13}. Each spectrum is shown in
the rest frame, and its phase in rest-frame days since $r$-band
maximum is reported next to it. For the SNe discovered later than 
$r$-band maximum, we instead report the Julian date next to each
spectrum (Figs.~\ref{spec_seq12} and \ref{spec_seq13}). All spectra in
these sequences have been corrected for MW and host galaxy extinction.

\subsection{Spectral temperature}
\label{sec:spectemp}
\begin{figure}
 \centering
$ \begin{array}{c}
\includegraphics[width=12cm,angle=0]{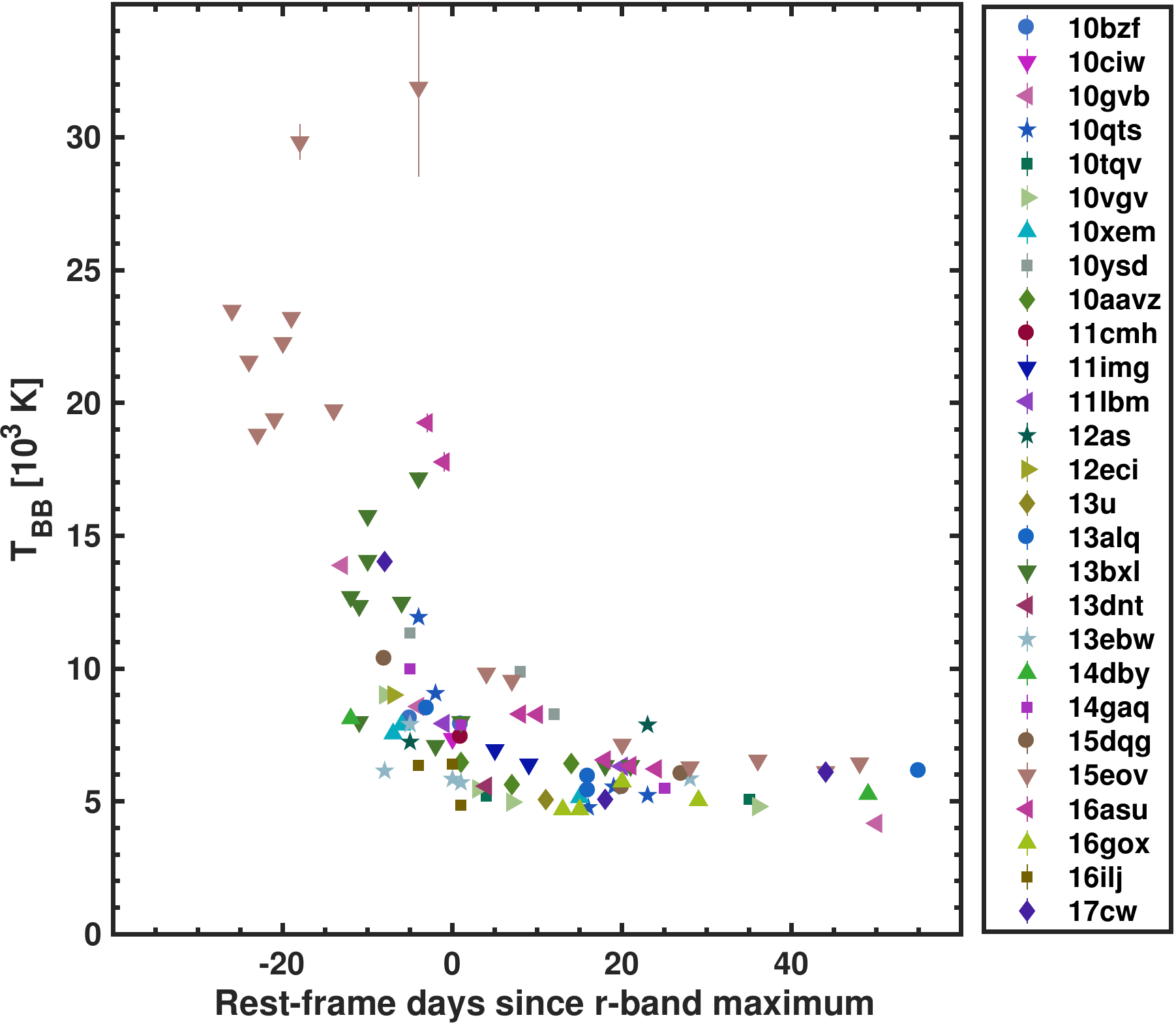}\\
\end{array}$
  \caption{\label{tfromspec}Black-body temperatures from the fit to the spectral sequences.}
 \end{figure}

After extinction correction, we fit a blackbody (BB) function to
the spectra at rest wavelengths longer than 4000~\AA. We do not
include the bluer parts since we want a temperature estimate that is
not affected by the suppression of blue flux due to line
blanketing. The best-fit temperatures are reported in
Table~\ref{tab:speclog} and plotted in Fig.~\ref{tfromspec}, where the
best fits are shown by dashed red lines in the spectral-sequence
figures. We fit only the spectra taken earlier than +60~d after $r$-band peak, when their shape still resembles that of a blackbody. 
Later spectra are dominated by strong emission lines. The BB
temperature evolution derived from the spectra is rather homogeneous
in our sample, as shown in Fig.~\ref{tfromspec}. A rapid cooling from
an initial $\sim$ (1.5--2) $\times 10^4$~K soon after explosion brings
the average temperature at peak down to
$\sim 7500\pm2500$~K around the epoch of maximum light. 
iPTF15eov and iPTF16asu are hotter than any
other SN in our sample, confirming their peculiarity.

\subsection{Photospheric velocities}
\label{sec:vel}
SNe~Ic-BL are characterized by high expansion velocities, as implied
by the broadness of their spectral lines. A good proxy for the
photospheric velocity is the \ion{Fe}{ii}~$\lambda$5169 velocity, as
measured by its maximum absorption position. We tried to estimate this
velocity using the method illustrated by \citet{modjaz16}. In fact, it
is not easy to measure the maximum absorption position of
\ion{Fe}{ii}~$\lambda$5169 in SNe~Ic-BL, given the strong line
blending in the blue part of the spectrum. The method proposed by
\citet{modjaz16} makes use of normal SN~Ic spectral templates that
are shifted and smoothed to match the SN~Ic-BL spectra in
the region around \ion{Fe}{ii}~$\lambda$5169. Before this fitting
procedure, which outputs the velocity shift as compared to the
expansion velocities of the SN~Ic templates, the spectra of our 
SNe~Ic-BL must be continuum-subtracted and smoothed. We therefore removed the
continuum using ``logwave," a software program included in the SNID package
\citep{snid}, and we smoothed the spectra with a fast Fourier transform (FFT) algorithm similar
to that used by \citet{modjaz16}. 
We first confirmed that
our smoothing procedure gave a similar result
to those reported by \citet{modjaz16} 
by applying it to a
demo spectrum already processed with the smoothing program
they used. After continuum-subtracting and smoothing our spectra, we fit the spectra with the
script provided by \citet{modjaz16} to obtain the velocity
shifts with respect to the SN~Ic templates at similar phases,
providing
the actual SN~Ic-BL \ion{Fe}{ii}~$\lambda$5169 velocities in the end. We obtained the SN~Ic template velocities by measuring the \ion{Fe}{ii}~$\lambda$5169 maximum velocity directly from the templates provided by \cite{modjaz16} and  added these velocities to the shifts obtained via the spectral fit to obtain the final \ion{Fe}{ii}~$\lambda$5169 velocities for our SNe. 

\begin{figure}
 \centering
$ \begin{array}{c}
\includegraphics[width=12cm,height=9cm,angle=0]{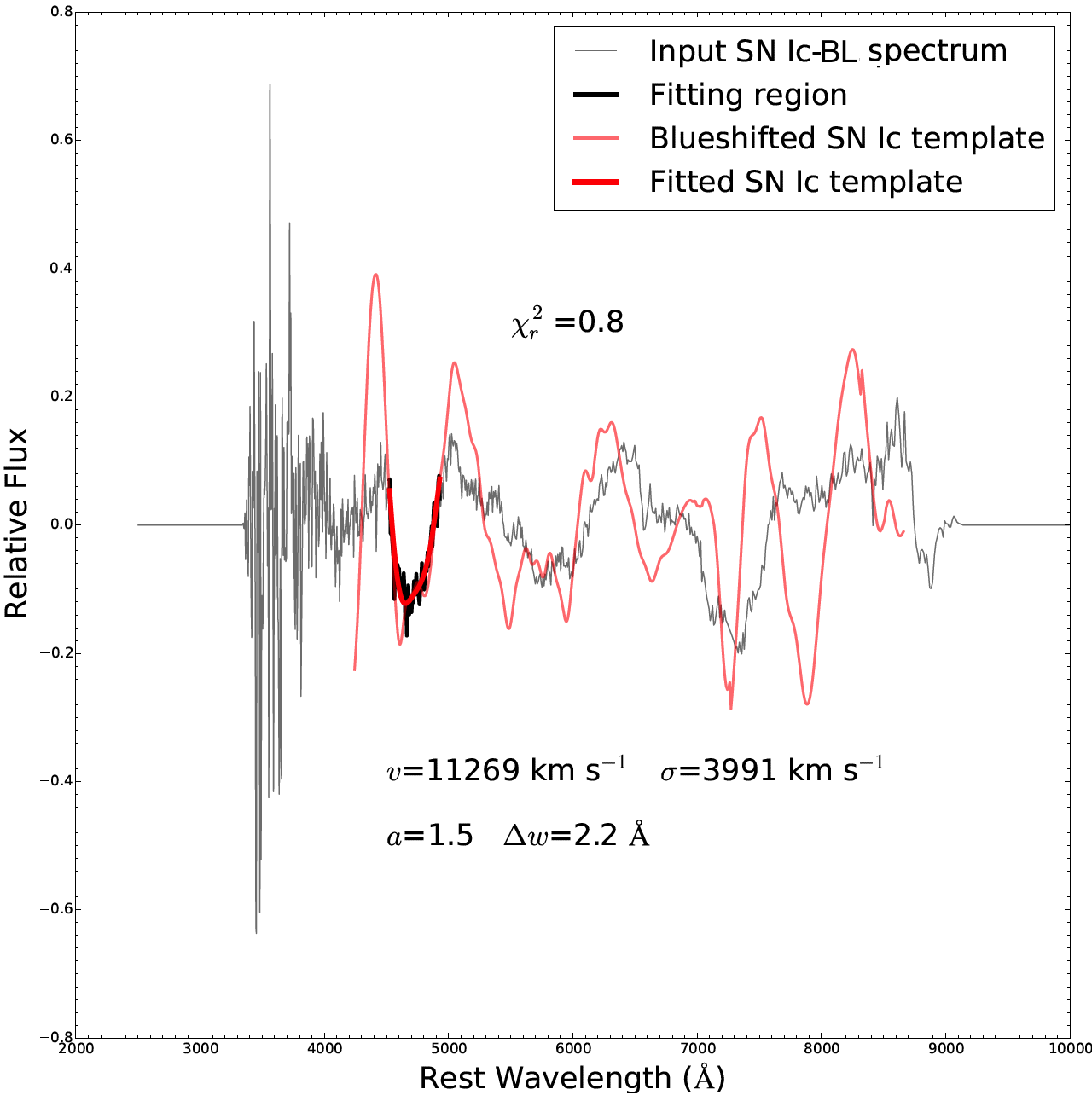}\\
\includegraphics[width=12cm]{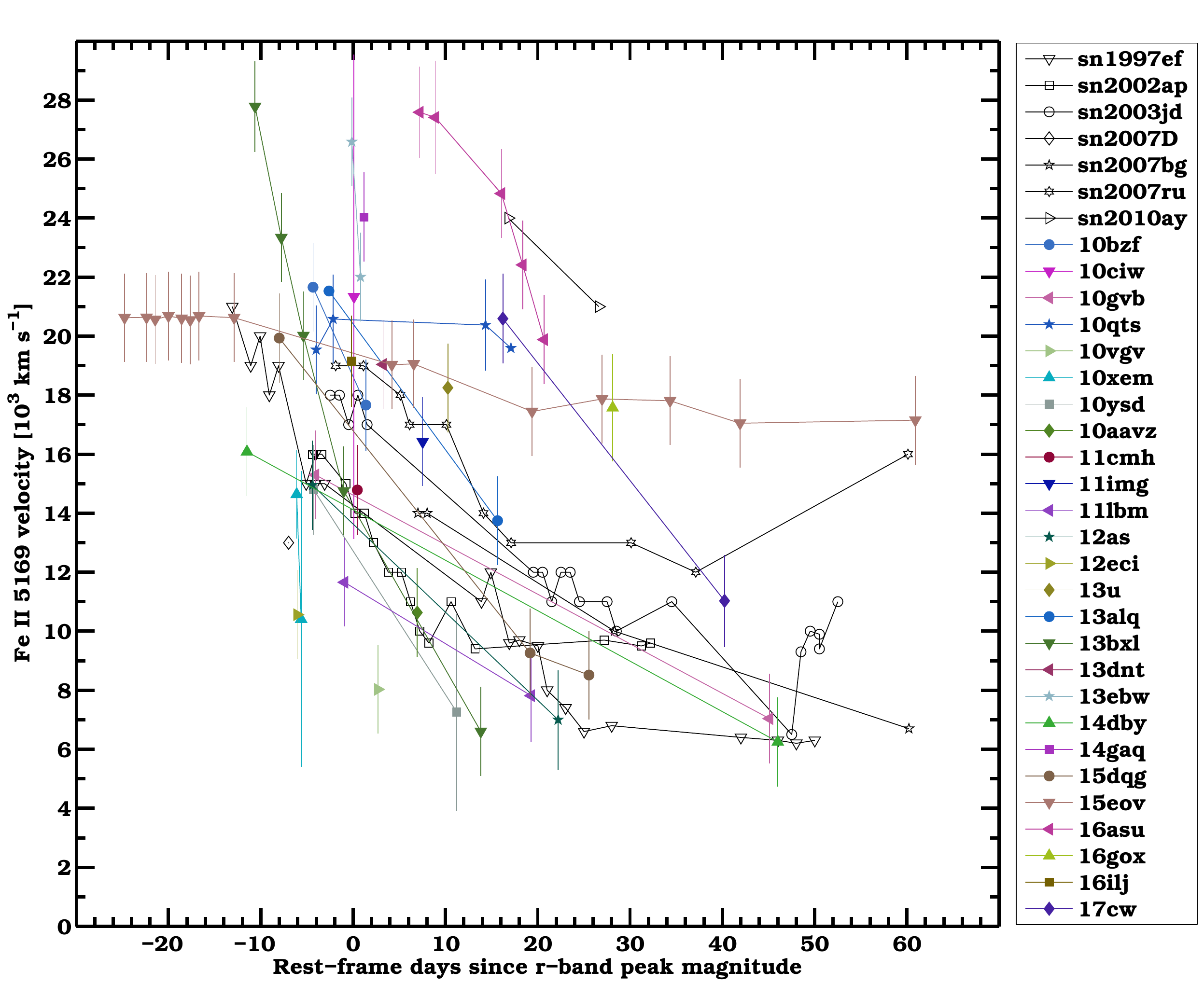}
\end{array}$
  \caption{(Top panel) Spectral template fit to a spectrum of PTF10bgz, as obtained from the \citet{modjaz16} routine. The SN~Ic template (thin red line) has been shifted and smoothed to match the spectrum of our SN (gray line) in the wavelength region around \ion{Fe}{ii}~$\lambda$5169 (black line). The best fit is shown as a thick red line. The shift velocity $v$ must be added to the known  \ion{Fe}{ii}~$\lambda$5169 velocity of the SN~Ic template to obtain the \ion{Fe}{ii}~$\lambda$5169 velocity of our SN. (Bottom panel) Velocity of our SN~Ic-BL sample as compared to the  SNe~Ic-BL (not associated with GRBs) from \citet{modjaz16}. \label{velfit}}
 \end{figure}

An example of a fit is provided in the top panel of Fig.~\ref{velfit}. 
Each \ion{Fe}{ii}~$\lambda$5169 velocity measured for each SN is shown in the
bottom panel of Fig.~\ref{velfit}, as well as in the lower subpanels
of Fig.~\ref{bolofit}. These velocities are also reported in
Table~\ref{tab:speclog}, along with the information about the spectra
of our SNe. In the bottom panel of Fig.~\ref{velfit}, our velocities
are compared to those of other SNe~Ic-BL from \citet{modjaz16}. With
the exception of iPTF15eov, which exhibits a high velocity for a relatively long
time, the velocity trend for our SNe is
similar to that found for the SNe discussed by \citet{modjaz16}. 

Since we need the velocities at peak brightness for deriving properties from
the light-curve models (see Sect.~\ref{sec:boloprop}),
we first fit the SN~Ic-template velocities with a polynomial
and then use this polynomial (with its
normalization left as a free parameter) to fit the velocity profiles
of our SNe.
The best fits are shown by solid lines in the bottom panels of Fig.~\ref{bolofit}, and the interpolated velocities at $r$-band peak are shown by empty diamonds. 
The uncertainty in the velocity at peak is estimated by taking the standard
deviation of the peak velocities derived from 100 Monte
Carlo-simulated velocity profiles (based on the uncertainties of each
velocity measurement)
fitted with the previous velocity template. 
The values of the photospheric velocities at the epoch of $r$-band peak
for our SNe are given in Table~\ref{tab:contardo_bolo}.

\subsection{Spectral signatures of different viewing angles of jet-related explosions}
\label{barnestest}
\citet{barnes17} have predicted that if the explosion of an SN~Ic-BL is
characterized by the presence of a central engine that produces a jet,
then the SN spectra and light curves should depend on the viewing angle.
In particular, a polar point of view would give higher velocities
at
early epochs and redder spectra at later epochs 
\citep[see][Fig. 5]{barnes17}, 
as compared to an equatorial viewing angle. 
In our SN sample, we found a few examples 
where a SN that displays higher velocities than another SN at early
epochs also turns out to be redder than that SN at late epochs,
in line with the models provided by \citet{barnes17}.
However, when we systematically test the late-time colors versus the
early-time velocity for the entire sample,
there is no clear evidence that the faster events are also redder at late epochs. 

\section{Bolometric properties}
\label{sec:boloprop}

We next want to determine the bolometric properties of our SNe to model them and derive progenitor and explosion parameters.
We therefore require estimates of the explosion epochs and
bolometric light curves, in addition to the photospheric velocity profiles, which we obtained in Sect.~\ref{sec:vel}.

\subsection{Explosion epochs and rise times}
\label{sec:explo}

In order to estimate the explosion epochs, we fit the $r$-band light
curves of our SNe with the $R$-band light curve of SN~1998bw, the
prototypical SN~Ic-BL \cite[see][for details about the fit to the light curve of
SN~1998bw.]{cano13}
For SN~1998bw, the epoch of explosion is set
equal to the time of GRB~980425.
The light curve of SN~1998bw is shifted in
magnitude and stretched in time until it fits our SN light curves at
early epochs; that is, until $+$30~d post peak. Since the epochs of
explosion  and  $R$-band maximum
for SN 1998bw are known, 
the temporal stretch of the best fit allows us to infer the explosion
epochs of our SNe. We check the estimates against the pre-explosion
upper limits and, with the exception of three cases, these estimates are consistent.
When not consistent, we assume the last nondetection as the explosion
epoch. We note that, in the special case of iPTF16asu \citep{whitesides17}, whose 
peculiar light-curve shape (characterized by a very fast rise) 
cannot be properly reproduced by stretching SN~1998bw, we
take a value from the literature.
We furthermore checked this method against iPTF13bxl, whose explosion epoch
is known thanks to the associated GRB, 
and the results are consistent within 3.6~d, but for this event we had to modify the light curve to subtract the bright afterglow (see Fig.~\ref{afterglow}), and this introduces more uncertainty in the light-curve fit. We adopt $\pm 2$~d as the uncertainty in the explosion epochs. 

\begin{figure}
 \centering
 \includegraphics[width=18cm,angle=0]{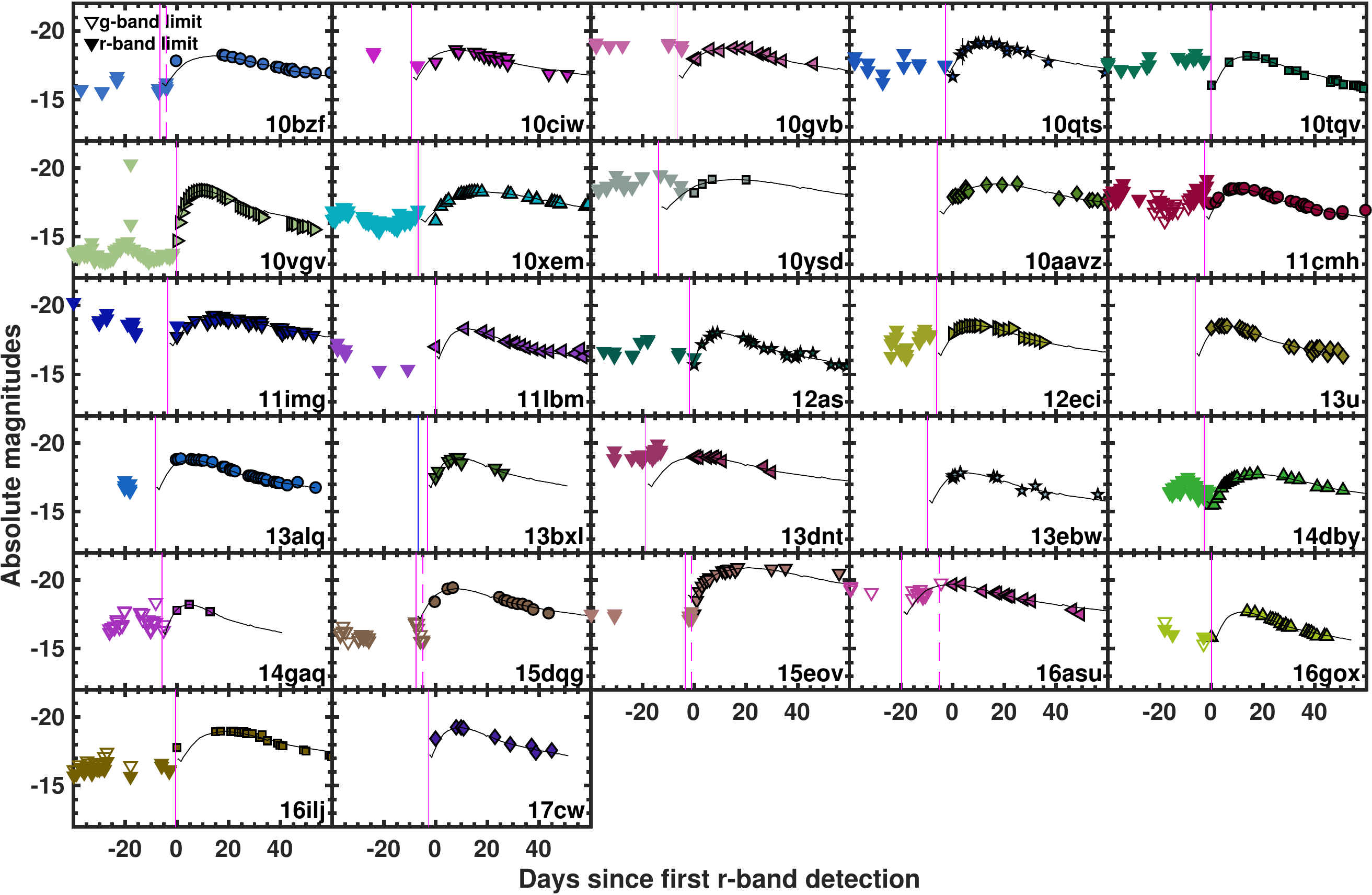}
  \caption{Fits of the $r$ light curves of our SNe~Ic-BL with the
    stretched light curve of SN 1998bw (black solid line). This fitting
    aims at determining the explosion epoch, marked by a vertical
    magenta line. The pre-explosion limits, marked by empty ($g$-band)
    and filled ($r$-band) triangles, are in agreement with the
    computed explosion epochs, except in four cases where we instead adopt the last nondetection as the explosion epoch and a value from the  literature  for iPTF16asu (magenta dashed lines). For iPTF13bxl, we adopt the time of the detected GRB as the explosion epoch (blue vertical line). \label{texplo}}
 \end{figure}

\begin{figure}
 \centering
 \includegraphics[width=12cm,angle=0]{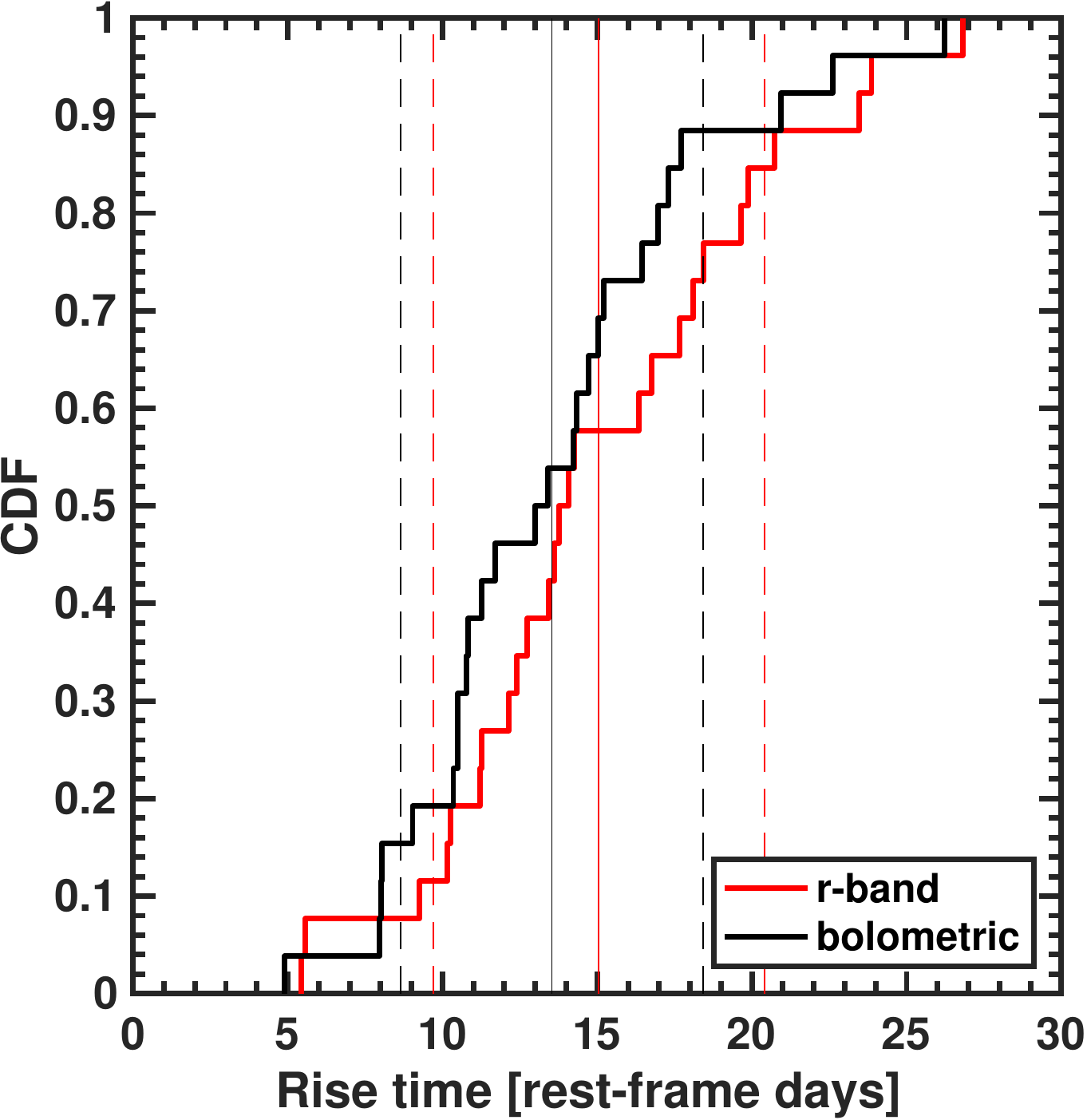}
  \caption{Rest-frame rise-time distributions for the $r$-band (red) and the bolometric (black) light curves. The averages are marked by solid vertical lines, and their standard deviations as dashed vertical lines.\label{risetime}}
 \end{figure}

The best fits and the obtained explosion epochs are shown
in Fig.~\ref{texplo}. The values of the inferred explosion epochs are reported in
Table~\ref{tab:contardo_r}. 
The explosion epochs and the epochs of $r$-band maximum allow us to
compute the rest-frame $r$-band rise time; the 
distribution is plotted in Fig.~\ref{risetime}. 
We note that the average $r$-band rise time ($15\pm6$~d) is consistent with that
computed by \citet{taddia15} for the SDSS SNe~Ic-BL (14.7~d). 
In the same figure, we also plot the rise times for the bolometric
light curves, which we discuss further in
Sect.~\ref{sec:bololightcuvre}. The bolometric rise times are similar to
those in the $r$ band and on average only 1.6~d shorter.
This was also observed in the SDSS SN~Ibc sample \citep{taddia15},
where the bolometric rise times are also slightly shorter than the
$r$ rise times. The $r$-band rise time obviously correlates with $\Delta m_{-10}$ but does not correlate with $\Delta m_{15}$ in the same band.

\subsection{Bolometric light curves}
\label{sec:bololightcuvre}

\begin{figure}
 \centering
 $\begin{array}{cc}
 \includegraphics[width=9cm,height=7cm,angle=0]{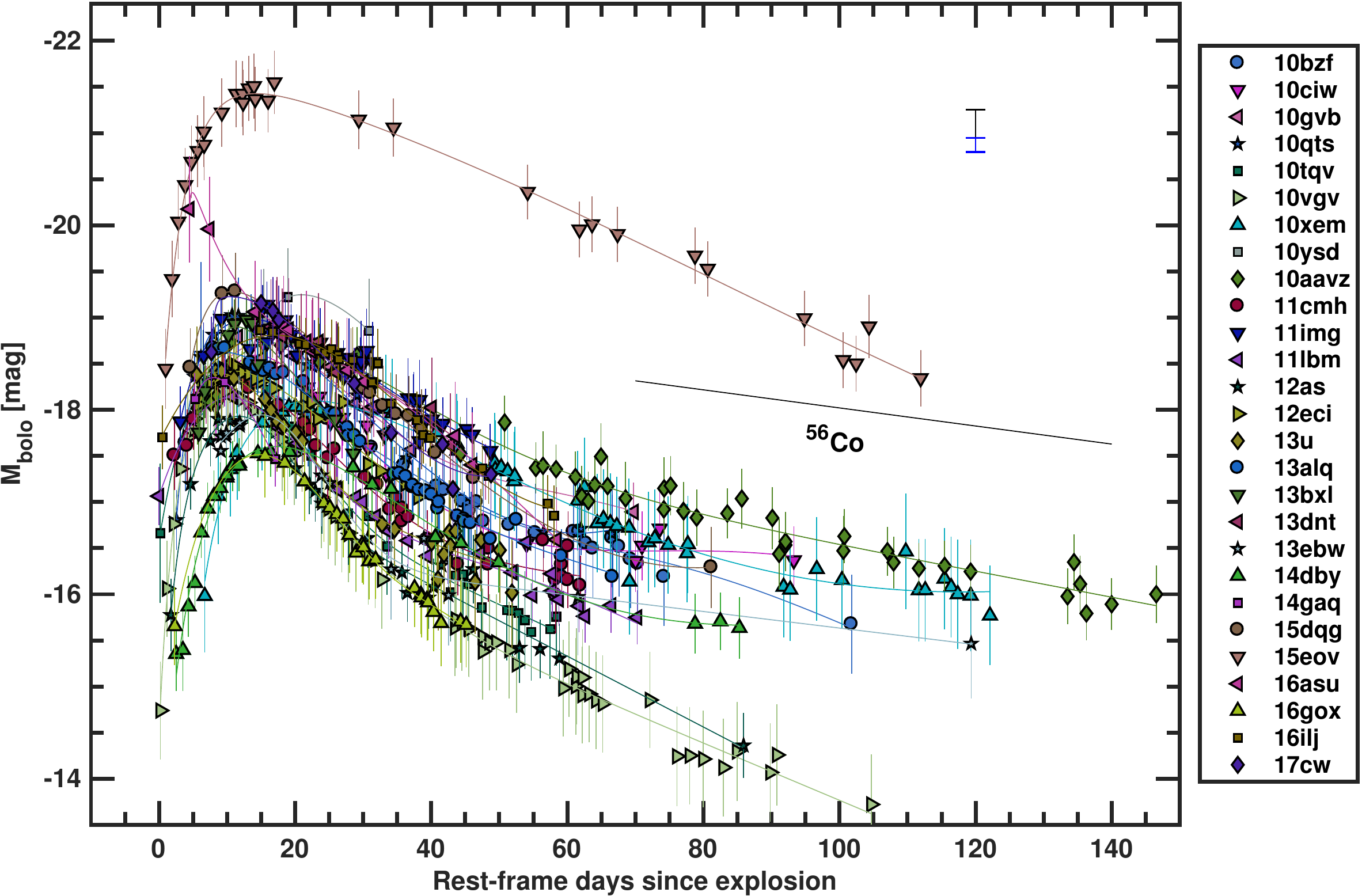} 
\includegraphics[width=9cm,height=7cm,angle=0]{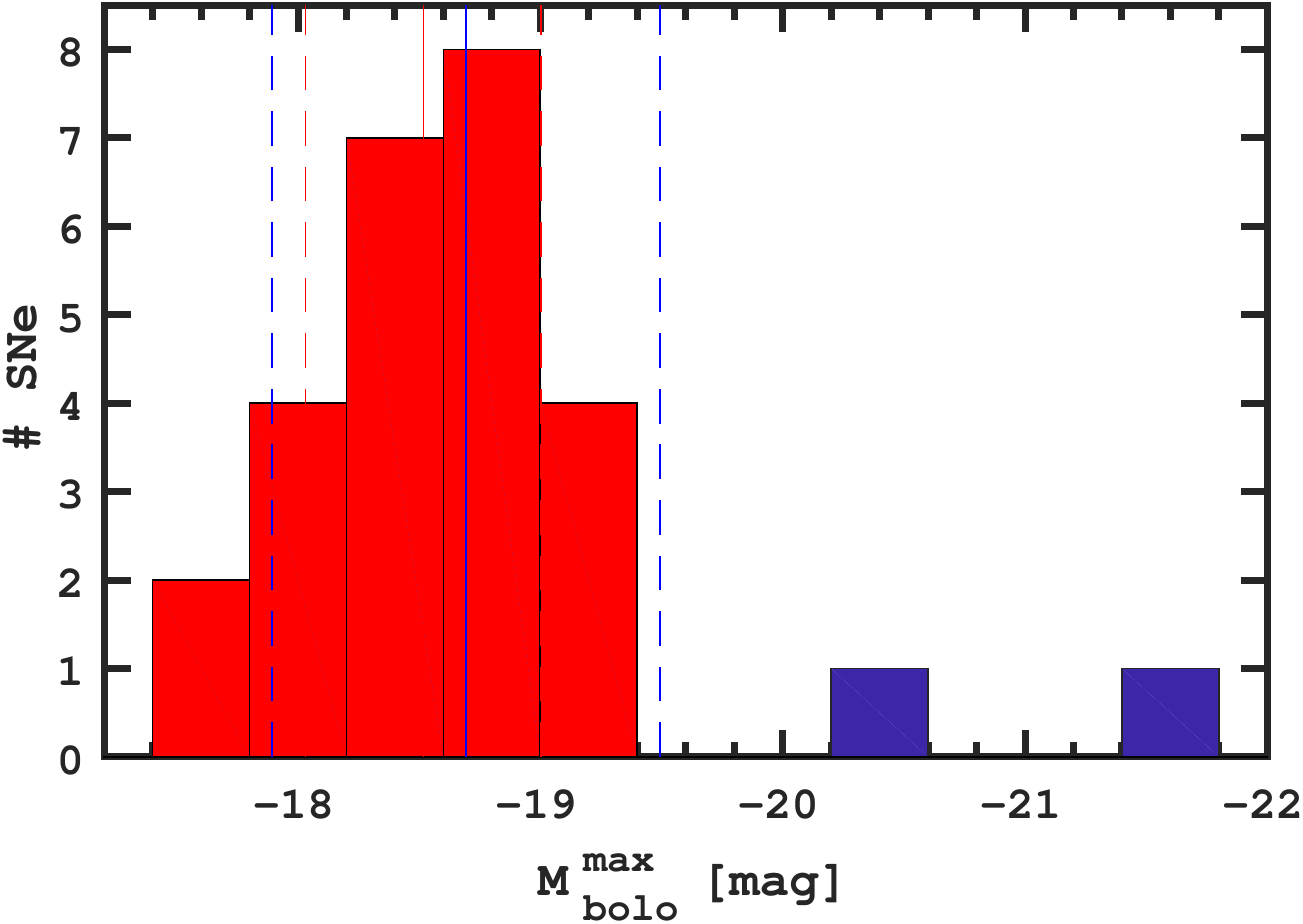}
\end{array}$ 
  \caption{(Left panel) Bolometric light curves, after (host + MW)
    extinction correction and K-correction. The solid lines are the
    best fits of the function also used to fit the $r$ band. The systematic uncertainties in the distance (black segment) and in the bolometric correction (blue segment) are shown in the upper-right corner. (Right panel)
    Distribution of the absolute bolometric peak magnitudes for our sample. iPTF15eov and iPTF16asu are marked in blue. Blue vertical lines mark the average and the standard deviation of the distribution when including all of the objects. 
    Red vertical lines mark the average and the standard deviation of the distribution when excluding iPTF15eov and iPTF16asu.\label{bolo}}
 \end{figure}

Given the lack of complete multiband coverage, especially at early
epochs, we resort to use the absolute $r$-band light curves
(Fig.~\ref{absmag_r}) and the individual fits to the $g-r$ colors shown
in Fig.~\ref{color_fit} to compute the bolometric light curves.
This is done using the bolometric corrections for SE~SNe presented by
\citet{lyman14}, 
as also tested by \citet{taddia18} on a large sample of SE~SNe
from the Carnegie Supernova Project.
In this way, we can estimate bolometric light curves covering the rising phase as well.
We present the final bolometric light curves in Fig.~\ref{bolo}
(left-hand panel), as a function of days since explosion.
The uncertainty associated with each bolometric light-curve point in the
figure does not include the
systematic error due to the bolometric correction (0.076 mag) or to the
uncertainty in the distance (0.04 mag), but it includes the uncertainties in
the photometry, extinction correction, and $g-r$ template.

 We fit the bolometric light curves with the same function used to fit
 the $r$ light curves and plot the best fits as solid lines.
 As in the case of the $r$ light curves, the best fits allow us to measure some properties of the bolometric light curves, such as the peak epoch and magnitude, $\Delta m_{15}$, $\Delta m_{-10}$, and the linear decay slope. In Table~\ref{tab:contardo_bolo}, we report all these parameters. 

Our SNe range in bolometric peak absolute magnitude between $-17.5$ and $-21.4$
mag; the range is greater than that in the $r$ band, as iPTF15eov has a large fraction of its bolometric emission in the bluer bands. 
In our SNe sample, iPTF15eov is the most luminous, followed by iPTF16asu. 
These two SNe have the longest and shortest bolometric rise times,
respectively. The distribution of the peak absolute bolometric magnitudes is
shown in the right-hand panel of Fig.~\ref{bolo}.
If we include iPTF15eov and iPTF16asu, the average bolometric peak is $-18.7\pm0.8$ mag,
whereas if we exclude these peculiar objects, the average bolometric
peak is $-18.5\pm0.5$~mag. 
As with $r$, we find no correlation between peak magnitude and
$\Delta m_{15}$. 
Like in the $r$ band, the bolometric rise time obviously correlates with $\Delta m_{-10}$. The correlation between  $\Delta m_{-10}$ and $\Delta m_{15}$ is present but weaker than in $r$, as we have fewer points. 
The individual bolometric light curves are also shown in the subpanels of Fig.~\ref{bolofit}. 

\subsection{Blackbody temperatures and radii}
\label{sec:temp_rad}

\begin{figure}
 \centering
 \includegraphics[width=18cm,angle=0]{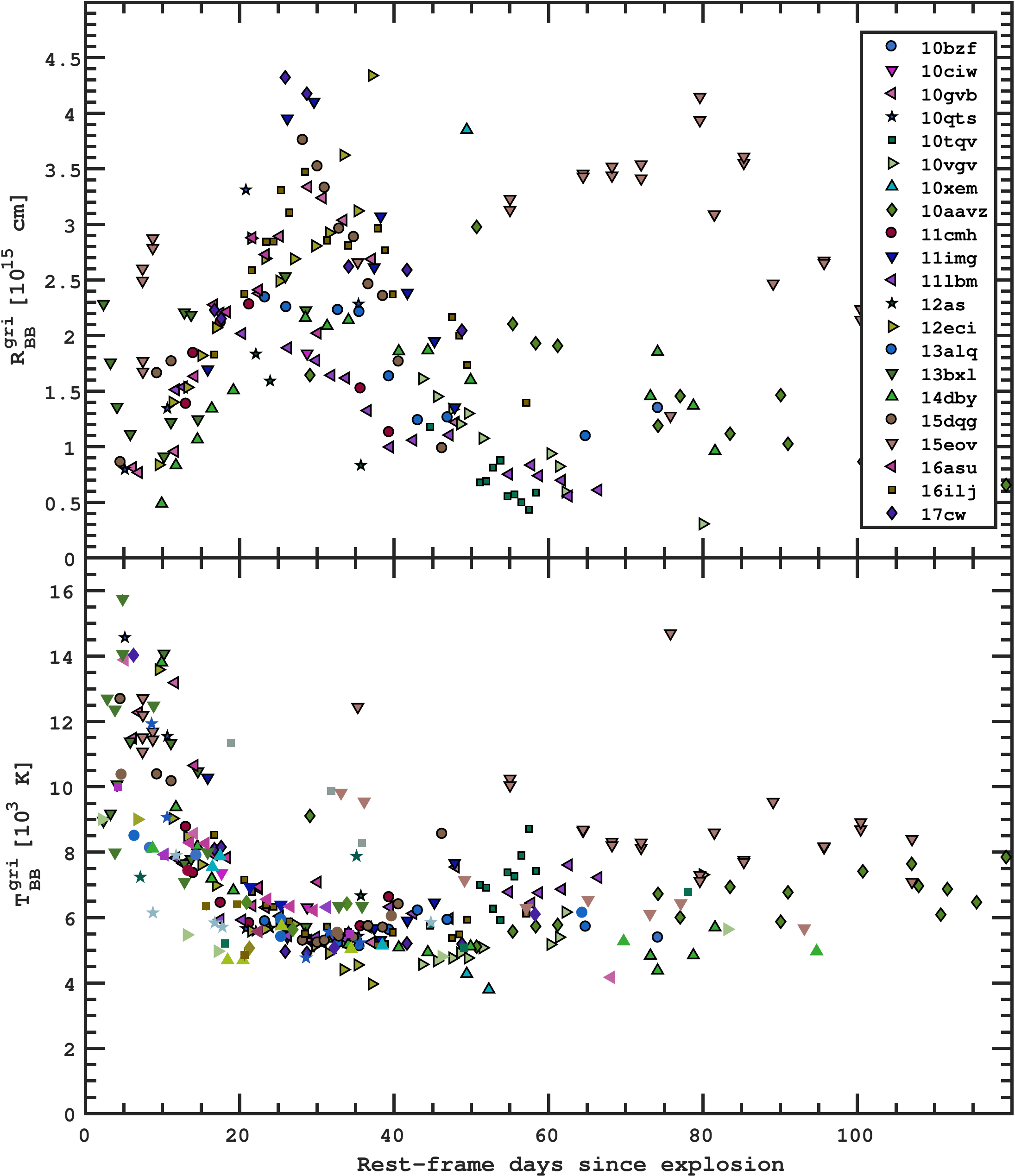}
  \caption{BB radius (top panel) and temperature (bottom panel) from the fit of the $gri$ SEDs of our SNe (symbols with black marker edges). We also report the temperatures from the spectral BB fits already reported in Fig.~\ref{tfromspec} (markers without black edges), which overall match the evolution of temperature from the $gri$ SED fits. \label{TR}}
 \end{figure}

\begin{figure}
 \centering
 \includegraphics[width=18cm,height=21cm,angle=0]{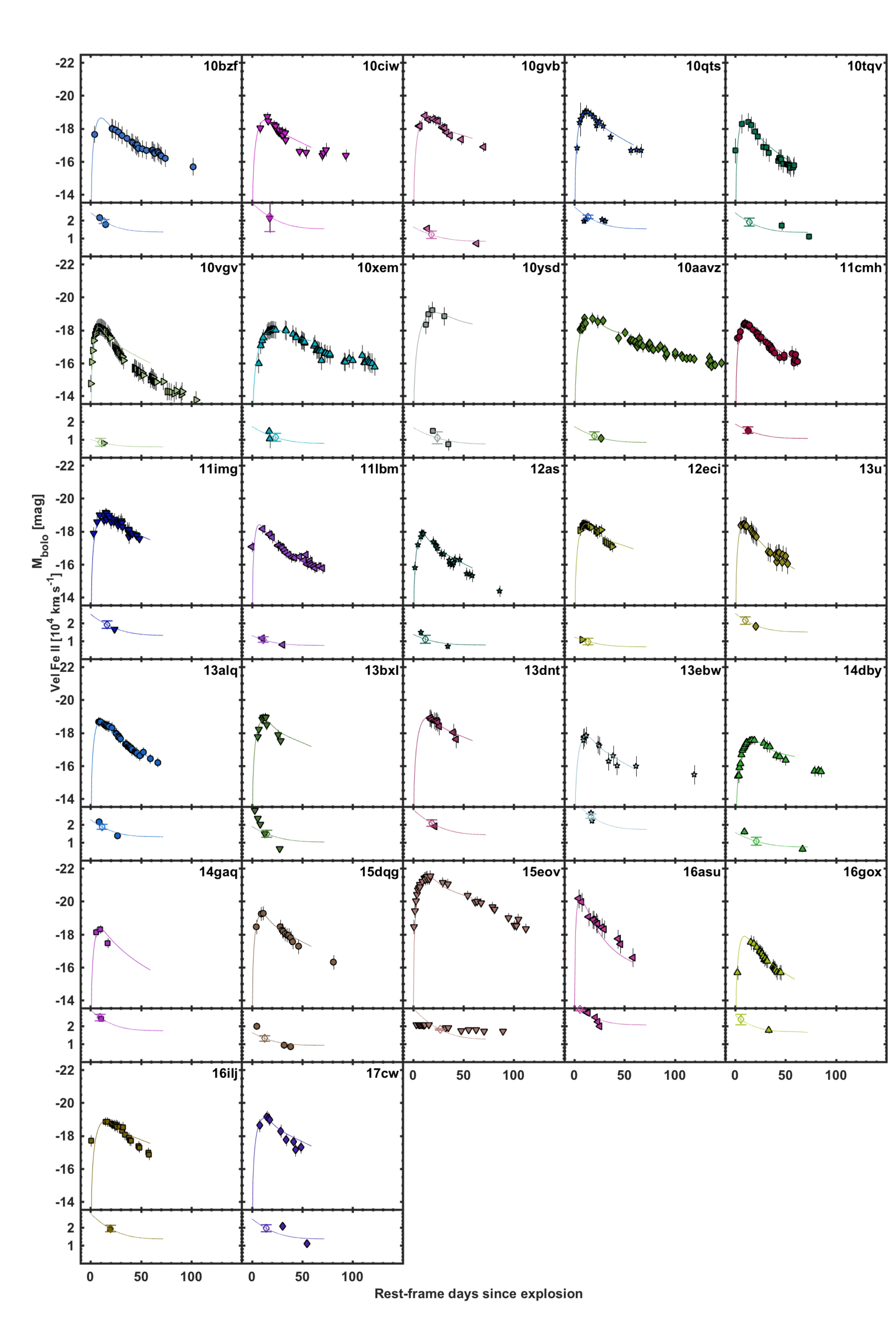}
  \caption{\label{bolofit}Bolometric light curves and velocity profiles for our sample. For the SNe without velocity measurements, we assumed the average velocity profile. The bolometric light curves were fit with the Arnett model, shown as a solid line. The velocity profiles were fit with a scaled polynomial that also reproduces the velocity profile of the SN Ic spectral templates provided by \citet{modjaz16}, and it is shown with solid lines in the velocity subpanels. The velocity at peak brightness is marked with an empty diamond.}
 \end{figure}

Having multiband photometry, we can investigate 
how the BB temperature corresponding to the spectral energy distribution (SED) changes
with time for each SN. We interpolate the K-corrected $g$ and $r$ photometry
to each epoch of $i$-band data. The extinction-corrected
$gri$ magnitudes are converted into fluxes at the effective wavelength of each
filter, and for each epoch we fit the SED with a BB
function. In the bottom panel of Fig.~\ref{TR}, the temperature shows a
progressive decline from the explosion date until $\sim 25$~d, when
it flattens out at $\sim 6000$~K, in line with what was found 
by \citet{taddia15,taddia18} for SE~SNe. We note that the temperature values also exhibit a relatively small spread as a consequence of how we computed the host-extinction corrections; that is, assuming the same $g-r$ colors between 0 and 20~d after $r$-band peak. 
With the BB fit, we also obtain the BB radius, which is shown in
the top panel of Fig.~\ref{TR}.
The radius increases up to $\sim 30$~d since explosion and then slowly decreases, again consistent with previous findings for SE~SNe \citep{taddia15,taddia18}.

\section{Progenitor and explosion parameters}
\label{sec:model}

We fit the bolometric light curves with an Arnett model
\citep{arnett82}, as done by \citet{taddia18}. 
This includes the possibility of escaping gamma rays.
We restrict our fits to the early epochs of the
light curves, when the SNe are in their photospheric phases
(i.e., $\lesssim 60$~d after peak brightness).
From this modeling, it is possible to obtain estimates of the
$^{56}$Ni mass, $M(^{56}$Ni),
kinetic energy of the explosion ($E_{K}$), and ejecta mass
($M_{\rm ej}$). 
We use $E/M = (3/10)\,v^2$ assuming the SN ejecta are spherical with a
uniform density,
where $v$ is equal to the velocity at $r$-band peak as measured
in Sect.~\ref{sec:vel}. 
We also assume constant opacity $\kappa=0.07$
cm$^2$~g$^{-1}$.
This is likely not completely correct, but it allows for comparisons with
other works on SE~SN samples. We also note that in
\citet{taddia18} such an assumption provided similar results to more
sophisticated hydrodynamical models.

The obtained values for $M_{\rm ej}$, $E_{K}$, and $M(^{56}$Ni) are reported in Table~\ref{tab:param}. 
The uncertainties in $M_{\rm ej}$ and $E_{K}$ depend mainly on the uncertainty in the
expansion velocity, whereas that in $M(^{56}$Ni) is mostly due to the
uncertainty in the extinction, but also in the SN distance; the systematic uncertainty due to the bolometric corrections used in Sect.~\ref{sec:bololightcuvre} is significant as well. We obtained averages
$<M_{\rm ej}> = 3.9\pm2.7$~M$_{\odot}$,
$<E_{K}> = (7.1\pm5.4) \times 10^{51}$~erg, 
and
$<M(^{56}{\rm Ni})> = 0.51\pm0.94$~M$_{\odot}$. 
If we remove iPTF16asu and iPTF15eov, we obtain
$<M_{\rm ej}> = 3.9\pm2.8$~M$_{\odot}$,
$<E_{K}> = (7.0\pm5.6) \times 10^{51}$~erg, 
and
$<M(^{56}{\rm Ni})> = 0.31\pm0.16$~M$_{\odot}$. 
If we also remove the GRB SN iPTF13bxl, considering only the regular SNe~Ic-BL, we obtain 
$<M_{ej}> = 4.0\pm2.9$~M$_{\odot}$, 
$<E_{K}> = (7.0\pm5.8) \times 10^{51}$~erg, 
and 
$<M(^{56}{\rm Ni})> = 0.31\pm0.17$~M$_{\odot}$. 
Removing the two outliers significantly reduces the average value of $M(^{56}$Ni). The averages of the other parameters are similar. The GRB SN parameters do not significantly alter the averages.

\begin{figure}
 \centering
 \includegraphics[width=18cm,angle=0]{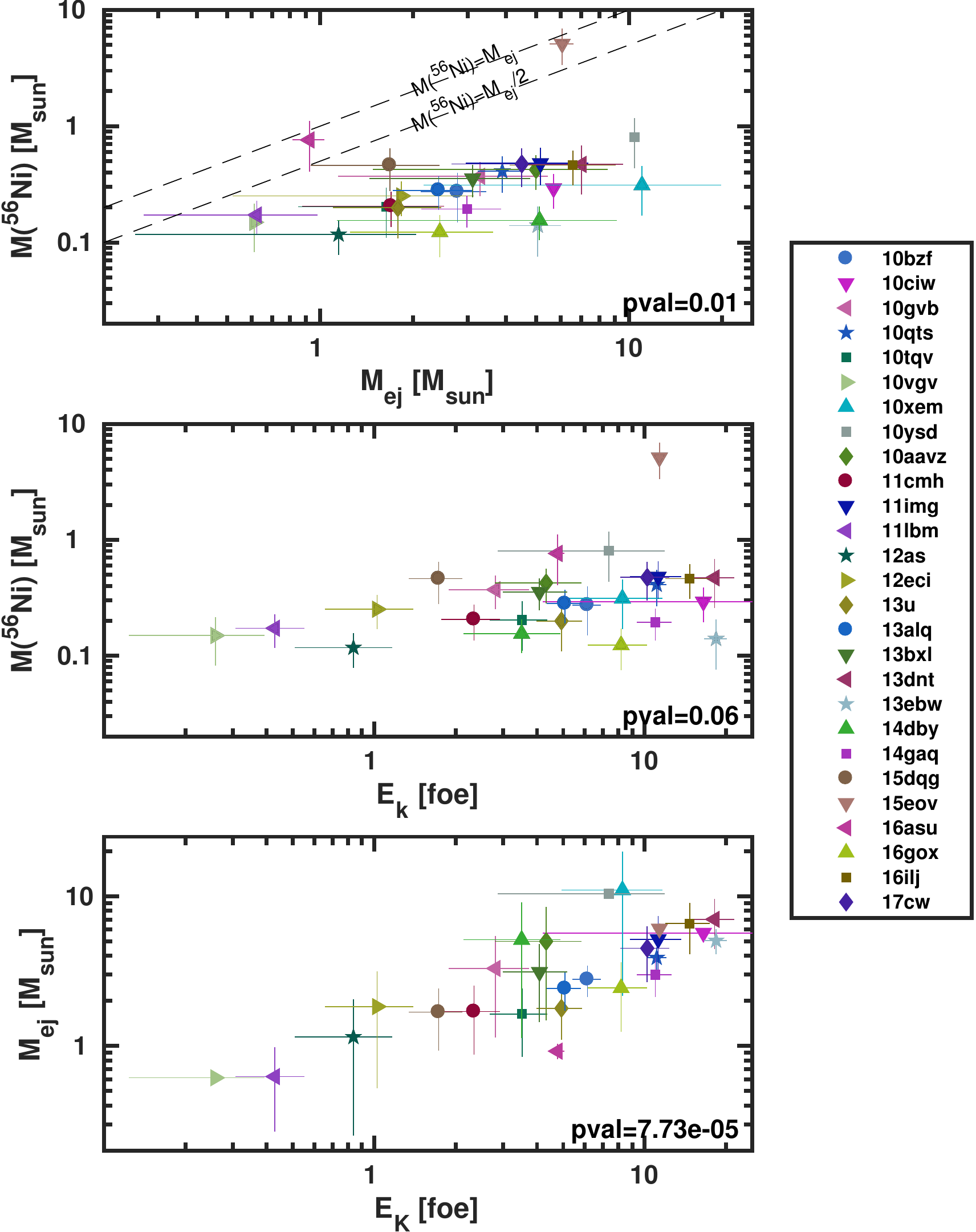}
  \caption{\label{param}Explosion and progenitor parameters plotted against each other for our SN sample.}
 \end{figure}

 In Fig.~\ref{param}, we plot each parameter against the others, identifying a correlation between 
$M_{\rm ej}$ and $E_{K}$ (see bottom panel), a weaker correlation
between $M_{\rm ej}$ and  $M(^{56}$Ni) (top panel),
and between  $M(^{56}$Ni) and $E_{K}$ (central panel). This was also
observed for SE~SNe by \citet{lyman16} and \citet{taddia18}.

iPTF15eov and iPTF16asu show a very large ratio ($>80$\%) of
$M(^{56}$Ni) over  $M_{\rm ej}$,
which indicates that they do not fit the traditional radioactive-powered
scenario, as already suggested by their peculiar light curves. As
such, the values derived for these objects from the above method are
not valid.

\begin{figure}
 \centering
 $\begin{array}{c}
 \includegraphics[width=18cm,angle=0]{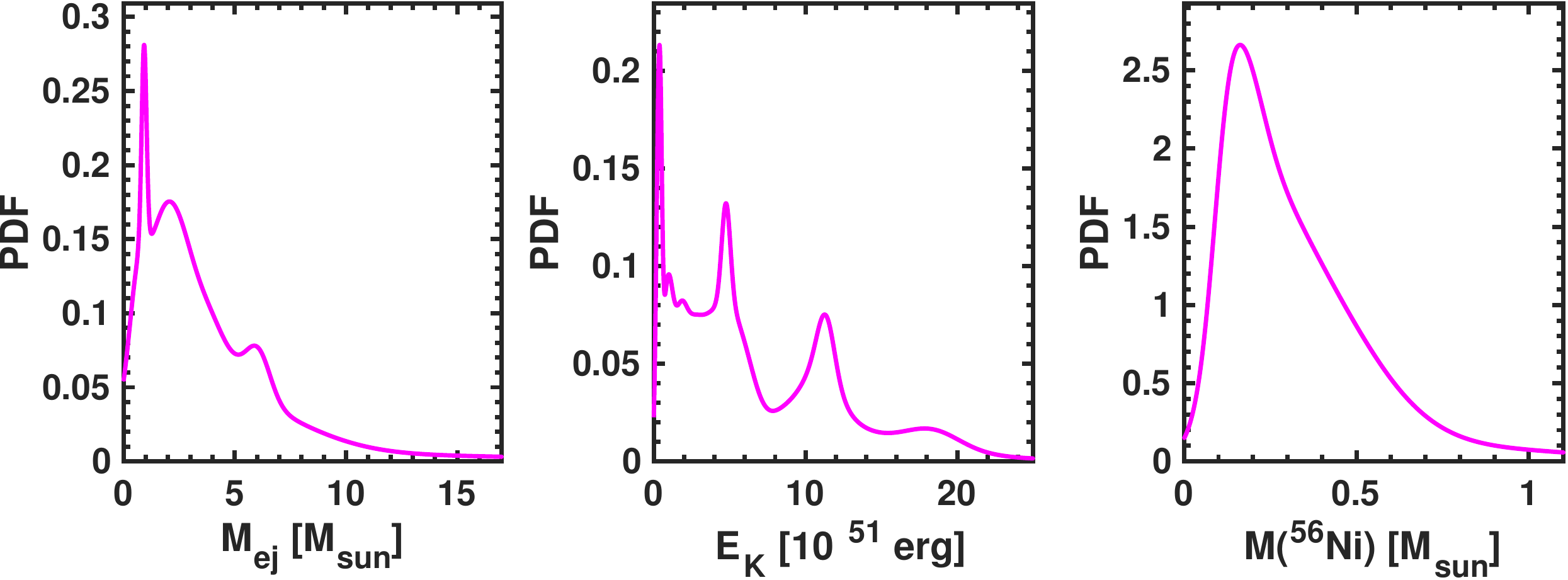} 
\end{array}$  
  \caption{\label{pdf}Probability density functions for the three main explosion parameters for our sample of SNe~Ic-BL, excluding GRB-SNe as well as the peculiar SNe iPTF15eov and iPTF16asu. }
 \end{figure}

The probability distributions of the three aforementioned parameters are plotted in
Fig.~\ref{pdf}.
We excluded GRB SNe, as well as iPTF15eov and iPTF16asu, for further comparison with literature data on regular SNe~Ic-BL.
The scatter of these three parameters is larger than the typical uncertainties for each individual SN, so there is an intrinsic range of values in the SN~Ic-BL family for $M_{\rm ej}$, $E_{K}$, and $M(^{56}$Ni). 

An interesting property of SNe powered by radioactive decay is the
distribution of the radioactive $^{56}$Ni. The degree of $^{56}$Ni
mixing in the ejecta affects the light-curve shape,
and it is possible to estimate its extent by comparison with
theoretical models.
Arnett's model assumes that all of the $^{56}$Ni is in the center of the ejecta, and therefore
we turned to more sophisticated hydrodynamical models to derive
scaling relations aimed to estimate the degree of $^{56}$Ni mixing.
We followed an approach similar to that used by \citet{taddia16bsg} to
determine the progenitor parameters of SN~1987A-like events from
(i)PTF.
We first built a realistic progenitor star for SNe~Ic-BL, which is a
moderately low final mass 
(7.34~ M$_{\odot}$), 
He-poor (0.56
M$_{\odot}$ of He), compact ($R = 10$~R$_{\odot}$) star.
This was carried out using MESA \citep{paxton11}, evolving a star with
initial mass 85~M$_{\odot}$, fast rotation velocity (350
km~s$^{-1}$), and at solar metallicity.
This star is similar to (though less massive than) that used to model
PTF11mnb by \citet{taddia17_11mnb},
and with much less He than the star used to model iPTF15dtg by \citet{taddia16_15dtg}. 
We then exploded this progenitor star with the hydrodynamical code
SNEC \citep{morozova15}
with a range of different explosion energies, $^{56}$Ni masses, and
$^{56}$Ni mixing parameters.

The model assumes an opacity floor that scales linearly with
metallicity, where $\kappa= 0.07$~cm$^2$~g$^{-1}$ at $Z=1$ and
$\kappa= 0.01$~cm$^2$~g$^{-1}$ at $Z=0.02$.
We obtained bolometric light curves and their rise times for each
model. We defined a first model with $E_{K}=7 \times 10^{51}$ erg, 
$M_{\rm ej} = 7.34-1.4 = 5.9$~M$_{\odot}$  
(the removed 1.4~M$_{\odot}$ from
the total mass of the star accounts for the compact remnant), 
$^{56}$Ni  mixing up to 80\% of the ejecta mass, and a
$^{56}$Ni mass of 0.3~M$_{\odot}$.
Then, we kept all of these parameters constant except the $^{56}$Ni mass,
which we also set to 0.1 and 0.5~M$_{\odot}$.
Subsequently, we varied the explosion energy with the values 2.0, 3.0,
4.5, 7.0, 9.5, 12, 14.5, 19.5, and 24.5~$\times 10^{51}$~erg , keeping 
all of the other parameters unchanged.
Finally, we varied the $^{56}$Ni mixing from 40\% to 100\%.

We plot all of the models and their rise times versus the mentioned
parameters in Fig.~\ref{hydromixing}. This shows how these
parameters correlate with the rise time.
We derived simple scaling relations between these parameters and the
rise time --- in particular, a linear relation with $^{56}$Ni  mixing and
$^{56}$Ni  mass,
and a power-law scaling with $E_{K}/M_{\rm ej}$.
The best-fit relations are shown in red in the figure. We calibrated
these relations using the known rise time of SN~1998bw (15.5~d) and
the progenitor parameters of SN~1998bw from the work of
\citet{chugai98bw} (see Fig.~\ref{hydromixing}). SN~1998bw is believed to
be fully mixed.
These calibrated scaling relations require the rise time of a
SN~Ic-BL, its $E_{K}/M_{\rm ej}$, and its $^{56}$Ni  mass to determine the
degree of $^{56}$Ni  mixing. We used the bolometric rise times reported in
Fig.~\ref{risetime} and the $E_{K}/M_{\rm ej}$ and  $^{56}$Ni mass
parameters from the Arnett fit in Table~\ref{tab:param}. 
We did this for all of the radioactively powered SNe in our sample,
obtaining the mixing reported in Table~\ref{tab:param} and shown in
Fig.~\ref{hydromixing}.
All of the SNe are strongly mixed ($>87$\%) and more than 70\% of these SNe are fully mixed. 
We note that $E_{K}/M_{\rm ej}$ in our models comes from the constraints
on the photospheric velocity,
so it is independent of the early light-curve shape, from which we estimate the $^{56}$Ni mixing with our scaling relations. 

\section{Discussion}
\label{sec:discussion}

\subsection{Literature comparison}
\label{sec:literature}

First, we emphasize that PTF and iPTF, as compared with the other SN~Ic-BL samples, provide a larger, homogeneous, untargeted set with good constraints on the explosion epoch. Furthermore, this work makes use of state-of-the-art host extinction and bolometric corrections, and spectral velocity measurements; moreover, in addition to deriving the classic explosion parameters, it also tests the role of $^{56}$Ni mixing.

A number of SNe~Ic-BL and GRB-SNe have been presented 
in the literature. Our PTF+iPTF sample of SNe~Ic-BL is mainly composed
of SNe not associated with GRBs. We note that we could only determine the explosion parameters for 1 GRB-SN of the 25 events. In the following we do not consider the parameters for iPTF15eov and iPTF16asu since those SNe are not radioactively powered.

We checked for any eventual GRB coincidence in time and space with our
SNe using the {\it Fermi} archive and the GRB interplanetary 
network\footnote{\href{https://heasarc.gsfc.nasa.gov/W3Browse/all/ipngrb.html}{https://heasarc.gsfc.nasa.gov/W3Browse/all/ipngrb.html}}. 
We did not find GRB associations aside from those already known from the literature of iPTF13bxl (and iPTF14bfu, for which we did not have enough photometry to perform an analysis).  

\begin{figure}
 \centering
 $\begin{array}{cc}
  \includegraphics[width=14cm,height=6cm,angle=0]{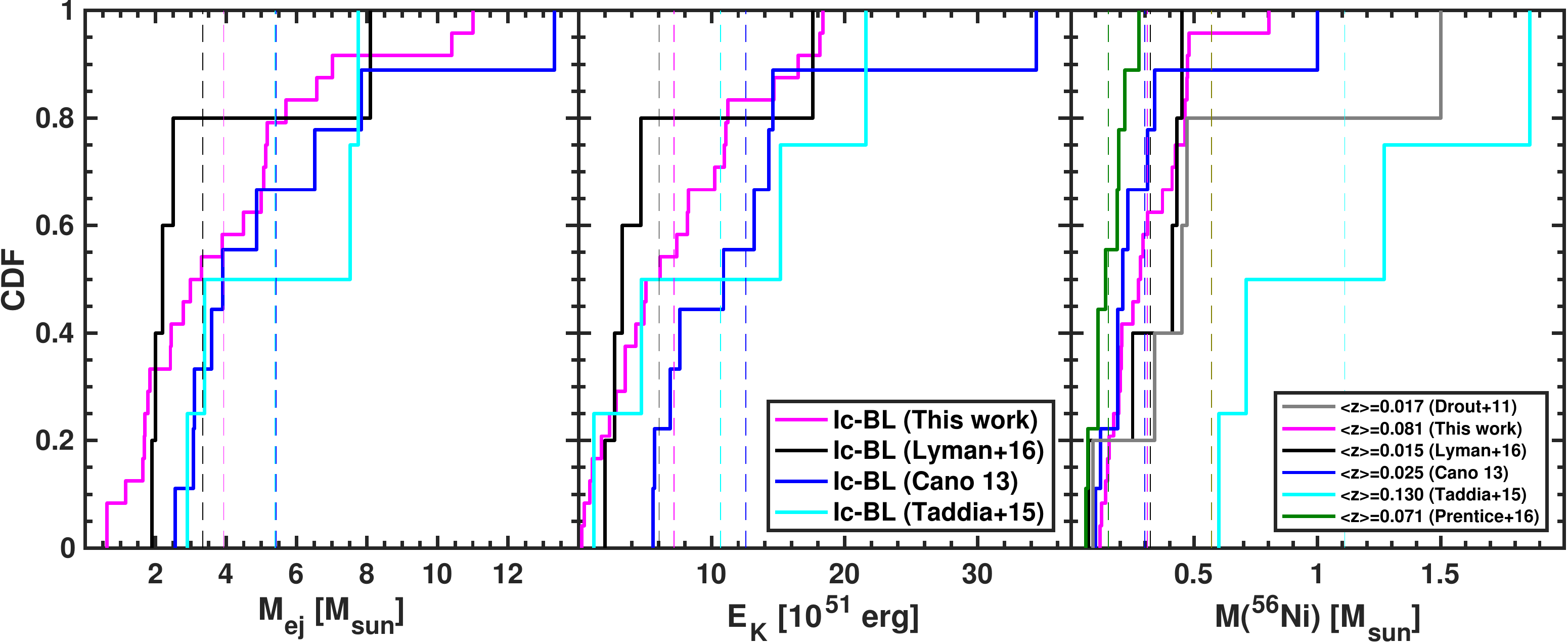} &
  \includegraphics[width=4cm,height=6cm,angle=0]{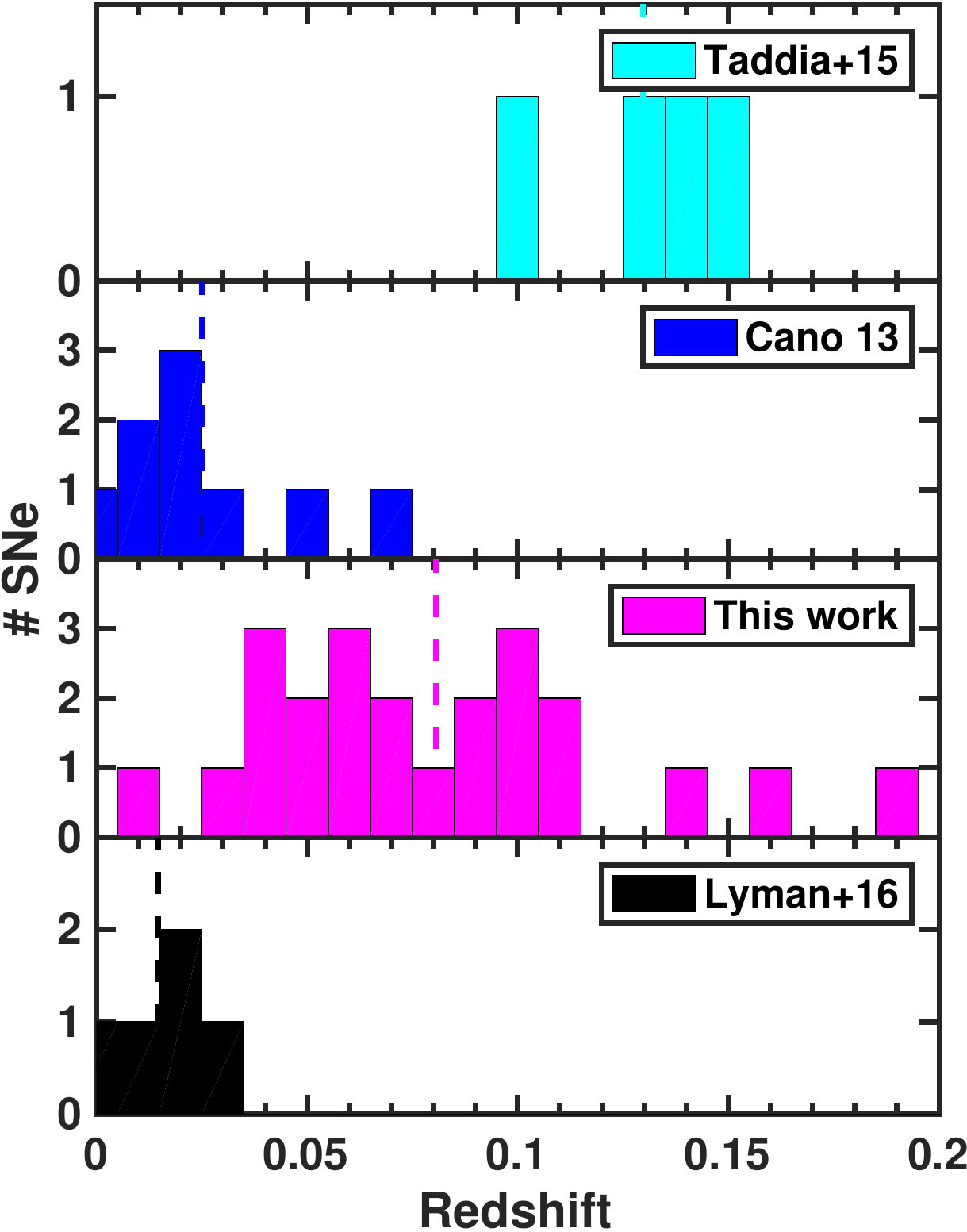} 
    \end{array}$
  \caption{\label{cdf}Cumulative probability functions for the three main explosion parameters as compared to those of other SN~Ic-BL samples in the literature.}
 \end{figure}

\begin{figure}
 \centering
 $\begin{array}{cc}
  \includegraphics[width=9cm,angle=0]{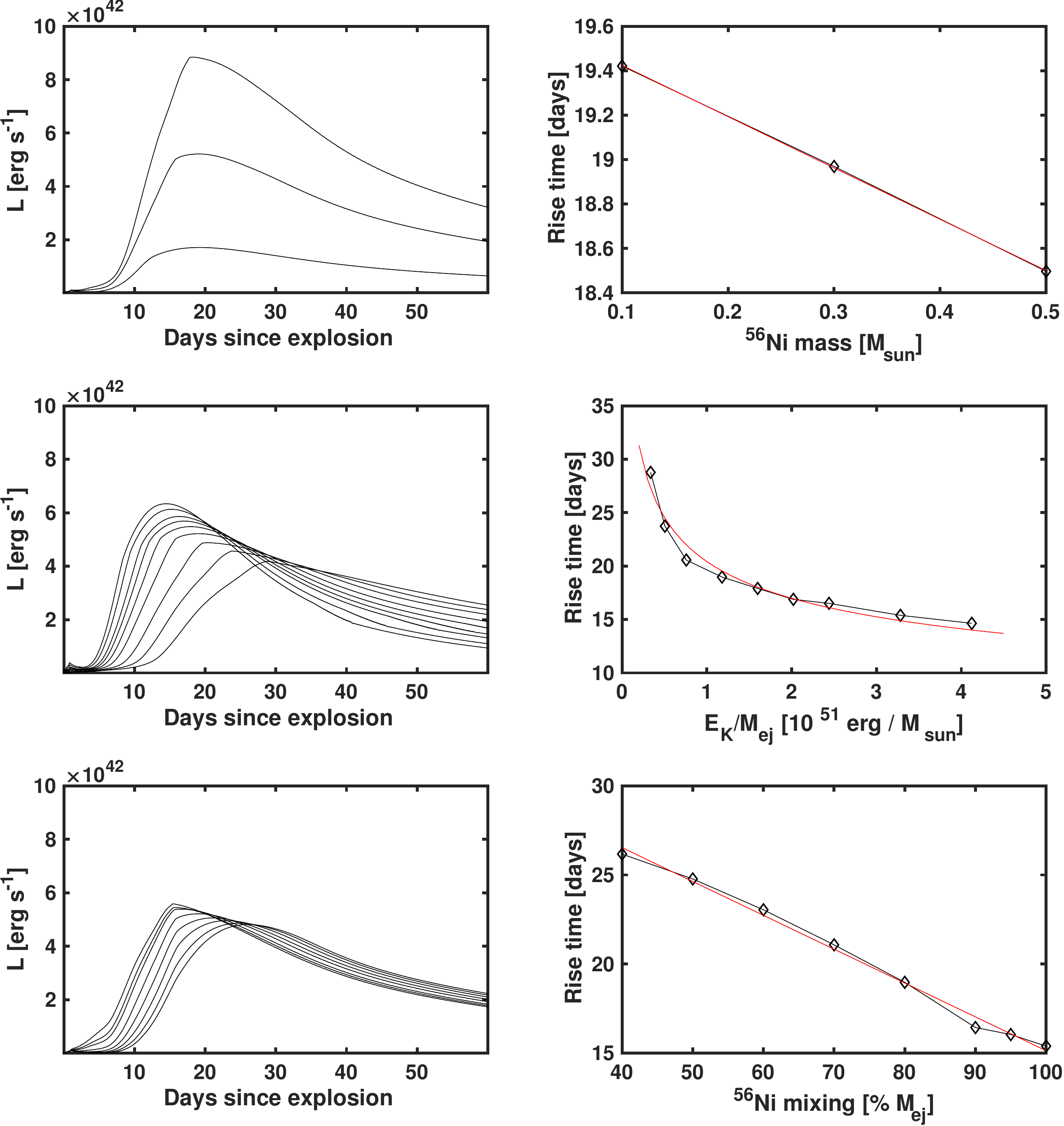} &
  \includegraphics[width=9cm,angle=0]{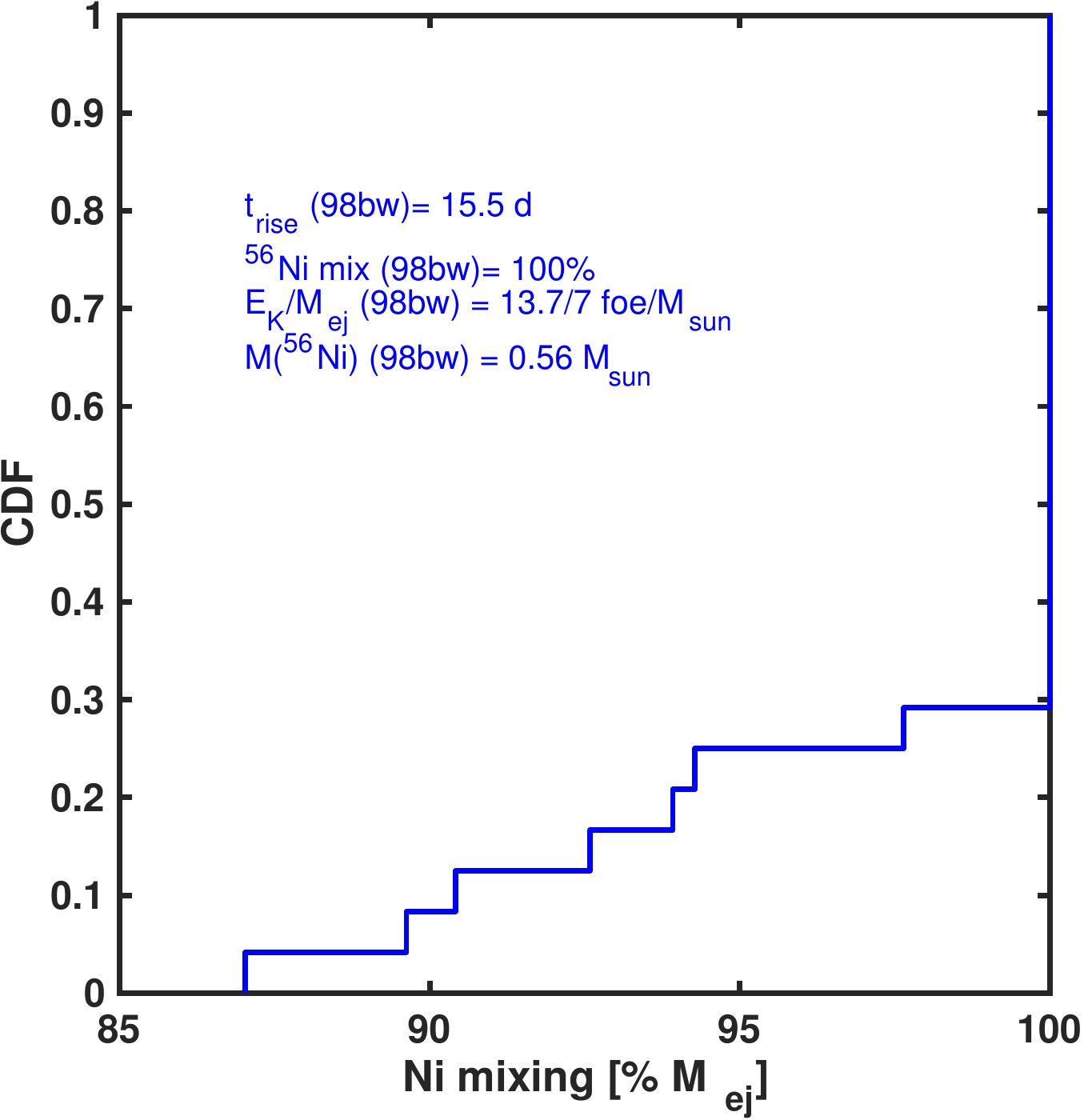} 
    \end{array}$
  \caption{\label{hydromixing}(Left panels) Light curves from the hydrodynamical models and their rise times as a function of three explosion and progenitor parameters ($^{56}$Ni mass, $E_{K}/M_{\rm ej}$, $^{56}$Ni mixing). The observed correlations are fit with two linear and one power-law functions. These functions are used to build scaling relations aimed at determining the $^{56}$Ni mixing of our SNe, relations that are calibrated against the parameters of SN~1998bw from \citet{chugai98bw}. (Right panel) $^{56}$Ni mixing obtained from the scaling relations determined from the hydrodynamical models and calibrated to SN~1998bw. All of the SNe are strongly ($>87$\%) mixed.}
 \end{figure}

Therefore, we start by comparing the explosion and progenitor
parameters of the 24 PTF+iPTF 
SNe~Ic-BL not associated
with a GRB to samples of other similar events, excluding GRB-SNe. \citet{cano13} presented explosion parameters for 9 SNe~Ic-BL
not associated with a GRB. \citet{taddia15} showed four events, whereas
\citet{lyman16} included 5 SNe~Ic-BL without GRB association.
\citet{taddia18} presented 2 additional SNe~Ic-BL, namely SN~2009bb and SN~2009ca. 
Our (i)PTF sample (24 objects) therefore more than doubles the sum of the samples
(20) for which explosion parameters have been derived. 
We also note that \citet{prentice16} estimated the $^{56}$Ni mass for a
collection of  12 SNe~Ic-BL not associated with GRBs (and of 10 GRB
SNe), but they provided only $M_{\rm ej}^3/E_{K}$. Also \citet{drout11} presented $^{56}$Ni mass values for 5 SNe~Ic-BL, and values for $M_{\rm ej}^{3/4}/E_{K}^{1/4}$.
 In Fig.~\ref{cdf}, we present the cumulative distribution functions of
 the main parameters, as compared to those obtained by the other
 mentioned studies in the literature.
 In Table~\ref{tab:literature} we report averages and standard deviations of the different samples. 

These works make use of similar assumptions regarding the opacity ($\kappa=0.05$--0.07~cm$^2$~g$^{-1}$) used in the Arnett models. \citet{cano13} did not assume $E/M~=~(3/10) v^2$ as in the other works, but $E/M = (1/2) v^2$. The velocities derived for our (i)PTF SNe are from \ion{Fe}{ii} and determined via the robust method proposed by \citet{modjaz16}. In the other works, the velocities are mainly from \ion{Si}{ii}.
The work by \citet{lyman16} is that with which the comparison is more direct owing to the most similar assumptions in the modeling.

 The ejecta masses turn out to be rather similar among the samples, confirming relatively low values for this parameter. Compared to \citet{lyman16} who found $3.3\pm2.7$~M$_{\odot}$, we found $4.0\pm2.9$~M$_{\odot}$. Our sample has a tail of the ejecta-mass probability distribution that extends to larger masses than that of  \citet{lyman16}; the implication of this is discussed in Sect.~\ref{sec:discussion}. 
 
In the case of \citet{cano13} (and \citealp{taddia15}, which has fewer events), the derived kinetic energies are higher than in our sample. The difference is  statistically significant ($p=0.037$ through a K-S test). However, we stress that in \citet{cano13} the assumption on $E_{K}/M_{\rm ej}$ relative to the photospheric velocity was different by a factor of 5/3 compared to the other works, implying higher energy.

 The estimated $^{56}$Ni masses are similar for all of the samples except
 \citet{taddia15}, where the mean value is considerably higher.
 However, this can be understood in terms of the average redshift of
 the sample shown by \citet{taddia15},
 which is much higher than in the other samples, and therefore biased
 toward brighter, more $^{56}$Ni-rich events (see the left panel in
 Fig.~\ref{cdf}).
 Moreover, in \citet{taddia15}, large extinction corrections were applied based on colors; the intrinsic color was taken from a sample of SE~SNe that also included SNe~Ib and Ic, not only SNe~Ic-BL.
 In \citet{prentice17}, which does not provide estimates for $E_K$ and $M_{\rm ej}$ since they do not make use of the SN spectra, some events were not corrected for host extinction, implying a lower average $^{56}$Ni mass. 

\begin{figure}
 \centering
 $\begin{array}{cc}
  \includegraphics[width=11cm,height=9cm,angle=0]{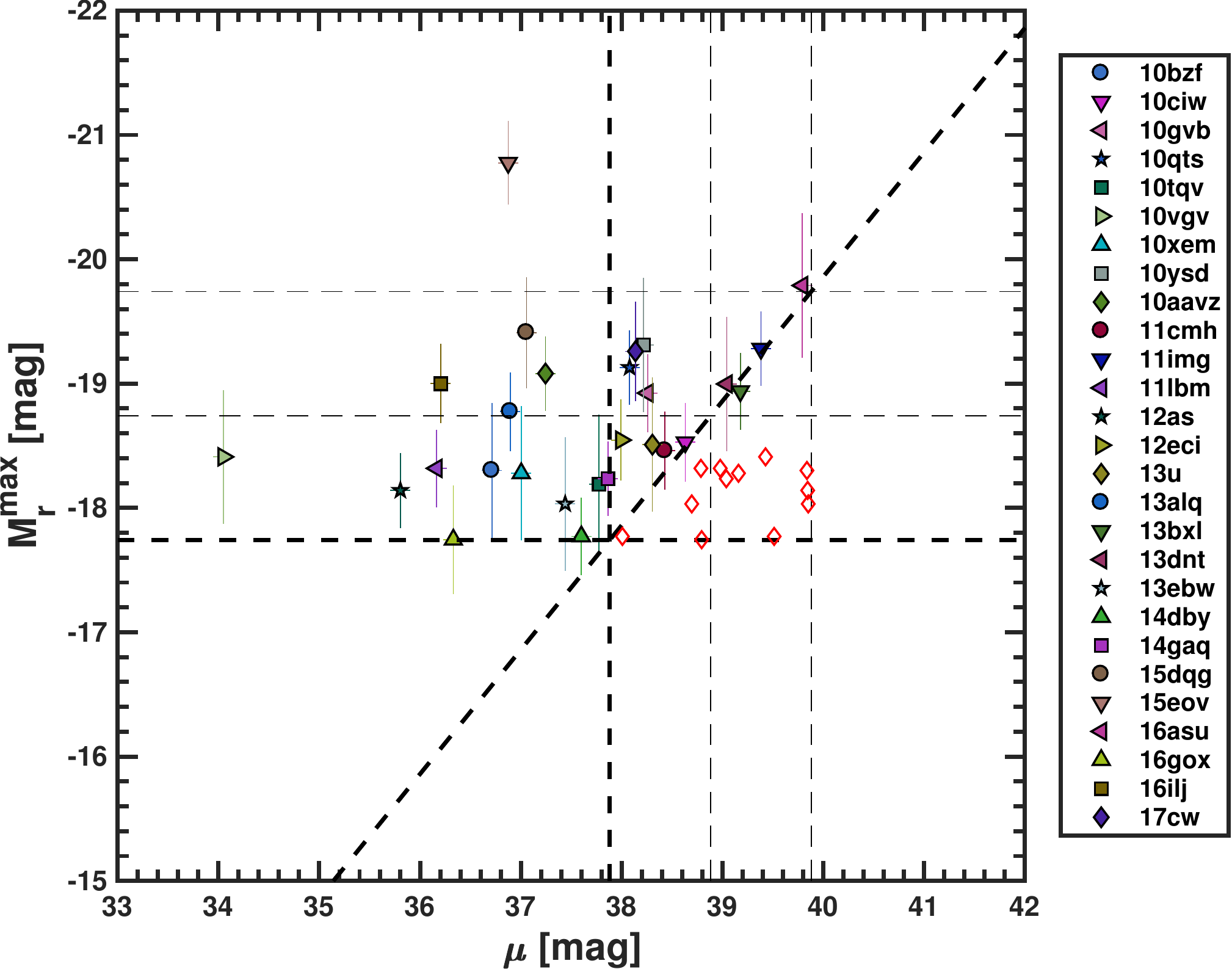} &
  \includegraphics[width=7cm,height=9cm,angle=0]{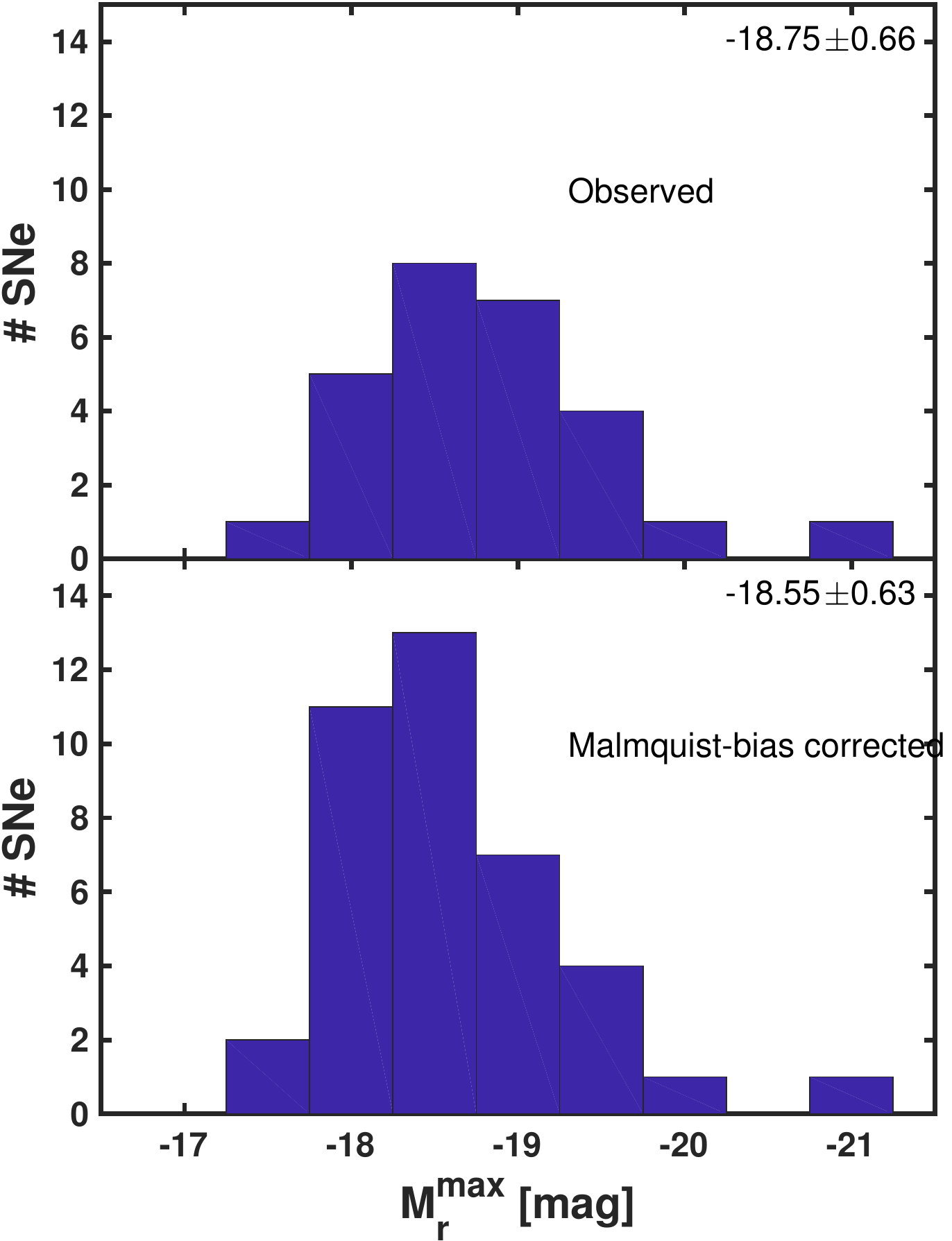} 
    \end{array}$
  \caption{\label{malmquist}(Left panel) Peak absolute magnitude vs. distance modulus. The red empty diamonds are the simulated SNe at large distance moduli ($>37.88$ mag), where the SN sample is not complete owing to the Malmquist bias (see diagonal dashed line). We followed the method by \citet[][see text in Sect.~\ref{sec:discussion}]{richardson14}. (Right panel) Peak magnitude distribution before and after Malmquist-bias correction. It implies a decrease of the $^{56}$Ni mass average of about 17\%.}
 \end{figure}

 We further investigated the effect of the Malmquist bias on the $^{56}$Ni-mass distribution for our SNe, by studying its
 effect on the peak absolute $r$-band magnitude distribution.
 Following the approach by \citet{richardson14}, we plot the peak
 absolute magnitudes versus their respective distance moduli in
 Fig.~\ref{malmquist}.
 We identify the minimum peak magnitude at $-17.74$ mag in
 accordance with the faintest objects in our sample. Given the
 apparent magnitude limit of iPTF (20.5 mag) and considering the extinction,
 this sets the distance modulus within which we observe the complete
 peak magnitude distribution to 37.88 mag.
 The fraction of SNe between $-17.74$ and $-18.74$ mag as well as between $-18.74$ and $-19.74$ mag within a distance modulus of 37.88 mag is considered intrinsic,
 and we reproduce the same fraction at larger distances (distance 
 moduli of 37.88--38.88 mag and 38.88--39.88 mag)
 by randomly simulating the magnitude of the 
 missing SNe in those bins.
 The distribution of the peak magnitudes after creating these new SNe
 has an average of $-18.55\pm0.63$ mag versus the previous
 $-18.75\pm0.66$ mag (see Fig.~\ref{malmquist}).  
 A difference of 0.2 mag in peak luminosity means that the
 Malmquist-bias corrected $^{56}$Ni mass distribution
 has an average that is $\sim 17$\% lower than that of the observed $^{56}$Ni mass distribution. Overall, this is a rather small effect. 

\begin{figure}
 \centering
 $\begin{array}{c}
  \includegraphics[width=12cm,angle=0]{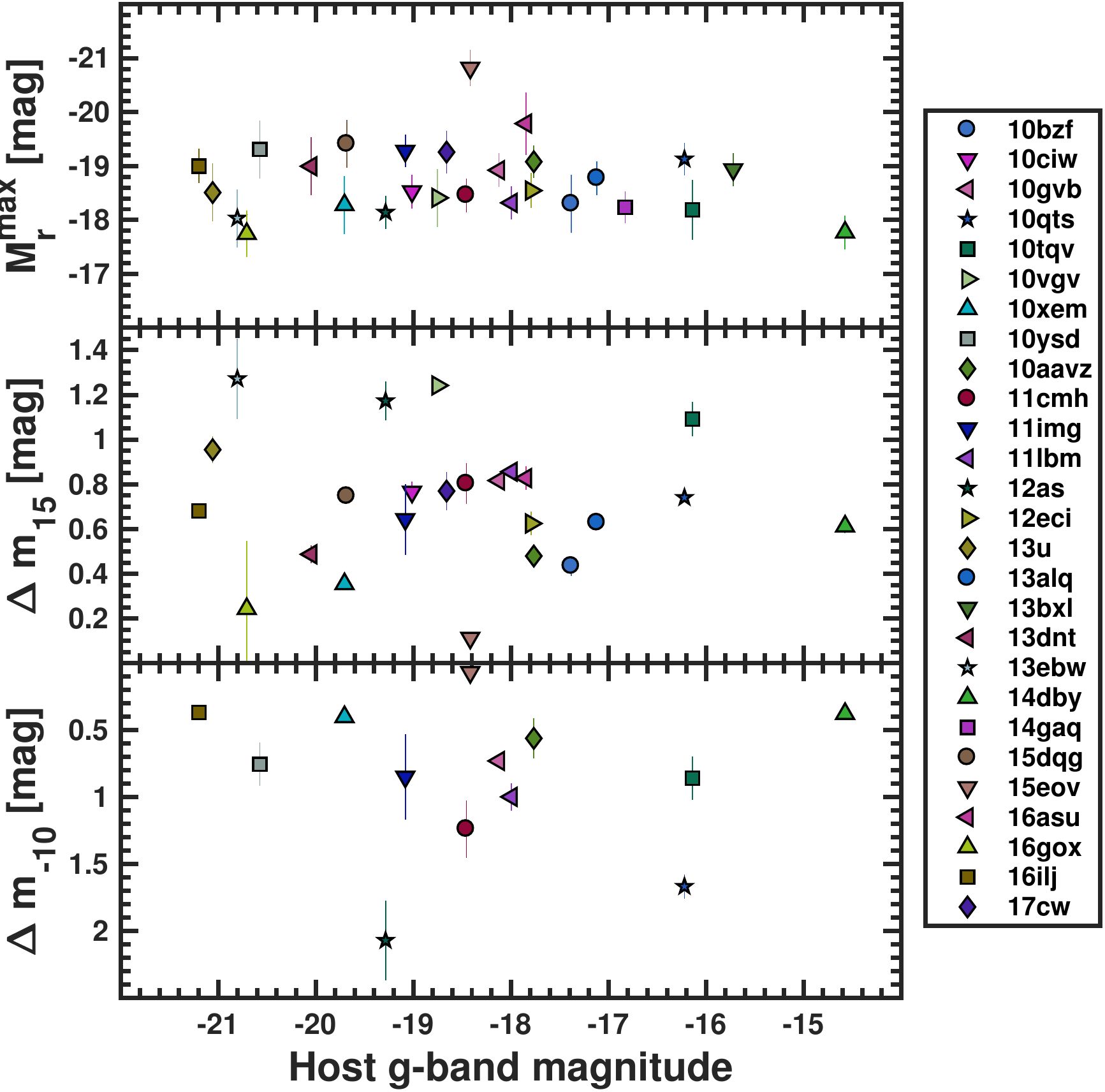}
    \end{array}$
  \caption{\label{untargeted} Light-curve properties in $r$ vs. the host-galaxy $g$ absolute magnitude. Over a range of more than 7~mag in the host-galaxy luminosity, we do not see a clear trend in the light-curve properties. }
 \end{figure}

Our sample is untargeted, so we found SNe in all kinds of
star-forming galaxies within a span of host galaxy magnitudes larger
than 7 mag in the $g$ band (see Table~\ref{tab:sample} and
Fig.~\ref{untargeted}).
In Fig.~\ref{untargeted}, we plot the $r$ light-curve properties
versus the host galaxy $g$ absolute magnitude;
despite the large range of host galaxy luminosity, we do not see clear differences between the populations at low and high host galaxy luminosity.

Finally, we compare the main parameters of our sample to those of 14 GRB-SNe from \citet[][see their Table 1]{cano13}. Their average ejecta mass and the kinetic energy are 6.34$\pm$4.40~M$_{\odot}$ and 2.49$\pm$1.77$\times10^{52}$ erg~s$^{-1}$, the average $^{56}$Ni mass 0.40$\pm$0.27~M$_{\odot}$. The averages from \citet{cano13} are higher than those of our sample (especially the kinetic energy), but still compatible within the uncertainties.
 
\subsection{Implications for progenitor stars and powering mechanism}
\label{sec:prog}

SNe~Ic-BL and GRB SNe have been suggested to come from massive stars,
like Wolf-Rayet (WR) stars, which have initial masses higher than
25--30~M$_{\odot}$, typically have fast rotation \citep{woosley06b}, and have low metallicity \citep[e.g.,][]{sanders12}. 
From the Arnett models, we derived progenitor and explosion parameters for 27 SNe~Ic-BL, assuming these SNe to be powered by radioactivity.
By looking at the top panel of Fig.~\ref{param}, it is clear that both
iPTF15eov and iPTF16asu would require a mass of $^{56}$Ni that is more
than half of their ejecta masses according to the Arnett model.
This suggests that the powering mechanism of these SNe is not (only)
radioactive decay,
but rather another mechanism such as the spindown of a magnetar \citep{kasen10}. 
The other 25 SNe are compatible with radioactive powering of their
light curves, and their derived parameters can be used to constrain
the nature of their progenitor stars.
The ejecta masses are characterized by a majority of events (64\%)
having 
$M_{\rm ej}< 4.5$~M$_{\odot}$ and the peak of the asymmetric
distribution around  2~M$_{\odot}$ (see Fig.~\ref{pdf}).
A non-negligible fraction (21\%) of the SNe have ejecta masses above 
5.5~M$_{\odot}$. 
This is larger than what was found by \citet{lyman16}, where the
probability of finding SNe~Ic-BL with ejecta masses above
5.5~M$_{\odot}$ was almost negligible.
Our SNe with ejecta masses higher than 5.5~M$_{\odot}$ could be
consistent with single massive ($M_{\rm ZAMS} > 28$~M$_{\odot}$) WR-star
progenitors, as shown by \citet{lyman16}. 
The other SNe exhibit lower ejecta masses, which might rather come from
binary stars with lower initial masses ($M_{\rm ZAMS}<$~28~M$_{\odot}$).

An alternative hypothesis to pre-supernova mass loss could be that the
observations of lower ejecta masses imply that some of the
progenitor-star material fell back onto the forming black hole and did
not participate to form the ejecta \citep{colgate71}. However, the large amount of $^{56}$Ni in SNe~Ic-BL and the late-time nebular abundances of core-collapse SNe \citep{jerkstrand18} might disfavor this scenario.

Another possibility is that we inferred low ejecta masses (and also
high energies and high $^{56}$Ni masses)
as a consequence of the assumption related to the symmetry of the ejecta. This is an important caveat to discuss. 
We assumed spherical symmetry in the modeling of our SN bolometric light curves, since we fit these with an Arnett model. 
However, we know that SN~1998bw was likely characterized by
asphericity, as discussed and suggested by many works
\citep[e.g.,][]{maeda06,tanaka07,dessart17}. This was concluded from the analysis
of the early-time and late-time emission of SN~1998bw,
which provides different and inconsistent results in terms of derived
ejecta masses, explosion energy, and $^{56}$Ni masses.
When assuming spherical symmetry, the early peak of the light curve
of SN~1998bw is compatible with low ($\sim 3$~M$_{\odot}$) ejecta
mass, large explosion energy (a few $10^{52}$ erg), and
0.5~M$_{\odot}$ of $^{56}$Ni \citep{dessart17}. 
However, the interpretation of the late-time light curve suggests larger ejecta mass ($\sim 10$~M$_{\odot}$), lower explosion energy ($10^{51}$ erg), and only
0.1~M$_{\odot}$ of $^{56}$Ni \citep{dessart17}. \citet{maeda03} tried to reconcile these conflicting results with a two-zone model in which two different optical depths and Ni masses characterized a slower inner core and a faster envelope. This model still assumes spherical symmetry, which is shown to be inappropriate by \citet{maeda06} and \citet{maeda08} in further light-curve and late-time spectra investigations of SN~1998bw and other GRB SNe.
\citet{dessart17} have declared that it is not possible to have a two-zone model with high density in both zones such as that assumed by
\citet{maeda03}. 

A solution to the discrepancy between early-time and late-time emission
from SN~1998bw could be asymmetric ejecta.  Recently,
\citet{barnes17} used a two-dimensional relativistic hydrodynamical and
radiative-transfer model to show that a central engine that triggers
long GRBs might also produce a SN~Ic-BL. In these models,
which account for asymmetry, it is shown that the effect of the
viewing angle is small on the early-time light curves. It could be more evident in the spectra; however, in Sect.~\ref{barnestest} 
we checked for signs of asymmetry in the spectra of our SNe Ic-BL, but did not find any.

If the ejecta of our SNe~Ic-BL are asymmetric, then the progenitor
parameters derived from the Arnett model might be substantially
different.
In particular, as in SN~1998bw, larger ejecta masses would be
possible, enlarging the fraction of SNe~Ic-BL with ejecta masses consistent with massive single WR stars. 

\begin{figure}
 \centering
 $\begin{array}{c}
  \includegraphics[width=12cm,angle=0]{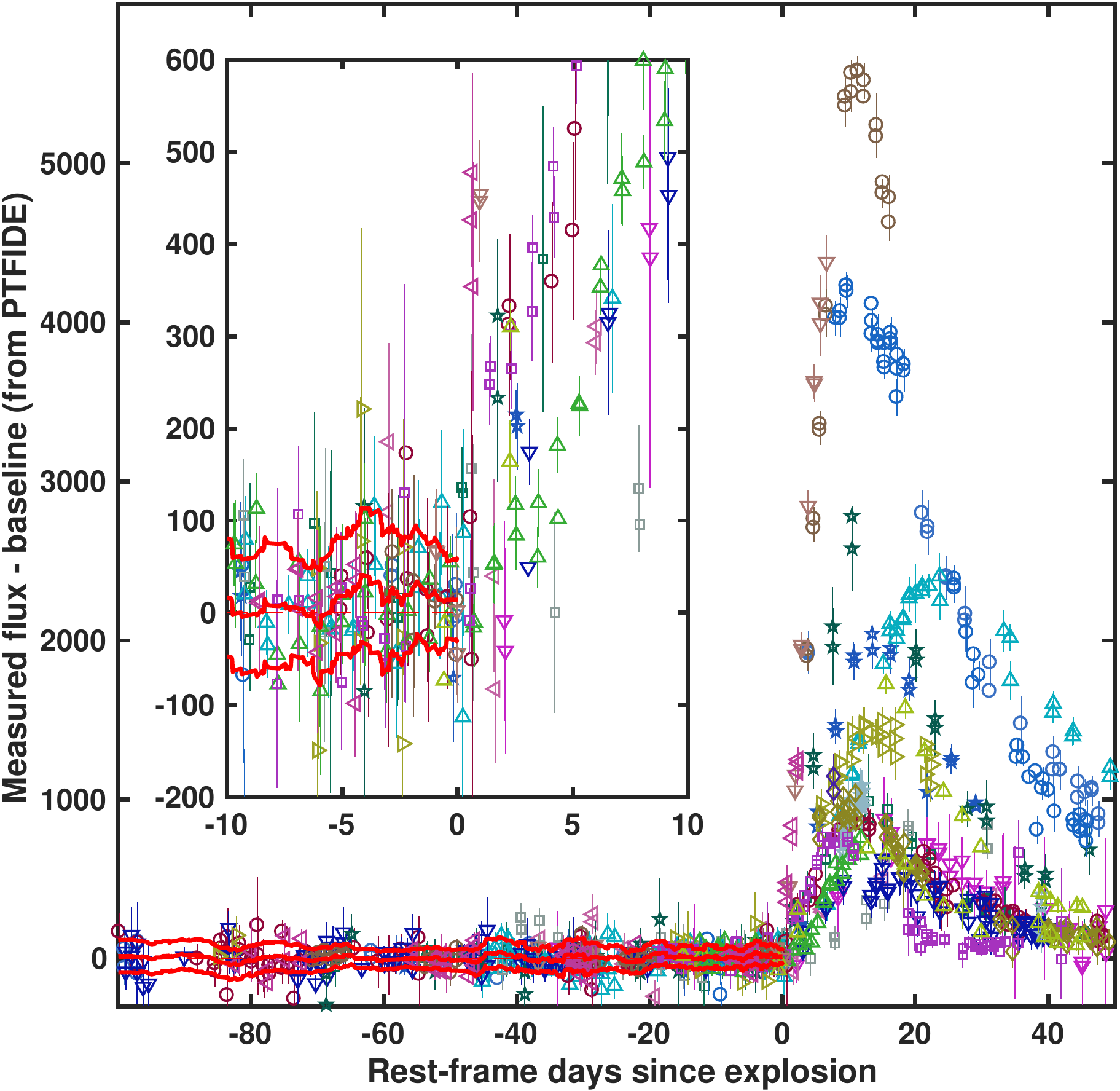}
    \end{array}$
  \caption{\label{baseline} Observed $r$-band ($g$-band in the cases of iPTF14gaq, iPTF15dqg, iPTF16asu, and iPTF16gox) fluxes from PTFIDE minus the zero-flux level (baseline) as a function of time since explosion for all of the SNe with derived explosion epoch and data available a few days before explosion. The inset gives a close-up view of the time range between $-10$ and +10~d. No pre-rise excess is observed between $-10$ and +0~d. Colors and symbols as in Fig.~\ref{param}. A moving mean over 20 points and its enveloping standard deviations are marked by thick solid red lines, and it is consistent with zero flux.}
 \end{figure}

Another potential issue is connected to the explosion epochs that we derived in
Sect.~\ref{sec:explo}. We assumed that our SNe can be represented
by a stretched version of SN~1998bw.
This particular SN did not display any low-luminosity phase between the
time of explosion and the light-curve rise, as predicted, for example, by \citet{dessart11} for a series of SN~Ibc models. 
In \citet{taddia15} we  showed that the four 
SNe~Ic-BL in the SDSS sample did not have such an early dark
phase either. However, we tested whether this early dark phase was indeed absent in our (i)PTF SNe. In Fig.~\ref{baseline}, we plot the fluxes from PTFIDE as a function of time since explosion for all of the SNe with 
estimated explosion epoch, and also with data taken just a few days earlier than the explosion epoch. From the flux obtained from PTFIDE, we subtracted the zero-flux level (baseline) taking the median of the prediscovery fluxes. We then checked whether in the range between $-10$ and 0~d before explosion there is an excess in the fluxes, but we found the data to be consistent with zero flux. Therefore, we conclude that our sample was not characterized by SNe with early light-curve dark plateau phases.
This implies that our explosion epoch estimates are reliable along with the explosion parameters derived using those explosion epochs, such as the large $^{56}$Ni mixing. This mixing implies that some mechanism in the SN explosion brings much inner material to the outer parts of the ejecta.

Yet another possibility that recent works have explored is that SNe~Ic-BL
are powered by magnetars or by the combination of a magnetar and
radioactivity \citep{mazzali14,wang17a,wang17b}. We have seen that
neither iPTF16asu nor iPTF15eov are compatible with solely radioactive
power, whereas the rest of the sample could well be.
(i)PTF has also found other SNe~Ic that are possibly powered by magnetars, such as iPTF15dtg \citep{taddia16_15dtg} and PTF11mnb \citep{taddia17_11mnb}. However, these SNe, such as iPTF16asu and iPTF15eov, showed very peculiar light curves, clearly distinct from those of normal SNe~Ic and SNe~Ic-BL. \citet{cano16magnetar} showed that magnetars cannot be soley responsible for the emission of SNe~Ic-BL. But as pointed out by \citet{dessart17}, the fact that the energetics of SNe~Ic-BL are consistent with a magnetar source does not necessarily mean magnetars are the actual powering source. 

\subsection{Future perspectives: SNe~Ic-BL from Zwicky Transient Facility}
With PTF/iPTF, we found 28 SNe~Ic-BL observed before peak brightness, with good
$r$-band coverage and additional $gBi$ observations, mainly
after the peak.
All of these SNe had good classification spectra and around three spectra per SN on average, mainly before and around peak. 

With the Zwicky Transient Facility (ZTF) we can discover up to $\sim10$
times more SNe, including SNe~Ic-BL. Only a fraction of these will be classified, as we have limited spectroscopic time at the different facilities we can use within the ZTF collaboration. With the P48 during the ZTF era, what
we can do better than iPTF from the observational point of view will be 
(i) obtain color information before peak, as ZTF will observe
both in $g$ and $r$ during the same night, and (ii) provide even tighter constrains on the explosion epoch, given the higher cadence we can reach with a larger field of view. 
Aside from P48, to improve our observations of SNe~Ic-BL, we will need
good spectroscopic and photometric follow-up observations until late epochs,
using the different facilities to which the ZTF collaboration has access. 
With such an improved dataset from ZTF, we will be able to build
more accurate bolometric light curves and velocity profiles
in order to quantify the progenitor properties with higher precision.
Late-epoch data will be of importance to study the asymmetry of the ejecta of these SNe and test alternative powering models.

\section{Summary}
\label{sec:conclusions}

We have presented the optical light curves of the largest existing
homogeneous sample of SNe~Ic-BL from an untargeted survey.
This sample is characterized by good constraints on the SN explosion epochs thanks to the high cadence of PTF and iPTF.
The light-curve shapes demonstrate that the decline parameter and rise parameter are correlated and that fast risers are also fast decliners.

The average peak absolute magnitude in the $r$ band is $-18.6\pm0.5$~mag.
Expansion velocities from the SN spectra are calculated with the technique from \citet{modjaz16}, and are consistent with what was obtained for the SNe~Ic-BL in their sample.

Arnett models fitted to the bolometric light curves and to the velocity profiles imply the following average explosion parameters:
  $<M_{\rm ej}> = 3.9\pm2.8$~M$_{\odot}$, 
  $<E_{K}> = (7.1\pm5.7) \times 10^{51}$~erg, 
  and $<M(^{56}{\rm Ni})> = 0.31\pm0.16$~M$_{\odot}$.
  These results generally agree with those previously
  reported in the literature, but our sample has a number of SNe~Ic-BL
  with large ($>5.5$~M$_{\odot}$) ejecta masses that is clearly
  higher than the numbers found by others \citep[e.g.,][]{lyman16}.

The implications for the progenitors of SNe~Ic-BL are that at
  least 21\% of these SNe are compatible with originating in very
  massive WR stars ($M_{\rm ZAMS} > 28$~M$_{\odot}$). Of the SNe, 64\% might
  come from close binary systems in which the progenitor was less massive
  ($M_{\rm ZAMS} < 20$~M$_{\odot}$). However, our simple models do not account
  for any asymmetry of the SN ejecta, which has been discussed by others
  \citep[e.g.,][]{dessart17}. This might push the actual ejecta masses to higher values for the SNe.
Our hydrodynamical models indicate that our SNe are strongly
  mixed and that the $^{56}$Ni occurs all the way to the outermost layers. 

Finally, we note that ZTF will allow us to observe $\sim 10$ times more SNe~Ic-BL,  will have better color information at early epochs, and tighter constraints on the 
explosion epoch. Substantial follow-up observations will be necessary to fully 
exploit these discoveries.

\begin{acknowledgements}
The Oskar Klein Centre is funded by the Swedish Research
Council. F.T. and J.S. gratefully acknowledge support from the
Knut and Alice Wallenberg Foundation.
D.A.H. and G.H. are supported by National Science
Foundation (NSF) grant AST-1313484.
A.V.F. is grateful for financial assistance from NSF grant AST-1211916, 
the TABASGO Foundation,
the Christopher R. Redlich Fund, and the Miller Institute for Basic
Research in Science (U.C. Berkeley).
A.G.-Y. is supported by the EU via ERC grant No. 725161, the Quantum Universe I-Core program, the ISF, the BSF Transformative program and by a Kimmel award.

The intermediate Palomar Transient Factory project is a scientific collaboration among the California Institute of Technology, Los Alamos National Laboratory, the University of Wisconsin, Milwaukee, the Oskar Klein Center, the Weizmann Institute of Science, the TANGO Program of the University System of Taiwan, and the Kavli Institute for the Physics and Mathematics of the Universe. LANL participation in iPTF is supported by the US Department of Energy as a part of the Laboratory Directed Research and Development program. 

We acknowledge contributions from the full PTF and iPTF collaborations that made it possible to discover and monitor the SE~SNe analyzed in this work.
This work was supported by the GROWTH project funded by  NSF under grant AST-1545949.

We thank the staff at the various observatories at which data were collected.
The William Herschel Telescope is operated on the island of La Palma by the Isaac Newton
Group of Telescopes in the Spanish Observatorio del Roque de
los Muchachos of the Instituto de Astrofísica de Canarias. 
The data presented herein were obtained in part with ALFOSC, which is provided by the Instituto de Astrofisica de Andalucia (IAA) under a joint agreement with the University of Copenhagen and NOTSA. 
This work is partly based on
observations made with DOLoRes@TNG. 
Our results made use
of the Discovery Channel Telescope (DCT) at Lowell Observatory. Lowell is a private, nonprofit institution dedicated to astrophysical research and public appreciation of astronomy, and
it operates the DCT in partnership with Boston University, the
University of Maryland, the University of Toledo, Northern Arizona University, and Yale University. The upgrade of the DeVeny
optical spectrograph has been funded by a generous grant from John and Ginger Giovale. 
This work makes use of the Las Cumbres Observatory network and the LCO Supernova Key Project. 
Research at Lick Observatory is partially supported by a generous gift from Google.

This research has made use of the NASA/IPAC Extragalactic Database (NED) which is operated by the Jet Propulsion Laboratory, California Institute of Technology, under contract with the National Aeronautics and Space Administration (NASA). Some of the data presented herein were obtained at the W. M. Keck Observatory, which is operated as a scientific partnership among the California Institute of Technology, the University of California, and NASA. The Observatory was made possible by the generous financial support of the W. M. Keck Foundation. The authors wish to recognize and acknowledge the very significant cultural role and reverence that the summit of Maunakea has always had within the indigenous Hawaiian community. We are most fortunate to have the opportunity to conduct observations from this mountain.

We thank Mark Sullivan for
contributing spectral data obtained under his observing programs. 
We acknowledge the large number
of observers and those responsible for data reduction who helped acquire the spectroscopic
database over the years. Among others, these include
S. Adams,
C. Badenes,
S. Ben-Ami,
N. Blagorodnova,
J. S. Bloom, 
J. Botyanszki, 
K. Burdge, 
Y. Cao, 
J. Choi,
N. Chotard, 
K. I. Clubb, 
D. Cook,
N. Cucchiara, 
A. De Cia,
A. Drake, 
G. Duggan,
M. Fraser, 
M. L. Graham,
T. Hashimoto, 
A. Ho,
I. Hook, 
A. Horesh, 
E. Hsiao,
T. Hung, 
M. Kandrashoff,
M. Kasliwal,
E. Kirby,
I. Kleiser, 
S. Knezevic,
T. Kupfer, 
K. Maguire, 
T. Matheson,
A. A. Miller, 
K. P. Mooley, 
E. Ofek,                                           
Y.-C. Pan, 
D. Polishook, 
D. Poznanski,  
V. Ravi, 
J. Rex,
A. Rubin,
B. Sesar, 
I. Shivvers, 
A. Sternberg,
N. Suzuki, 
D. Tal,
C. Theissen, 
J. van Roestel.                          
P. Vreeswijk,
E. Walker,
A. Waszczak, 
D. Xu,
O. Yaron, and
B. Zackay.
We also thank the staff at the observatories where data
were obtained.
We acknowledge Umaa Rebbapragada and her group for the work with the machine-learning algorithm that allowed iPTF to discover transients.
We thank Maryam Modjaz for the productive discussion on the SN classification.

\end{acknowledgements}

\bibliographystyle{aa}

\onecolumn

\clearpage
\begin{deluxetable}{lllclllllccc}
\rotate
\tabletypesize{\scriptsize}
\tablewidth{0pt}
\tablecaption{PTF/iPTF sample of 34 SNe~Ic-BL.\label{tab:sample}}
\tablehead{
\colhead{SN} &
\colhead{$\alpha$ (J2000.0)} &
\colhead{$\delta$} &
\colhead{Galaxy} &
\colhead{Redshift} &
\colhead{$\mu$} &
\colhead{Distance} &
\colhead{$E(B-V)^{\rm MW}$} &
\colhead{$M_g^{\rm gal}$} &
\colhead{$E(B-V)^{\rm host}$} &
\colhead{$E(B-V)^{\rm host}$}&
\colhead{Other work}  \\
\colhead{} &
\colhead{(hh:mm:ss)} &
\colhead{(dd:mm:ss)} &
\colhead{} &
\colhead{} &
\colhead{(mag)} &
\colhead{(Mpc)} &
\colhead{(mag)} &
\colhead{(mag)} &
\colhead{from $g-r$} &
\colhead{from \ion{Na}{i}~D}&
\colhead{}\\
\colhead{} &
\colhead{} &
\colhead{} &
\colhead{} &
\colhead{} &
\colhead{} &
\colhead{} &
\colhead{} &
\colhead{} &
\colhead{(mag)} &
\colhead{(mag)}&
\colhead{}}
\startdata
 PTF09sk                   & 13:30:51.147 & +30:20:04.88 &  SDSS J133051.17+302002.1     &   0.0355   &  36.07  &  163.6(11.6)   & 0.0100 &  -18.12  &  \ldots     &    \ldots       &    \\
 PTF10cs                   & 02:33:41.620 & +34:22:17.10 &  GALEXASC J023341.18+342218.8 &   0.1660   &  39.48  &  788.4(55.9)   & 0.0526 &  -19.75  &  \ldots     &    0.04(0.20)   &    \\
 PTF10bzf                  & 11:44:02.988 & +55:41:27.64 &  SDSS J114402.97+554122.8     &   0.0498*  &  36.76  &  224.8(15.9)   & 0.0098 &  -17.38  &  \ldots     &    \ldots       & {\bf C11}, {\bf M13}, C13, M16, C16 \\
 PTF10ciw                  & 12:33:25.460 & +14:25:37.29 &  SDSS J123325.58+142538.4     &   0.1150   &  38.66  &  540.0(38.3)   & 0.0321 &  -19.01  &  0.00(0.10) &    \ldots       &    \\
 PTF10gvb                  & 12:15:32.281 & +40:18:09.45 &  SDSS J121532.31+401810.6     &   0.0980   &  38.29  &  454.1(32.2)   & 0.0212 &  -18.12  &  0.00(0.10) &    \ldots       &    Pr16\\
 PTF10qts          & 16:41:37.598 & +28:58:21.08 &  SDSS J164137.53+285820.3     &   0.0907   &  38.11  &  419.6(29.8)   & 0.0249 &  -16.22  &  0.02(0.10) &    \ldots       &    {\bf W14}, C16, M16\\
 PTF10tqv                  & 22:46:55.059 & +17:47:29.40 &  SDSS J224654.99+174728.5     &   0.0795   &  37.76  &  356.7(25.3)   & 0.0610 &  -16.14  &  \ldots     &    \ldots       & \\
 PTF10vgv                  & 22:16:01.169 & +40:52:03.30 &  LCSB S2641N                  &   0.0150   &  34.10  &   66.2(4.7)    & 0.1416 &  -18.75  &  \ldots     &    \ldots       &    {\bf C12} , P13, Pr16, M16\\
 PTF10xem                  & 01:47:06.884 & +13:56:28.75 &  SHOC 084                     &   0.0566   &  36.96  &  246.9(17.5)   & 0.0467 &  -19.70  &  \ldots     &    \ldots       &    C16 \\
 PTF10ysd                  & 23:41:04.609 & +24:10:04.86 &  2MASX J23410465+2410044      &   0.0963   &  38.20  &  436.3(30.9)   & 0.0493 &  -20.57  &  \ldots     &    0.40(0.20)   &    \\
 PTF10aavz                 & 11:20:13.357 & +03:44:45.24 &  SDSS J112013.37+034442.6     &   0.0630   &  37.30  &  288.5(20.5)   & 0.0479 &  -17.77  &  0.00(0.10) &    0.08(0.20)   & C16    \\
 PTF11cmh                  & 13:10:21.743 & +37:52:59.56 &  SDSS J131021.75+375257.7     &   0.1055   &  38.46  &  491.7(34.9)   & 0.0121 &  -18.46  &  0.00(0.11) &    \ldots       & C16 \\
 PTF11img                  & 17:34:36.304 & +60:48:50.64 &  GALEXMSC J173436.35+604851.7 &   0.1580   &  39.39  &  755.3(53.6)   & 0.0382 &  -19.08  &  0.12(0.10) &    \ldots       & C16    \\
 PTF11lbm                  & 23:48:03.195 & +26:44:33.49 &  SDSS J234803.25+264429.6     &   0.0390   &  36.14  &  168.7(12.0)   & 0.0547 &  -17.99  &  0.14(0.11) &    \ldots       & C16     \\
 PTF12as                   & 10:01:34.049 & +00:26:58.41 &  SDSS J100133.74+002655.2     &   0.0332   &  35.89  &  150.8(10.7)   & 0.0213 &  -19.28  &  0.25(0.10) &    \ldots       & C16 \\
 PTF12eci                  & 10:09:57.812 & +46:15:02.84 &  SDSS J100957.90+461501.9     &   0.0874   &  38.01  &  400.1(28.4)   & 0.0068 &  -17.79  &  0.00(0.11) &    \ldots       &    \\
 PTF12grr                  & 16:25:14.523 & +24:23:17.82 &  SDSS J162514.46+242317.8     &   0.16**   &  39.43  &  769.9(54.6)   & 0.0417 &  -16.71  &  \ldots     &    \ldots       &    \\
iPTF13u                    & 15:58:51.205 & +18:13:53.13 &  SDSS J155851.42+181359.5     &   0.0991   &  38.34  &  466.1(33.1)   & 0.0387 &  -21.06  &  \ldots     &    \ldots       &  C16  \\
iPTF13alq                  & 11:48:02.088 & +54:34:38.18 &  SDSS J114802.12+543438.8     &   0.0540   &  36.94  &  244.2(17.3)   & 0.0118 &  -17.12  &  0.00(0.11) &    \ldots       &  C16   \\
iPTF13bxl$^{\rm GRB}$          & 14:29:14.780 & +15:46:26.45 &  Anon.                        &   0.1450   &  39.20  &  693.2(49.2)   & 0.0368 &  -15.72  &  0.00(0.10) &    \ldots       &    {\bf S13},{\bf D15},{\bf T16},Pr16,M16,{\bf V17}\\
iPTF13dnt                  & 21:15:21.033 & +12:25:43.56 &  SDSS J211520.92+122546.7     &   0.1370   &  39.04  &  643.1(45.6)   & 0.0624 &  -20.05  &  \ldots     &    \ldots       &    \\
iPTF13ebw                  & 08:17:15.884 & +56:34:41.59 &  2MASX J08171533+5634457      &   0.0686   &  37.45  &  308.6(21.9)   & 0.0550 &  -20.80  &  \ldots     &    \ldots       &   C16  \\
iPTF14bfu$^{\rm GRB}$          & 21:52:29.970 & +32:00:50.59 &  Anon.                        &   0.3840   &  41.58  &  2069.9(146.8) & 0.1022 &  -16.98  &  \ldots     &    \ldots       &    {\bf C15}\\
iPTF14dby                  & 15:17:06.287 & +25:21:11.44 &  SDSS J151706.24+252111.5     &   0.0736   &  37.65  &  339.3(24.1)   & 0.0464 &  -14.57  &  0.00(0.10) &    \ldots       &   C16, Pr16  \\
iPTF14gaq                  & 21:32:54.077 & +17:44:35.60 &  PSO J213254.044+174433.543   &   0.0826   &  37.86  &  373.5(26.5)   & 0.0979 &  -16.83  &  0.00(0.10) &    \ldots       &   C16  \\
iPTF15dld                  & 00:58:13.277 & -03:39:50.30 &  SDSS J005813.27-033950.0     &   0.0470   &  36.54  &  203.4(14.4)   & 0.0269 &  -18.37  &  \ldots     &    0.04(0.20)   &  C16, {\bf P17}   \\
iPTF15dqg                  & 00:09:21.682 & -00:04:06.60 &  SDSS J000921.56-000413.2     &   0.0577   &  37.03  &  254.3(18.0)   & 0.0530 &  -19.68  &  0.00(0.16) &    \ldots       &    \\
iPTF15eov                  & 04:05:41.568 & +28:56:40.67 &  PSO J040541.612+285641.698   &   0.0535   &  36.84  &  233.3(16.5)   & 0.5275 &  -18.42  &  0.00(0.12) &    \ldots       &    \\
iPTF16asu                  & 12:59:09.277 & +13:48:09.19 &  SDSS J125909.30+134808.8     &   0.1874   &  39.81  &  917.0(65.0)   & 0.0277 &  -17.85  &  0.00(0.21) &    \ldots       &    {\bf W17}\\
iPTF16coi                  & 21:59:04.138 & +18:11:11.03 &  UGC 11868                    &   0.0036   &  31.37  &  18.8(1.4)     & 0.0731 &  -16.85  &  \ldots     &    0.24(0.20)   &  {\bf K17,Pr18}  \\
iPTF16gox                  & 00:12:28.234 & +32:45:02.24 &  SDSS J001227.58+324509.8     &   0.0420   &  36.30  &  181.9(12.9)   & 0.0377 &  -20.71  &  0.16(0.15) &    \ldots       &    \\
iPTF16ilj                  & 00:55:16.307 & +12:19:11.20 &  CGCG 435-021                 &   0.0397   &  36.16  &  170.6(12.1)   & 0.0544 &  -21.20  &  0.29(0.11) &    \ldots       &    \\
iPTF17cw                   & 09:03:38.377 & +43:05:50.28 &  SDSS J090338.46+430551.7     &   0.0930   &  38.15  &  425.8(30.2)   & 0.0167 &  -18.66  &  0.03(0.14) &    \ldots       &    {\bf C17}\\
iPTF17axg                  & 12:36:30.748 & +51:38:40.02 &  SDSS J123630.71+513838.3     &   0.0590   &  37.14  &  268.0(19.0)   & 0.0128 &  -17.18  &  \ldots     &    \ldots       &    \\
\enddata
\tablecomments{(*) Redshift from \citet{corsi11}. (**) Redshift from best SNID fit. PTF10bzf is also known as SN~2010ah. iPTF13alq is also known as SN2013bn. iPTF13bxl is also known as  SN~2013dx. iPTF16coi is also known as  ASASSN-16fp. SNe which are associated with GRBs are marked by $^{\rm GRB}$. The uncertainty in the distance modulus is 0.154~mag; it is dominated by the uncertainty in H$_0$.
The last column includes references to other works on our (i)PTF SNe~Ic-BL. Those in boldface are single-object papers. The following abbreviations are used: 
C11$=$\citet{corsi11}; 
C12$=$\citet{corsi12}; 
C13$=$\citet{cano13}; 
C15$=$\citet{cano15}; 
C16$=$\citet{corsi16};  
C17$=$\citet{corsi17}; 
D15$=$\citet{delia15}; 
K17$=$\citet{kumar17};
M13$=$\citet{mazzali13}; 
M16$=$\citet{modjaz16}; 
P13$=$\citet{piro13}.
P17$=$\citet{pian17}; 
Pr16$=$\citet{prentice16}; 
Pr18$=$\citet{prentice17b}; 
S13$=$\citet{singer13}; 
T16$=$\citet{toy16}; 
V17$=$\citet{volvonova17}; 
W14$=$\citet{walker14}; 
W17$=$\citet{whitesides17}.}
\end{deluxetable}

\begin{deluxetable}{cccccccc}
\tabletypesize{\scriptsize}
\tablewidth{0pt}
\tablecaption{The $r$-band light-curve properties and explosion epochs$^a$\label{tab:contardo_r}}
\tablehead{
\colhead{SN}&
\colhead{$t(r_{\rm max})$}&
\colhead{$r_{\rm max}$}&
\colhead{$\Delta m_{15}$}&
\colhead{$\Delta m_{-10}$}&
\colhead{Linear slope}&
\colhead{$M_{r}^{\rm max}$}&
\colhead{$t_{\rm explo}^b$}\\
\colhead{}&
\colhead{(JD)}&
\colhead{(mag)}&
\colhead{(mag)}&
\colhead{(mag)}&
\colhead{(mag~d$^{-1}$)}&
\colhead{(mag)}&
\colhead{(JD)}}
\startdata
    PTF10bzf  &  2455261.03(1.91) & 18.48(0.06) & 0.43(0.05) & \ldots     & 0.019(0.001) &  -18.30(0.75) & 2455246.94 \\
    PTF10ciw  &  2455262.38(0.26) & 20.22(0.11) & 0.77(0.05) & \ldots     & 0.060(0.005) &  -18.53(0.41) & 2455242.69 \\
    PTF10gvb  &  2455335.97(0.16) & 19.42(0.06) & 0.82(0.01) & 0.73(0.01) & \ldots       &  -18.92(0.41) & 2455316.10 \\
    PTF10qts  &  2455425.84(0.04) & 19.10(0.05) & 0.74(0.01) & 1.67(0.09) & \ldots       &  -19.13(0.39) & 2455412.10 \\
    PTF10tqv  &  2455437.88(1.12) & 19.73(0.15) & 1.09(0.08) & 0.86(0.16) & \ldots       &  -18.19(0.76) & 2455422.67 \\
    PTF10vgv  &  2455463.78(0.02) & 16.06(0.01) & 1.24(0.01) & \ldots     & 0.067(0.013) &  -18.41(0.74) & 2455453.47 \\
    PTF10xem  &  2455486.44(0.16) & 18.80(0.04) & 0.35(0.01) & 0.40(0.01) & 0.032(0.007) &  -18.28(0.75) & 2455461.65 \\
    PTF10ysd  &  2455491.16(0.38) & 20.05(0.05) & \ldots     & 0.75(0.16) & \ldots       &  -19.31(0.75) & 2455465.00 \\
    PTF10aavz &  2455529.12(0.54) & 18.34(0.04) & 0.48(0.05) & 0.56(0.15) & 0.016(0.001) &  -19.08(0.39) & 2455507.98 \\
    PTF11cmh  &  2455682.98(0.07) & 20.03(0.03) & 0.80(0.09) & 1.24(0.21) & \ldots       &  -18.46(0.42) & 2455669.28 \\
    PTF11img  &  2455767.27(1.07) & 20.53(0.04) & 0.64(0.16) & 0.85(0.32) & \ldots       &  -19.28(0.39) & 2455748.34 \\
    PTF11lbm  &  2455805.47(0.10) & 18.32(0.01) & 0.86(0.01) & 1.00(0.10) & \ldots       &  -18.32(0.41) & 2455793.76 \\
    PTF12as   &  2455933.55(0.50) & 18.45(0.06) & 1.17(0.09) & 2.07(0.30) & 0.045(0.005) &  -18.14(0.39) & 2455921.01 \\
    PTF12eci  &  2456067.34(0.13) & 19.48(0.02) & 0.63(0.05) & \ldots     & \ldots       &  -18.55(0.43) & 2456052.37 \\
   iPTF13u    &  2456330.19(0.22) & 19.93(0.04) & 0.95(0.06) & \ldots     & \ldots       &  -18.51(0.75) & 2456318.90 \\
   iPTF13alq  &  2456398.28(0.10) & 18.19(0.02) & 0.63(0.01) & \ldots     & \ldots       &  -18.77(0.42) & 2456386.46  \\
   iPTF13bxl  &  2456492.01(0.60) & 20.36(0.09) & \ldots  & \ldots     & \ldots       &  -18.94(0.40) & 2456475.00 \\
   iPTF13dnt  &  2456566.01(0.63) & 20.21(0.06) & 0.49(0.04) & \ldots     & \ldots       &  -19.00(0.75) & 2456545.04 \\
   iPTF13ebw  &  2456630.13(0.19) & 19.56(0.01) & 1.27(0.18) & \ldots     & 0.007(0.004) &  -18.03(0.74) & 2456612.21 \\
   iPTF14dby  &  2456850.34(0.29) & 20.00(0.07) & 0.61(0.03) & 0.38(0.01) & \ldots       &  -17.77(0.40) & 2456828.08 \\
   iPTF14gaq  &  2456930.24(0.22) & 19.88(0.03) & \ldots     & \ldots     & \ldots       &  -18.23(0.40) & 2456920.21 \\
   iPTF15dqg  &  2457342.21(0.20) & 17.75(0.01) & 0.75(0.03) & \ldots     &\ldots &  -19.41(0.61) & 2457328.73 \\
iPTF15eov &  2457386.96(0.45) & 17.38(0.02) & 0.11(0.01) & 0.07(0.01) & \ldots &  -20.82(0.45) & 2457358.72    \\
   iPTF16asu  &  2457525.65(0.09) & 20.09(0.10) & 0.83(0.05) & \ldots     & \ldots       &  -19.79(0.80) & 2457519.03 \\
   iPTF16gox  &  2457647.26(4.92) & 19.06(0.15) & 0.24(0.30) & \ldots     & \ldots       &  -17.74(0.58) & 2457641.59 \\
   iPTF16ilj  &  2457721.67(0.21) & 18.05(0.03) & 0.68(0.01) & 0.37(0.03) & \ldots       &  -19.00(0.42) & 2457701.25 \\
   iPTF17cw   &  2457767.98(1.20) & 19.00(0.07) & 0.77(0.09) & \ldots     & \ldots       &  -19.26(0.54) & 2457752.40 \\  
\enddata

\tablenotetext{a}{For the 27 SNe~Ic-BL observed before and after $r$ maximum brightness.}
\tablenotetext{b}{Obtained by stretching the $R$-band light curve of  SN~1998bw to the $r$-band light curves of each of our SNe. Typical uncertainty of $\pm2$~d (see text).}
\end{deluxetable}

\begin{deluxetable}{ccccccc}
\tabletypesize{\scriptsize}
\tablewidth{0pt}
\tablecaption{Bolometric light curve properties and expansion velocities of the SNe listed in Table~\ref{tab:contardo_r}.\label{tab:contardo_bolo}}
\tablehead{
\colhead{SN}&
\colhead{$t(\rm bolo)_{\rm max}$}&
\colhead{$M(\rm bolo)_{\rm max}$}&
\colhead{$\Delta m^{\rm bolo}_{15}$}&
\colhead{$\Delta m^{\rm bolo}_{-10}$}&
\colhead{Linear slope (bolo)}&
\colhead{\ion{Fe}{ii} velocity at $r$-band peak}\\
\colhead{}&
\colhead{(JD)}&
\colhead{(mag)}&
\colhead{(mag)}&
\colhead{(mag)}&
\colhead{(mag~d$^{-1}$)}&
\colhead{(km~s$^{-1}$)}}
\startdata
  PTF10bzf  &  2455260.57(2.96) & -18.20(0.34)   & 0.47(0.11)  & \ldots     & 0.032(0.005)  &   19237(1154)   \\
  PTF10ciw  &  2455258.68(0.14) & -18.59(0.11)   & 0.90(0.08)  & \ldots     & 0.069(0.004)  &      22036(8208)   \\
  PTF10gvb  &  2455332.79(0.79) & -18.72(0.11)   & 0.70(0.05)  & \ldots     & \ldots       &       11952(1964)   \\
  PTF10qts  &  2455423.26(0.37) & -19.00(0.10)   & 0.78(0.09)  & \ldots     & \ldots       &       21830(923)   \\
  PTF10tqv  &  2455433.84(1.35) & -18.49(0.26)   & 1.30(0.05)  & 1.75(0.01) & \ldots        &      18985(2298)   \\
  PTF10vgv  &  2455462.64(0.45) & -18.23(0.06)   & 1.19(0.10)  & \ldots     & 0.031(0.003)  &       8401(2187)   \\
  PTF10xem  &  2455485.54(2.89) & -18.10(0.14)   & 0.33(0.02)  & 0.45(0.03) & \ldots &      11229(2268)   \\
  PTF10ysd  &  2455487.95(1.34) & -19.25(0.10)   & \ldots      & \ldots     & \ldots       &       10900(3336)   \\
  PTF10aavz &  2455526.81(0.29) & -18.72(0.28)   & 0.41(0.04)  & 0.50(0.09) & 0.013(0.001)  &       12040(2129)   \\
  PTF11cmh  &  2455681.25(0.36) & -18.34(0.06)   & 0.84(0.19)  & \ldots     & \ldots       &       15195(1852)   \\
  PTF11img  &  2455763.86(2.38) & -18.98(0.21)   & 0.49(0.19)  & 1.02(0.09) & \ldots       &       19088(2049)   \\
  PTF11lbm  &  2455804.66(2.54) & -18.13(0.12)   & 0.91(0.15)  & 0.96(0.13) & \ldots       &       10729(1537)   \\
  PTF12as   &  2455933.10(0.66) & -17.92(0.11)   & 1.16(0.10)  & \ldots     & 0.037(0.001) &      11055(2170)   \\
  PTF12eci  &  2456080.89(7.83) & -18.70(0.18)   & \ldots      & 0.41(0.50) & \ldots       &        9678(1741)   \\
 iPTF13u    &  2456327.74(0.44) & -18.41(0.04)   & 0.97(0.21)  & \ldots     & \ldots       &     21499(2066)   \\
 iPTF13alq  &  2456394.86(2.28) & -18.64(0.04)   & 0.55(0.13)  & \ldots     & \ldots       &       18751(1365)   \\
 iPTF13bxl  &  2456488.06(0.13) & -18.90(0.10)   & 1.12(0.09)  & \ldots     & \ldots       &       14820(1997)   \\
 iPTF13dnt  &  2456563.75(0.93) & -18.90(0.06)   & 0.59(0.19)  & \ldots     & \ldots       &       20821(1911)   \\
 iPTF13ebw  &  2456628.26(0.60) & -17.94(0.10)   & 1.28(0.15)  & \ldots     & 0.008(0.014)  &      24670(1182)   \\
 iPTF14dby  &  2456846.66(2.28) & -17.60(0.12)   & 0.48(0.11)  & 0.76(0.47) & \ldots       &       10708(2098)   \\
 iPTF14gaq  &  2456928.89(0.19) & -18.35(0.07)   & \ldots      & \ldots     & \ldots       &       24782(1800)   \\
 iPTF15dqg  &  2457340.13(0.56) & -19.28(0.03)   & 0.73(0.08)  & \ldots     & \ldots         &     13126(1467)   \\
 iPTF15eov  &  2457373.72(1.48) & -21.43(0.25)   & 0.29(0.05)  & 0.93(0.01) & \ldots        &      17695(410)   \\
 iPTF16asu  &  2457524.85(0.68) & -20.36(0.30)   & \ldots      & \ldots     & \ldots       &       29389(882)   \\
 iPTF16gox  &  2457656.93(0.47) & -17.51(0.31)   & 1.00(0.09)  & 1.07(0.68) & \ldots       &       23705(2911)   \\
 iPTF16ilj  &  2457718.90(2.06) & -18.81(0.11)   & 0.54(0.07)  & 0.37(0.66) & \ldots       &       19341(1821)   \\
 iPTF17cw   &  2457763.88(0.75) & -19.22(0.04)   & 0.74(0.07)  & \ldots     & \ldots       &      19526(1999)   \\
 \enddata
\end{deluxetable}

 \begin{deluxetable}{lcccccccc}
\tablewidth{0pt}
\tabletypesize{\scriptsize}
\tablecaption{\label{tab:speclog}Spectral log and \ion{Fe}{ii} velocity measurements.}
\tablehead{
\colhead{SN}       &
\colhead{Ic-$<N>^{**}$}&
\colhead{JD}&
\colhead{UT Date}&
\colhead{Telescope}&
\colhead{Wavelength range}&
\colhead{Phase [Rest-frame days]}&
\colhead{\ion{Fe}{ii} Velocity}&
\colhead{$T_{\rm BB}$}\\
\colhead{}&
\colhead{}&
\colhead{}&
\colhead{[yyyymmdd]}&
\colhead{}&
\colhead{[$\text{\AA}$]}&
\colhead{[since $r$-band maximum]}&
\colhead{[km~s$^{-1}$]}&
\colhead{[K]}}
\startdata
  10bzf  & 3        & 2455256.5   & 20100301 & GeminiN &    3501--9652 & -5 & 21660(1501)        &  8121(32)      \\
  10bzf  &          & 2455262.5   & 20100307 & Keck 1 &    3050--10200 & 1 & 17664(1541)          &  7895(111)     \\\hline
  10ciw  & 2-3      & 2455262.5   & 20100307 & Keck 1 &    3342--10239 & 0 & 21334(8208)          &  7362(57)     \\\hline
  10gvb  & 3        & 2455322.9   & 20100506 & Keck 1 &    3500--10000 & -13 & \ldots             &  13885(38)      \\
  10gvb  &          & 2455331.5   & 20100515 & Keck 1 &    3300--10180 & -4 & 15300(1505)         &  8573(33)     \\
  10gvb  &          & 2455385.5   & 20100708 & Keck 1 &    3120--10040 & 50 & 7038(1525)          &  4173(43)     \\\hline
  10qts  & 2-3      & 2455421.5   & 20100813 & P200 &    3505--10100 & -4 & 19540(1504)          &  11934(82)      \\
  10qts  &          & 2455423.5   & 20100815 & Lick 3~m &    3482--10012 & -2 & 20579(1506)       &  9069(42)     \\
  10qts  &          & 2455441.5   & 20100902 & P200 &    3440--9850 & 16 & 20375(1546)           &  4770(33)     \\
  10qts  &          & 2455444.5   & 20100905 & P200 &    3440--9850 & 19 & 19595(1985)           &  5551(33)     \\
  10qts  &          & 2455448.5   & 20100909 & KPNO 4~m &    3622--8144 & 23 & \ldots              &  5231(33)     \\
  10qts  &          & 2455678.5   & 20110427 & Keck 1 &    3100--11000 & 253 & \ldots             &  \ldots     \\\hline
10tqv & 3 & 2455441.5   & 20100902 & P200 &    3440--9850 & 4 & \ldots& 5208(22)\\
10tqv & & 2455472.5   & 20101003 & Keck 1 &    3090--10240 & 35 & 17137(2570) & 5080(31)\\
10tqv & & 2455501.5   & 20101101 & Keck 1 &    3500--7500 & 64 & 10944(1641)& \ldots\\\hline
  10vgv  & 2-3      & 2455455.8   & 20100916 & Lick 3~m &    3440--10128 & -8 & \ldots            &  9008(20)     \\
  10vgv  &          & 2455466.5   & 20100927 & HET &    4232--10356 & 3 & 8030(1500)             &  5471(80)     \\
  10vgv  &          & 2455470.8   & 20101001 & Lick 3~m &    3440--10700 & 7 & \ldots             &  4971(20)     \\
  10vgv  &          & 2455499.5   & 20101030 & P200 &    3340--9500 & 36 & \ldots                &  4806(31)     \\
  10vgv  &          & 2455536.5   & 20101206 & P200 &    3375--9760 & 73 & \ldots                &  \ldots     \\\hline
  10xem  & 2-3      & 2455479.9   & 20101010 & Lick 3~m &    3500--10000 & -7 & 14640(1506)       &  7544(21)     \\
  10xem  &          & 2455480.5   & 20101011 & KPNO 4~m &    3840--8400 & -6 & 10410(5015)         &  7877(29)     \\
  10xem  &          & 2455501.5   & 20101101 & Keck 1 &    3500--7500 & 15 & \ldots               &  5145(57)     \\
  10xem  &          & 2455593.5   & 20110201 & Keck 1 &    3195--10200 & 107 & \ldots             &  \ldots    \\
  10xem  &          & 2455600.5   & 20110208 & P200 &    3440--9800 & 114 & \ldots               & \ldots     \\\hline
  10ysd  & 2        & 2455486.5   & 20101017 & P200 &    3450--9900 & -5 & 14800(1522)           &  11344(101)      \\
  10ysd  &          & 2455499.5   & 20101030 & WHT &    3500--9498 & 8 & \ldots                  &  9877(38)     \\
  10ysd  &          & 2455503.5   & 20101103 & KPNO 4~m &    3330--8460 & 12 & 7259(3336)          &  8278(34)     \\\hline
  10aavz & 4        & 2455530.5   & 20101130 & WHT &    3100--9348 & 1 & \ldots                 &  6466(20)      \\
  10aavz &          & 2455536.5   & 20101206 & P200 &    3505--10000 & 7 & 10630(1506)          &  5631(21)      \\
  10aavz &          & 2455543.5   & 20101213 & P200 &    3500--9900 & 14 & \ldots               &  6425(55)      \\
  10aavz &          & 2455678.5   & 20110427 & Keck 1 &    3100--10200 & 149 & \ldots            &  \ldots      \\\hline
  11cmh  & 4        & 2455683.5   & 20110502 & WHT &    3015--11180 & 1 & 14784(1521)            &  7422(34)     \\\hline
  11img  & 3-4      & 2455772.5   & 20110730 & WHT &    3501--9499 & 5 & \ldots                  &  6939(80)     \\
  11img  &          & 2455775.9   & 20110802 & Keck 1 &    3404--10200 & 9 & 16420(1502)          &  6395(35)     \\\hline
  11lbm  & 3        & 2455804.5   & 20110831 & WHT &     3500--9500 & -1 & 11659(1503)           &  7931(45)     \\
  11lbm  &          & 2455825.5   & 20110921 & P200 &     3360--9600 & 20 & 7815(1560)           &  6320(26)     \\\hline
  12as   & 3        & 2455929.0   & 20120102 & APO &    3316--9864 & -5 & 14950(1516)             &  7239(75)      \\
  12as   &          & 2455956.5   & 20120130 & P200 &    3500--10200 & 23 & 6995(1688)            &  7879(42)       \\
  12as   &          & 2456044.8   & 20120427 & Keck 1 &    3702--9995 & 111 & \ldots               &  \ldots       \\\hline
  12eci  & 3        & 2456060.8   & 20120517 & Keck 1 &    3424--10100 & -7 & 10560(1516)         &  9005(28)     \\\hline
  13u    & $<=$4    & 2456341.5   & 20130218 & P200 &    3210--10100 & 11 & 18249(1504)            &  5067(25)         \\\hline
  13alq  & 3        & 2456395.5   & 20130413 & P200 &    3600--9900 & -3 & 21530(1503)           &  8492(41)     \\
  13alq  &          & 2456414.5   & 20130502 & P200 &    3840--10450 & 16 & 13745(1507)          &  5932(21)     \\
  13alq  &          & 2456414.7   & 20130502 & TNG &    3220--7989 & 16 & 13745(1507)            &  5403(22)     \\
  13alq  &          & 2456452.8   & 20130609 & Keck 1 &    3080--10284 & 55 & \ldots              &  6139(23)     \\\hline
  13bxl  & 3        & 2456476.5   & 20130703 & MagellanBaade &    3900--9500 & -12 & \ldots      &  12700(61)      \\
  13bxl  &          & 2456476.7   & 20130703 & P200 &    3700--10000 & -11 & \ldots              &  12362(63)      \\
  13bxl  &          & 2456476.8   & 20130703 & UH88 &    3301--9701 & -11 & \ldots               &  7993(26)     \\
  13bxl  &          & 2456477.7   & 20130704 & P200 &    3700--10000 & -10 & \ldots              &  15750(137)      \\
  13bxl  &          & 2456478.5   & 20130705 & P200 &    3400--8933 & -10 & 27780(1532)          &  14062(112)      \\
  13bxl  &          & 2456481.7   & 20130708 & P200 &    4450--9625 & -6 & 23340(1501)           &  12486(67)      \\
  13bxl  &          & 2456484.5   & 20130711 & Keck 2 &    4905--10131 & -4 & 20020(1502)         &  17169(213)      \\
  13bxl  &          & 2456486.5   & 20130713 & Keck 2 &    3700--10840 & -2 & \ldots              &  7096(63)     \\
  13bxl  &          & 2456489.5   & 20130716 & P200 &    4425--9640 & 1 & 14754(1508)            &  7997(36)     \\
  13bxl  &          & 2456506.5   & 20130802 & Keck 2 &    3066--10259 & 18 & 6605(1518)          &  6344(27)     \\
  13bxl  &          & 2456508.8   & 20130804 & Keck 1 &    3066--10259 & 21 & \ldots              &  6344(27)     \\\hline
  13dnt  & 4        & 2456569.7   & 20131004 & P200 &    3200--10200 & 4 & 19040(1502)           &  5575(13)          \\\hline
  13ebw  & 2        & 2456622.5   & 20131126 & P200 &     3400--10580 & -8 & \ldots              &  6148(11)     \\
  13ebw  &          & 2456625.5   & 20131129 & Keck 2 &     4905--10131 & -5 & \ldots             &  7903(45)     \\
  13ebw  &          & 2456630.1   & 20131203 & Keck 1 &     3200--10265 & -0 & 26584(1505)        &  5841(15)     \\
  13ebw  &          & 2456631.0   & 20131204 & Keck 1 &     3017--10237 & 1 & 22004(1505)         &  5711(21)     \\
  13ebw  &          & 2456658.5   & 20140101 & Keck 2 &     4909--10140 & 28 & \ldots             &  5855(18)     \\\hline
  14dby  & 4        & 2456837.9   & 20140629 & Keck 1 &    3057--10288 & -12 & 16085(1506)        &  8115(24)     \\
  14dby  &          & 2456899.8   & 20140830 & Keck 1 &    3099--10290 & 49 & 6248(1516)          &  5272(22)     \\
  14dby  &          & 2456924.7   & 20140924 & Keck 2 &    4600--9599 & 74 & \ldots               &  \ldots     \\
  14dby  &          & 2457045.1   & 20150122 & Keck 1 &    3069--10287 & 195 & \ldots             &  \ldots    \\\hline
  14gaq  & 2-4      & 2456924.9   & 20140924 & Keck 2 &    4600--9600 & -5 & \ldots               &  9992(23)     \\
  14gaq  &          & 2456931.5   & 20141001 & P200 &    3000--10239 & 1 & 24034(1512)           &  7881(27)     \\
  14gaq  &          & 2456955.7   & 20141025 & Keck 2 &    5000--9998 & 25 & \ldots               &  5491(16)     \\\hline
  15dqg  & 2-$<=$5  & 2457333.8   & 20151107 & Keck 2 &    4500--9601 & -8 & 19935(1519)          &  10364(24)      \\
  15dqg  &          & 2457362.5   & 20151206 & Keck 1 &    3076--10231 & 20 & 9265(1505)          &  5527(20)     \\
  15dqg  &          & 2457369.3   & 20151212 & TNG &    3381--10436 & 27 & 8515(1502)            &  6029(39)     \\\hline
  15eov  & 2-4      & 2457362.5   & 20151206 & Keck 1 &    3069--10233 & -24 & 20625(1503)        &  23481(104)       \\
  15eov  &          & 2457365.0   & 20151208 & FTN &    3249--10000 & -22 & 20635(1507)          &  21563(180)       \\
  15eov  &          & 2457365.9   & 20151209 & FTN &    3201--10000 & -21 & 20565(1508)          &  18815(143)       \\
  15eov  &          & 2457367.5   & 20151211 & P200 &    3091--10199 & -19 & 20685(1505)         &  19400(108)       \\
  15eov  &          & 2457368.9   & 20151212 & FTN &    3200--10000 & -18 & 20605(1507)          &  22267(215)       \\
  15eov  &          & 2457370.0   & 20151213 & FTN &    3200--10001 & -17 & 20545(1505)          &  23209(207)       \\
  15eov  &          & 2457371.1   & 20151214 & FTS &    3350--10000 & -16 & 20685(1505)          &  29823(681)       \\
  15eov  &          & 2457374.9   & 20151218 & FTN &    3221--10001 & -12 & 20625(1505)          &  19727(192)       \\
  15eov  &          & 2457384.8   & 20151228 & FTN &    4000--9300 & -2 & \ldots                 &  31875(3372)         \\
  15eov  &          & 2457392.9   & 20160105 & FTN &    3300--9999 & 6 & 19030(1507)             &  9819(29)      \\
  15eov  &          & 2457395.5   & 20160108 & Keck 2 &    4501--9641 & 9 & 19060(1506)           &  9549(18)      \\
  15eov  &          & 2457408.9   & 20160121 & FTN &    4000--10000 & 22 & 17445(1506)           &  7147(38)      \\
  15eov  &          & 2457416.9   & 20160129 & FTN &    3500--10000 & 30 & 17872(1505)           &  6292(23)      \\
  15eov  &          & 2457424.8   & 20160206 & FTN &    3300--10001 & 38 & 17812(1506)           &  6543(21)      \\
  15eov  &          & 2457432.8   & 20160214 & FTN &    3800--10000 & 46 & 17048(1504)           &  6113(27)      \\
  15eov  &          & 2457436.7   & 20160218 & FTN &    5000--9300 & 50 & \ldots                 &  6438(81)        \\
  15eov  &          & 2457452.8   & 20160305 & FTN &    3999--10001 & 66 & 17150(1503)           &  \ldots     \\\hline
  16asu  & $<=$3    & 2457522.5   & 20160514 & P200 &    3101--9199 & -3 & \ldots                &  19253(360)      \\
  16asu  &          & 2457524.6   & 20160516 & NOT &    3478--9662 & -1 & \ldots                 &  17776(370)      \\
  16asu  &          & 2457533.5   & 20160524 & TNG &    3315--10330 & 8 & 27590(1546)            &  8287(42)     \\
  16asu  &          & 2457535.5   & 20160527 & P200 &    3600--10237 & 10 & 27414(1924)          &  8271(52)     \\
  16asu  &          & 2457543.9   & 20160604 & Keck 2 &    4550--9649 & 18 & 24835(1504)          &  6560(24)     \\
  16asu  &          & 2457546.9   & 20160607 & Keck 1 &    3072--10285 & 21 & 22415(1508)         &  6343(18)     \\
  16asu  &          & 2457549.5   & 20160610 & Keck 1 &    3101--10290 & 24 & 19882(1513)         &  6221(20)     \\\hline
  16gox  & $<=$3    & 2457660.5   & 20160929 & P60 &    3807--10456 & 13 & \ldots                &  4694(111)     \\
  16gox  &          & 2457662.5   & 20161001 & P200 &    3100--10236 & 15 & \ldots               &  4693(21)     \\
  16gox  &          & 2457667.5   & 20161006 & P200 &    3101--10236 & 20 & \ldots               &  5735(16)     \\
  16gox  &          & 2457676.6   & 20161015 & NOT &    3702--9711 & 29 & 17572(1820)            &  5039(34)     \\\hline
  16ilj  & 3        & 2457717.5   & 20161125 & P60 &    3807--10456 & -4 & \ldots                &  6352(132)     \\
  16ilj  &          & 2457721.5   & 20161128 & NOT &    3463--9715 & -0 & 19154(1549)            &  6406(62)     \\
  16ilj  &          & 2457722.5   & 20161130 & P60 &    3807--9187 & 1 & \ldots                  &  4859(179)     \\\hline
  17cw   & $<=$3    & 2457760.5   & 20170107 & P200 &    3088--10261 & -8 & \ldots                &  14029(83)      \\
  17cw   &          & 2457785.9   & 20170201 & DCT &    3536--7996 & 18 & 20595(1528)             &  5076(34)     \\
  17cw   &          & 2457811.9   & 20170227 & Keck 1 &    3068--10266 & 44 & 11028(1563)          &  6104(28)         \\      
  \hline
  \hline                                                          
  09sk  &      & 2455008.8   & 20090626 & Keck 1  & 3027--9833      &     &    &     \\\hline
  10cs  &      & 2455205.5   & 20100109 & P200 &    3242--10224    &     &    &   \\
  10cs  &      & 2455262.5   & 20100307 & Keck 1 &    3342--10264   &     &    &   \\\hline
  12grr &      & 2456129.5   & 20120721 & P200 &    3200--10500   &     &    &   \\
  12grr &      & 2456135.5   & 20120727 & P200 &    3300--10300   &     &    &   \\\hline
  14bfu &      & 2456816.1   & 20140607 & Keck 2 &    4550--9599   &     &    &   \\
  14bfu &      & 2456833.0   & 20140624 & Keck 1 &    3116--10279  &     &    &   \\
  14bfu &      & 2456838.1   & 20140629 & Keck 1 &    3234--10289  &     &    &   \\\hline
  15dld &      & 2457333.8   & 20151107 & Keck 2 &    4500--9601   &     &    &   \\
  15dld &      & 2457399.7   & 20160112 & Keck 1 &    3064--10257  &     &    &   \\\hline
  16coi &      & 2457626.5   & 20160826 & P60 &    3807--9187     &     &    &   \\
  16coi &      & 2457656.6   & 20160925 & NOT &    3407--9647     &     &    &   \\\hline
  17axg &      & 2457809.5   & 20170225 & P200 &    3500--10924   &     &    &   \\
  17axg &      & 2457809.9   & 20170225 & P60 &    3807--10456    &     &    &   \\
  \enddata
\tablecomments{The double horizontal line separates the spectra for which the phase is known, which were therefore used to measure the \ion{Fe}{ii} velocity, from those for which it is not.}
\tablenotetext{**}{Number of absorption features in the spectra before and around maximum brightness, following the classification method by \citet{prentice17}.}

\end{deluxetable}

\begin{deluxetable}{lcccc}
\tablewidth{0pt}
\tabletypesize{\scriptsize}
\tablecaption{\label{tab:param}Explosion parameters from Arnett models and from hydrodynamical model scaling relations.}
\tablehead{
\colhead{Supernova}       &
\colhead{$M_{\bf ej}$}&
\colhead{$E_{K}$}&
\colhead{$M(^{56}$Ni)}&
\colhead{$^{56}$Ni mixing}\\
\colhead{}&
\colhead{($\rm M_\odot$)}&
\colhead{($10^{51}$ erg)}&
\colhead{($\rm M_\odot$)}&
\colhead{($\% M_{\rm ej}$)}}
\startdata
    PTF10bzf & 2.8(0.7)  & 6.2(0.7)   & 0.274(0.125)  &    100      \\
    PTF10ciw & 5.7(8.4)  & 16.5(12.3) & 0.292(0.097)  &    100      \\
    PTF10gvb & 3.3(2.2)  & 2.8(0.9)   & 0.371(0.120)  &    100      \\
    PTF10qts & 2.9(0.5)  & 8.3(0.7)   & 0.392(0.137)  &    100      \\
    PTF10tqv & 1.6(0.8)  & 3.5(0.9)   & 0.203(0.093)  &    100      \\
    PTF10vgv & 0.6(0.6)  & 0.3(0.1)   & 0.149(0.067)  &    100      \\
    PTF10xem & 11.1(8.9) & 8.3(3.4)   & 0.312(0.142)  &  93     \\
    PTF10ysd & 10.4(12.7)& 7.4(4.5)   & 0.802(0.370)  &  94     \\
    PTF10aavz& 5.0(3.5)  & 4.3(1.5)   & 0.424(0.141)  &    100      \\
    PTF11cmh & 1.7(0.8)  & 2.3(0.6)   & 0.205(0.070)  &    100      \\
    PTF11img & 5.2(2.2)  & 11.3(2.4)  & 0.479(0.170)  &    100      \\
    PTF11lbm & 0.6(0.4)  & 0.4(0.1)   & 0.172(0.056)  &    100      \\
    PTF12as  & 1.2(0.9)  & 0.8(0.3)   & 0.117(0.039)  &    100      \\
    PTF12eci & 1.9(1.3)  & 1.0(0.4)   & 0.253(0.083)  &  87     \\
   iPTF13u   & 1.8(0.7)  & 4.9(0.9)   & 0.199(0.091)  &    100      \\
   iPTF13alq & 2.4(0.7)  & 5.1(0.7)   & 0.281(0.090)  &    100      \\
   iPTF13bxl & 3.1(1.7)  & 4.1(1.1)   & 0.354(0.109)  &   \dots     \\
   iPTF13dnt & 7.0(2.6)  & 18.2(3.3)  & 0.469(0.211)  &    90   \\
   iPTF13ebw & 5.3(1.0)  & 19.1(1.8)  & 0.142(0.065)  &    94   \\
   iPTF14dby & 5.1(4.0)  & 3.5(1.4)   & 0.155(0.050)  &    100      \\
   iPTF14gaq & 3.0(0.9)  & 11.0(1.6)  & 0.195(0.060)  &    100      \\
   iPTF15dqg & 1.7(0.8)  & 1.7(0.4)   & 0.461(0.184)  &    100      \\
   iPTF15eov & 6.1(0.5)  & 11.3(0.5)  & 5.114(1.773)  &   \ldots    \\   
   iPTF16asu & 0.9(0.1)  & 4.8(0.3)   & 0.763(0.355)  &   \ldots    \\
   iPTF16gox & 2.4(1.2)  & 8.2(2.0)   & 0.123(0.049)  &    98   \\
   iPTF16ilj & 6.6(2.5)  & 14.8(2.8)  & 0.462(0.154)  &    90   \\
   iPTF17cw  & 4.5(1.8)  & 10.2(2.1)  & 0.473(0.174)  &    100      \\
    \hline                                           
   \hline
     $<$Ic-BL$>$  & 3.9(2.7) & 7.1(5.4) & 0.506(0.938)    \\
       Median Ic-BL  & 3.1 & 5.1 & 0.292\\
      $<$Ic-BL$>^a$ & 
      3.9(2.8) & 7.0(5.6) & 0.310(0.160)    \\
      $<$Ic-BL$>^b$ & \textbf{3.9(2.8)} &        \textbf{7.1(5.7)}        & \textbf{0.309(0.164)} 
\enddata
\tablecomments{Uncertainties in the averages are the standard deviations.}
\tablenotetext{a}{Computed without iPTF15eov and iPTF16asu.}
\tablenotetext{b}{Computed without iPTF15eov, iPTF16asu, and GRB SN iPTF13bxl.}
\end{deluxetable}

\begin{deluxetable}{lccc}
\tablewidth{0pt}
\tabletypesize{\scriptsize}
\tablecaption{\label{tab:literature}Averages of supernova explosion parameters.$^a$}
\tablehead{
\colhead{Sample (size)}    &
\colhead{$M_{\rm ej}$}&
\colhead{$E_{K}$}&
\colhead{$M(^{56}$Ni)}\\
\colhead{}&
\colhead{(M$_\odot$)}&
\colhead{($10^{51}$ erg)}&
\colhead{(M$_\odot$)}}
\startdata
{\bf\citealp{drout11} (5)}&\ldots & \ldots & 0.570(0.541)\\
{\bf\citealp{cano13} (9)}&5.4(3.4) & 12.6(8.9) & 0.299(0.274)\\
{\bf\citealp{taddia15} (4)}& 5.4(2.6) & 10.7(9.4) & 1.110(0.580)\\
{\bf\citealp{lyman16} (5)}&3.3(2.7) & 6.1(6.5) & 0.322(0.162)\\
{\bf\citealp{prentice16} (9)}&\ldots & \ldots & 0.151(0.073)\\
{\bf This work (24)}& 3.9(2.8) & 7.1(5.7) & 0.309(0.164) \\
  \enddata
\tablenotetext{a}{From Arnett semi-analytic models of SNe~Ic-BL (not associated with GRBs) from the literature and this work.}
\tablecomments{Uncertainties in the averages are the standard deviations. We removed the GRB SN iPTF13bxl and the peculiar iPTF15eov and iPTF16asu from our sample when computing the average. \citet{prentice16} do not include extinction corrections for some SNe~Ic-BL. }
\end{deluxetable}

\clearpage
 
\clearpage

 \clearpage
 
\clearpage

\clearpage
 
\clearpage

\clearpage
\begin{appendix}
\section{Spectral sequences}
\label{appendix}
\begin{figure}[h]
 \centering
 $\begin{array}{c}
  \includegraphics[width=9cm]{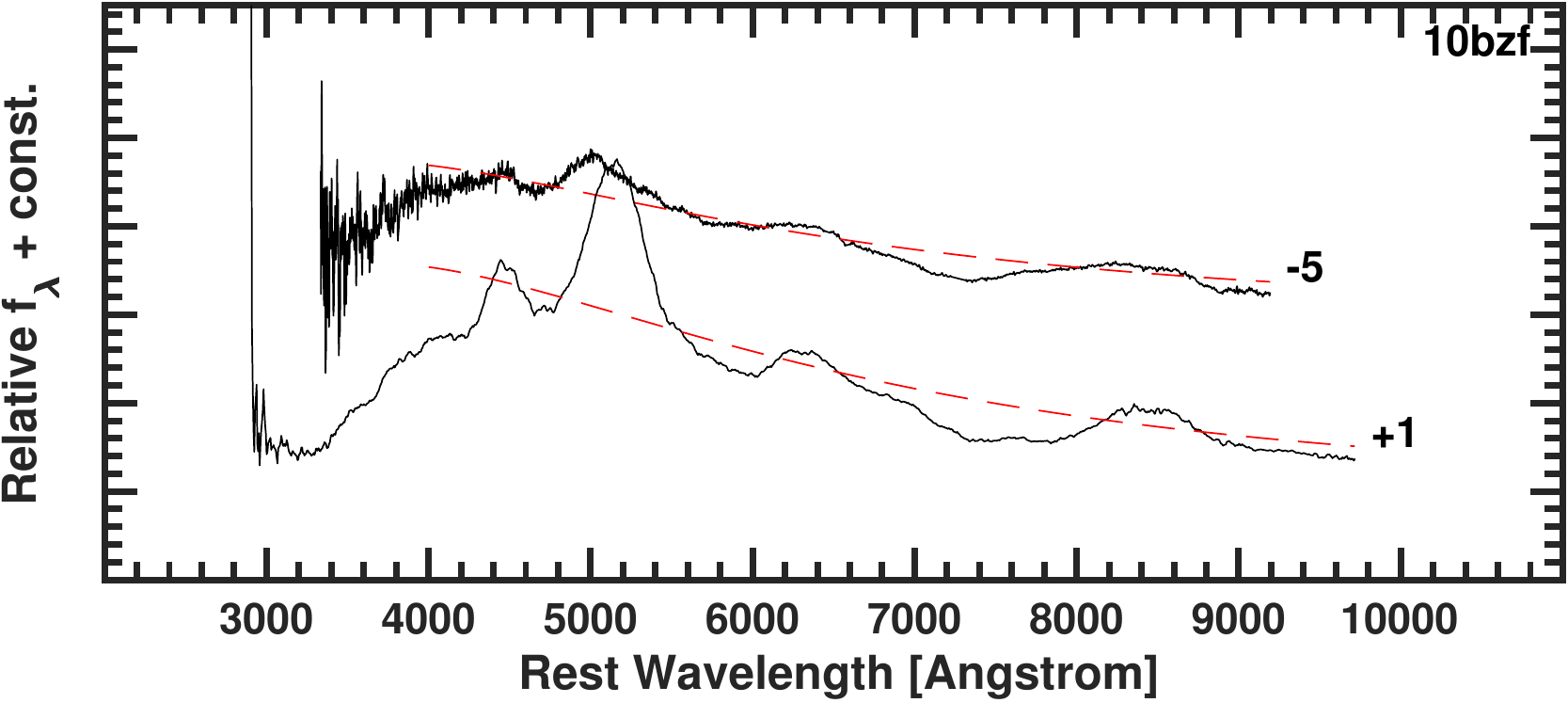} \\
  \includegraphics[width=9cm]{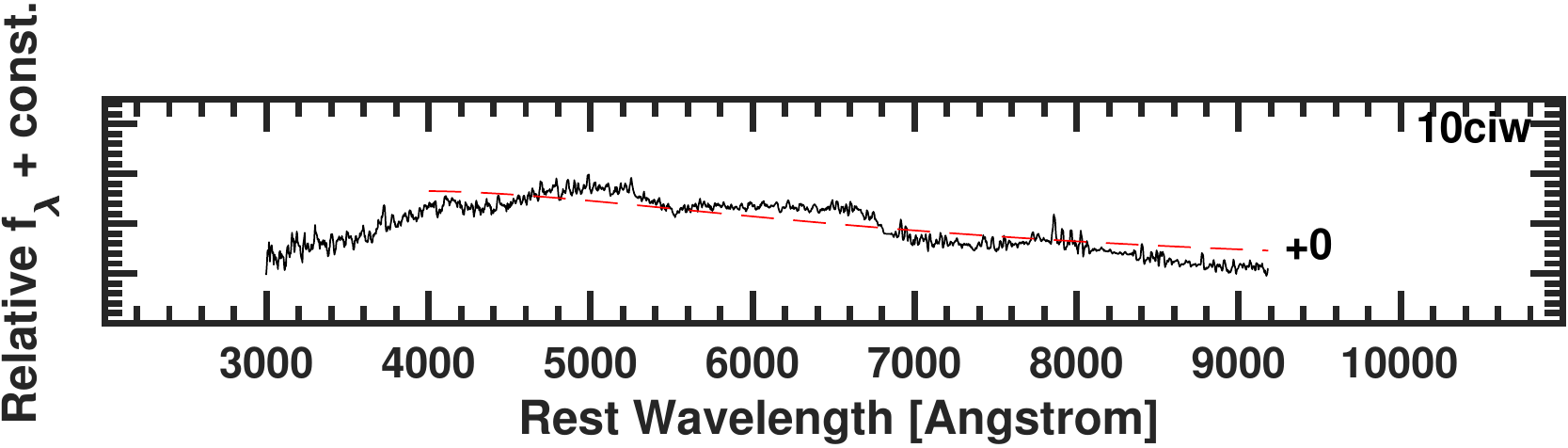} \\
  \includegraphics[width=9cm]{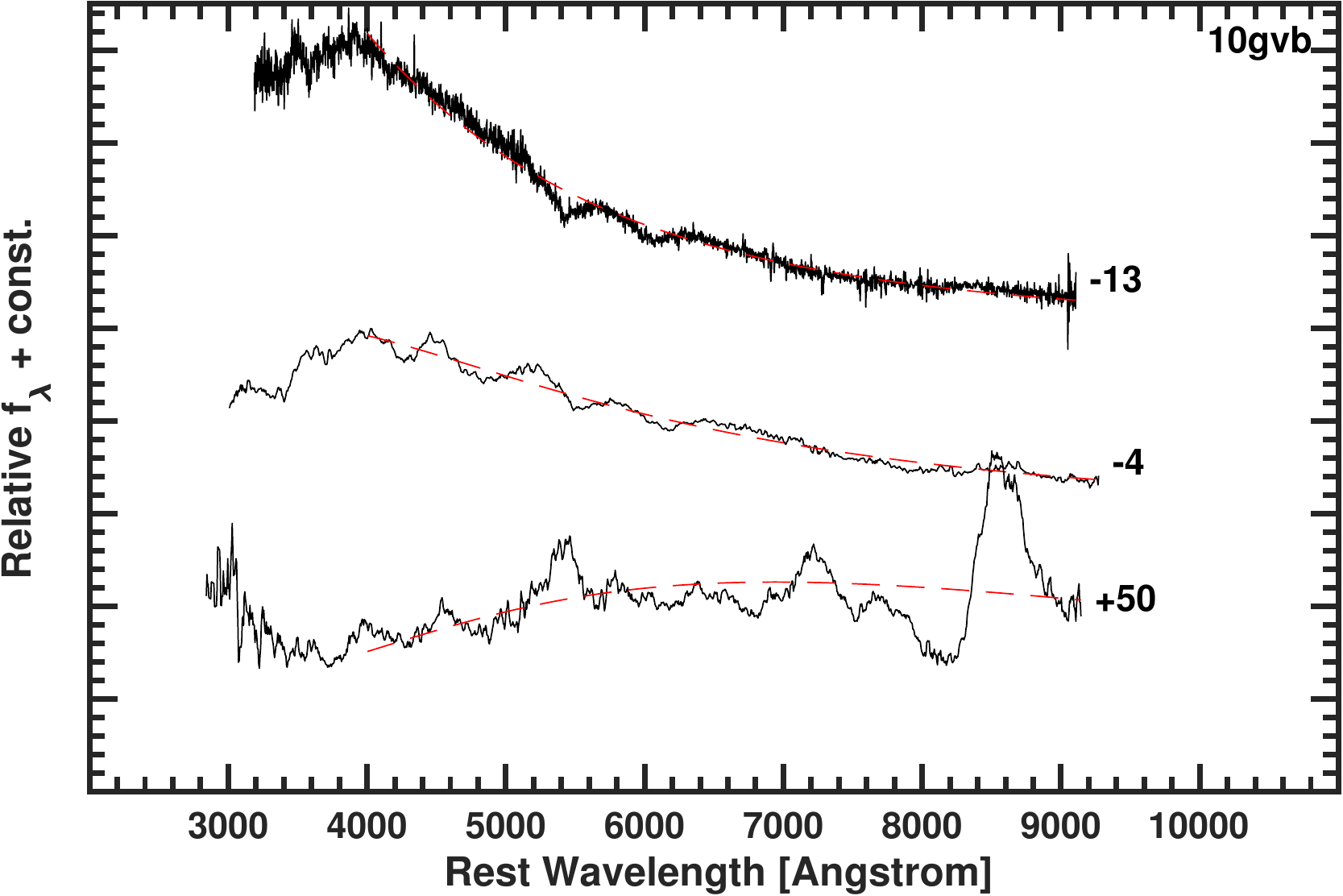} \\
    \end{array}$
  \caption{\label{spec_seq1}Spectral sequences of PTF10bzf, PTF10ciw, and PTF10gvb. The reported phases next to each spectrum are in rest-frame days since $r$-band maximum, and the dashed red lines are black-body fits. This format is followed for Figs. \ref{spec_seq2}$-$\ref{spec_seq11}.}
 \end{figure}

\begin{figure}
 \centering
 $\begin{array}{c}
      \includegraphics[width=9cm]{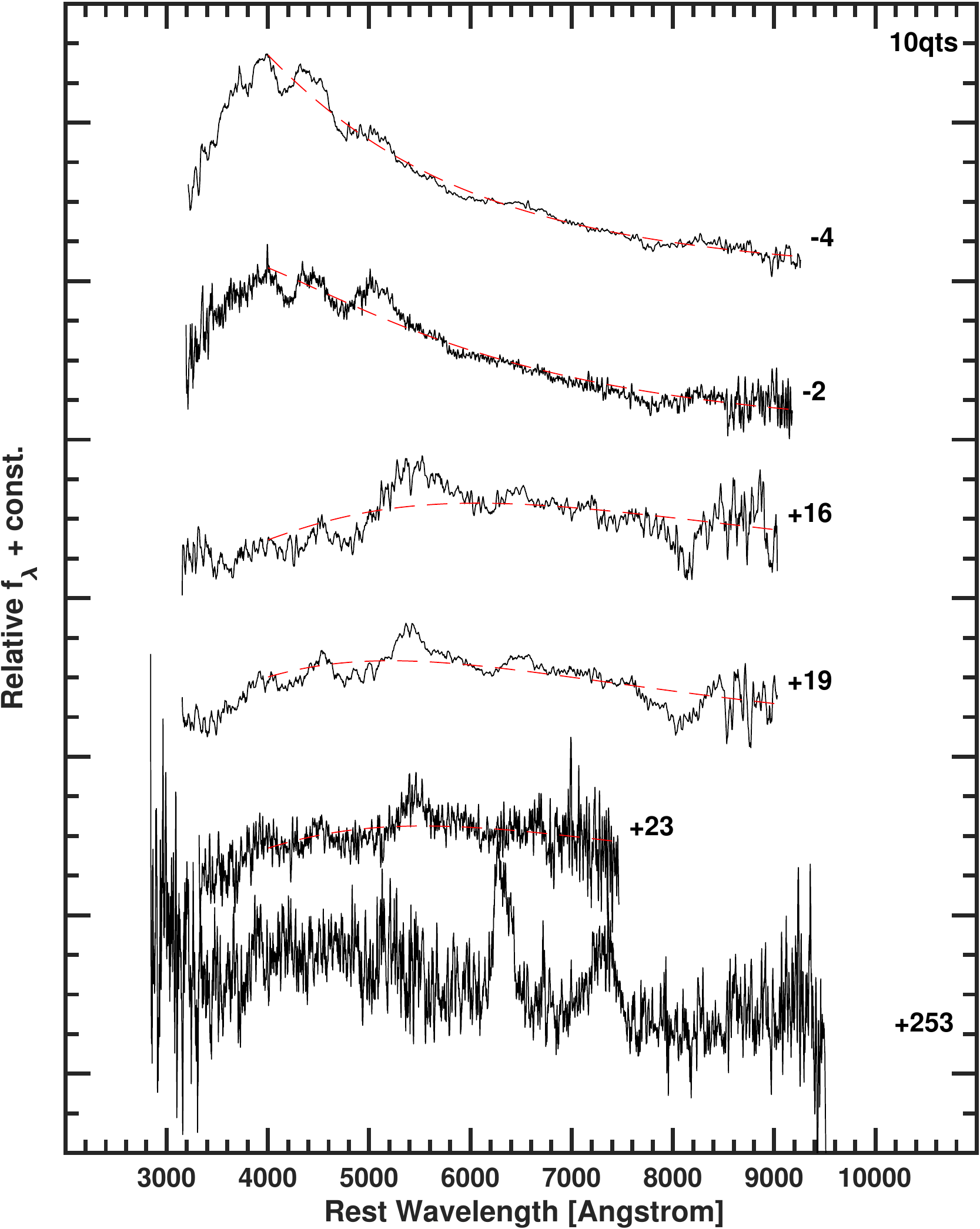} \\
  \includegraphics[width=9cm]{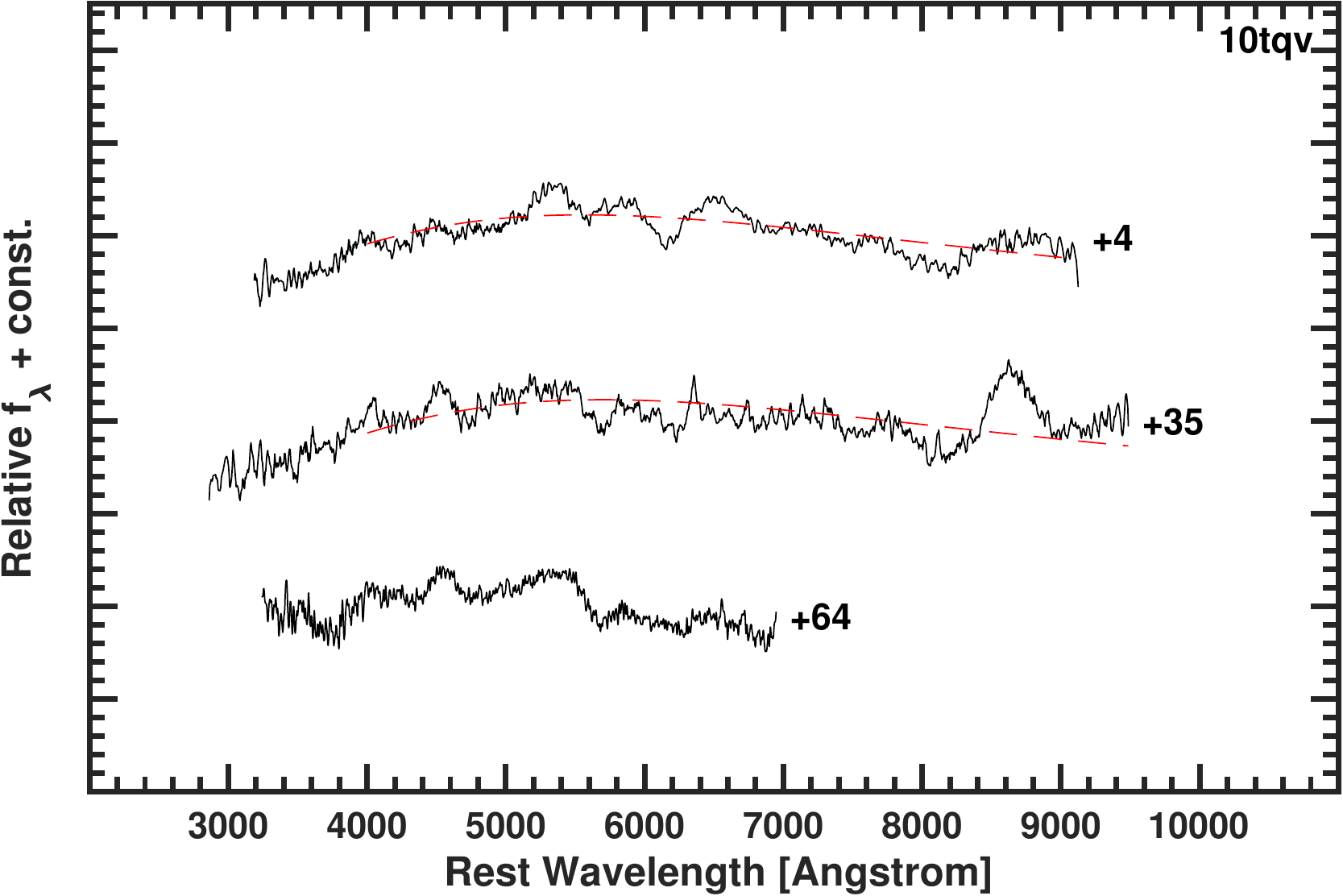} \\
    \end{array}$
  \caption{\label{spec_seq2}Spectral sequences of PTF10qts and PTF10tqv.}
 \end{figure}
 
 \begin{figure}
 \centering
 $\begin{array}{c}
  \includegraphics[width=9cm]{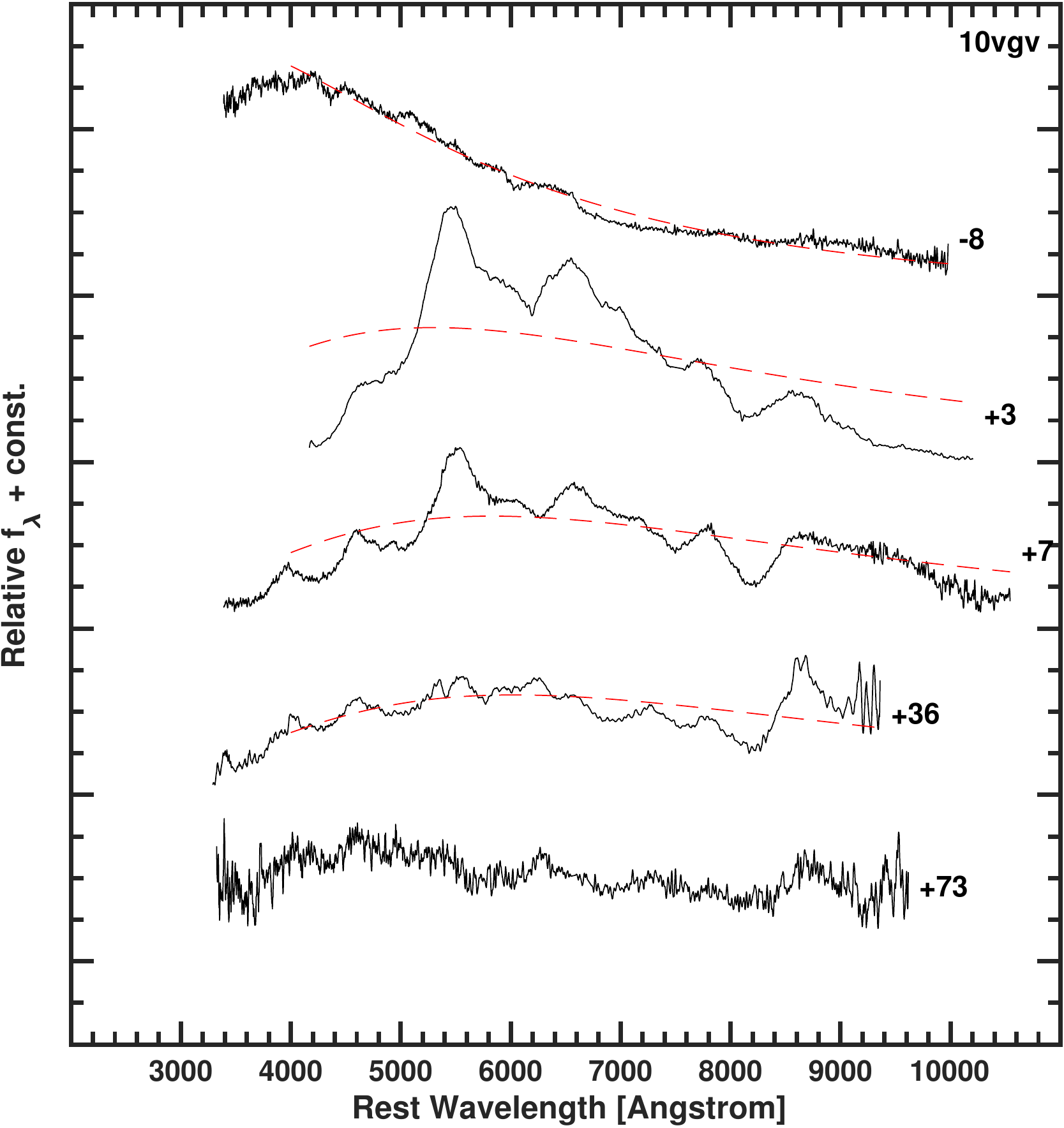} \\
    \end{array}$
  \caption{\label{spec_seq3}Spectral sequence of PTF10vgv.}
 \end{figure}
 \clearpage
 
 \begin{figure}
 \centering
 $\begin{array}{c}
   \includegraphics[width=9cm]{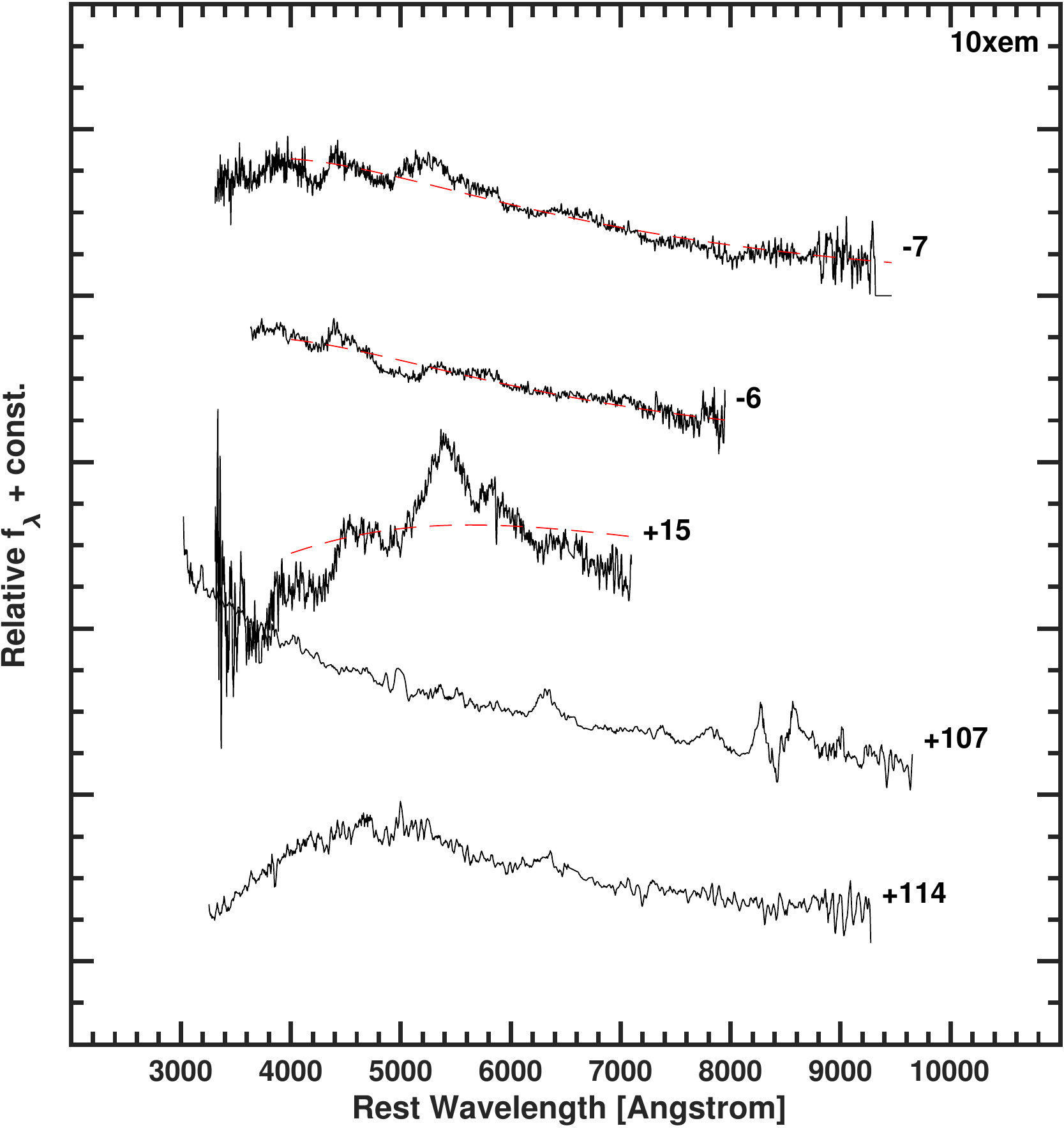} \\
  \includegraphics[width=9cm]{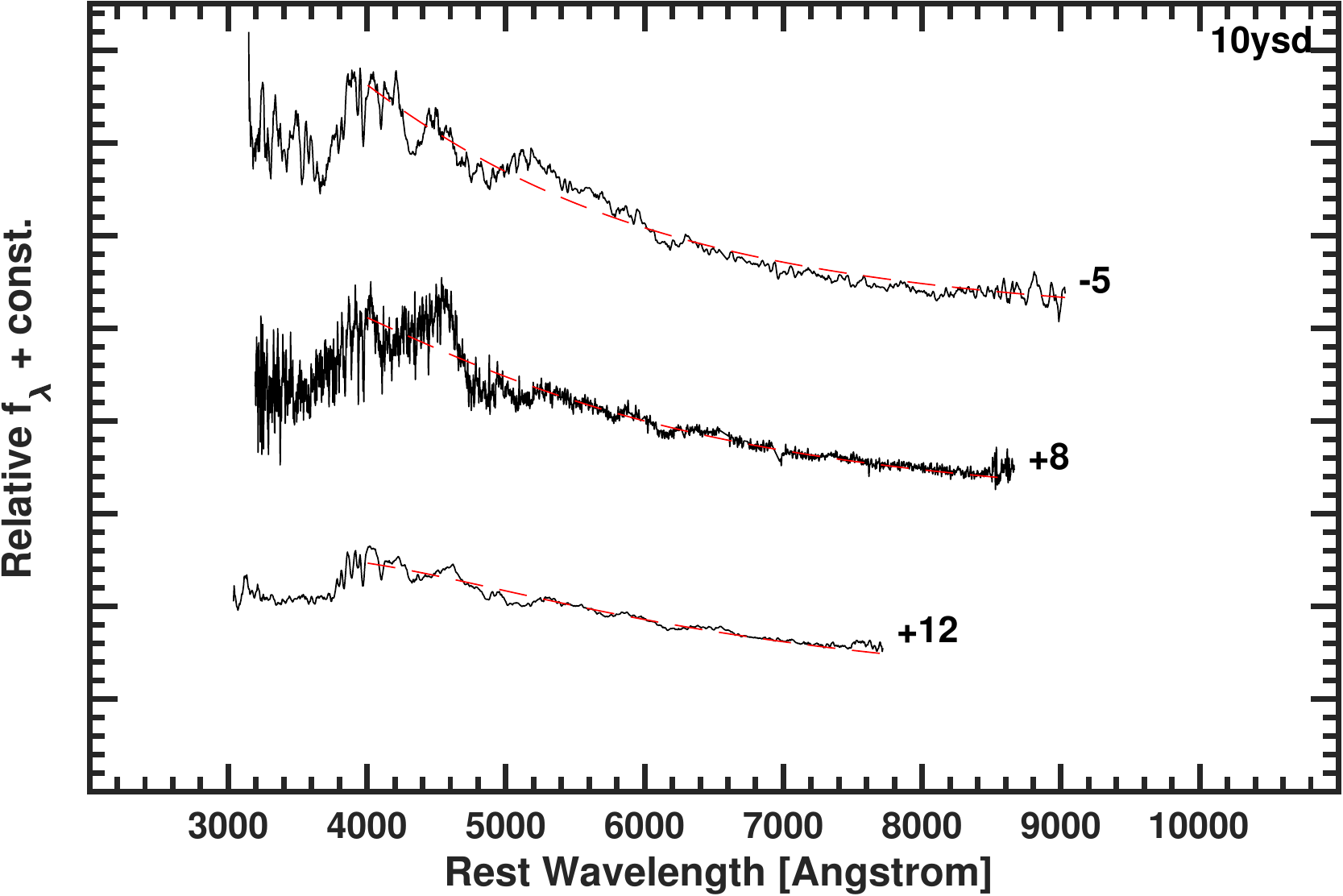} \\
  \includegraphics[width=9cm]{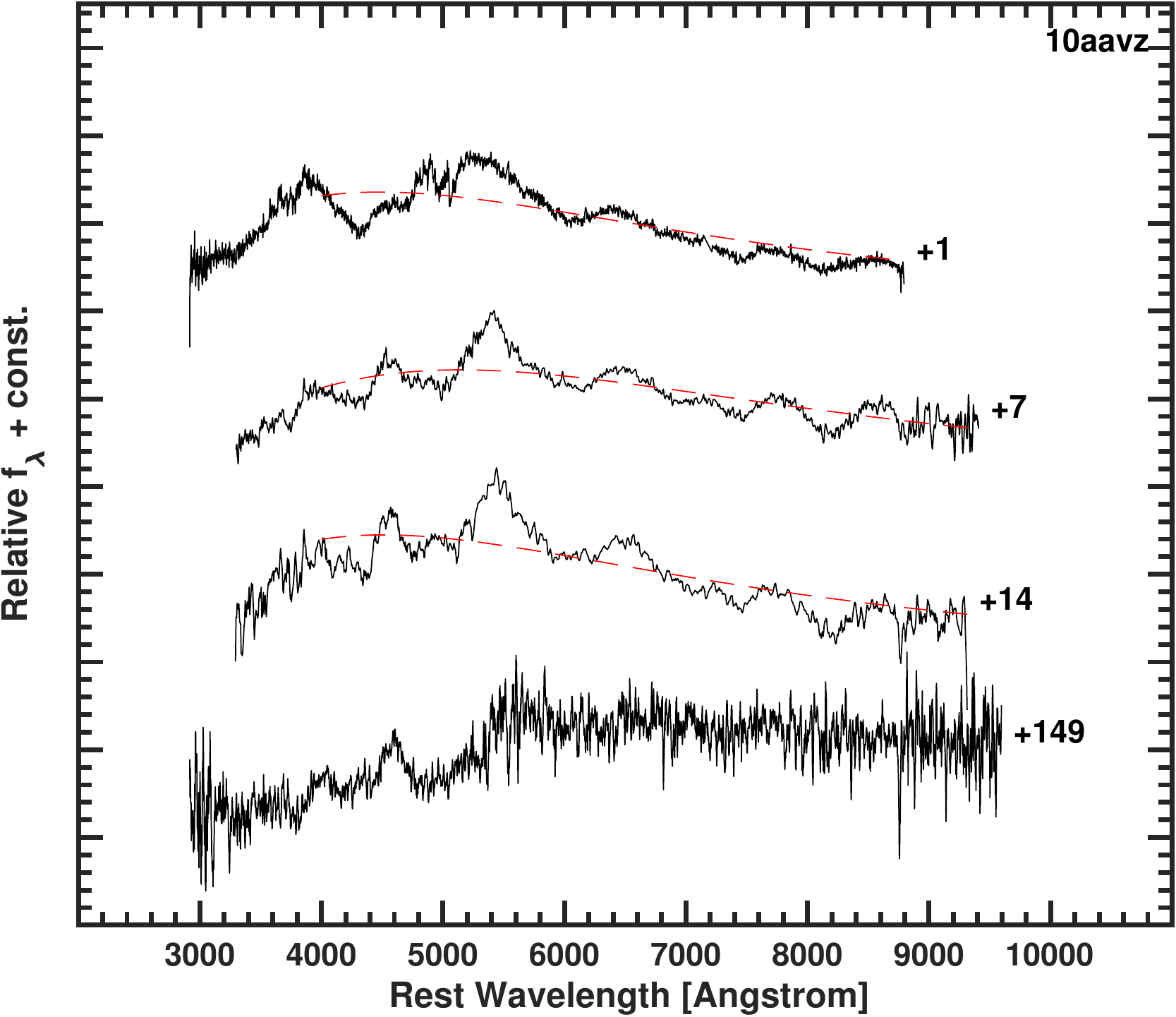}\\
    \end{array}$
  \caption{\label{spec_seq4}Spectral sequences of PTF10xem, PTF10ysd, and PTF10aavz.}
 \end{figure}

 \begin{figure}
 \centering
 $\begin{array}{c}
   \includegraphics[width=9cm]{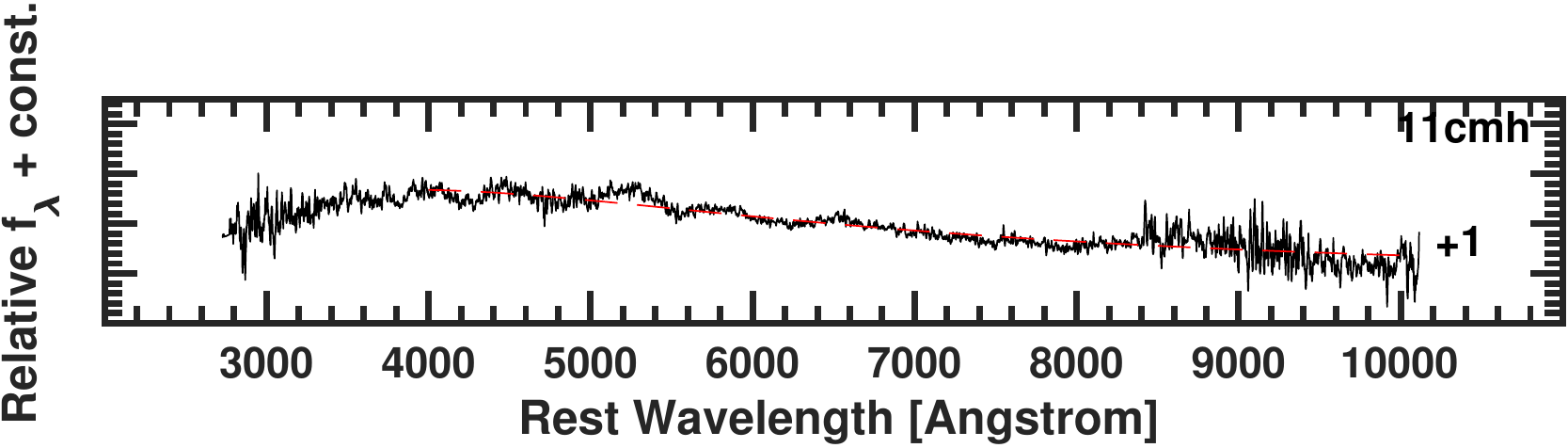} \\
 \includegraphics[width=9cm]{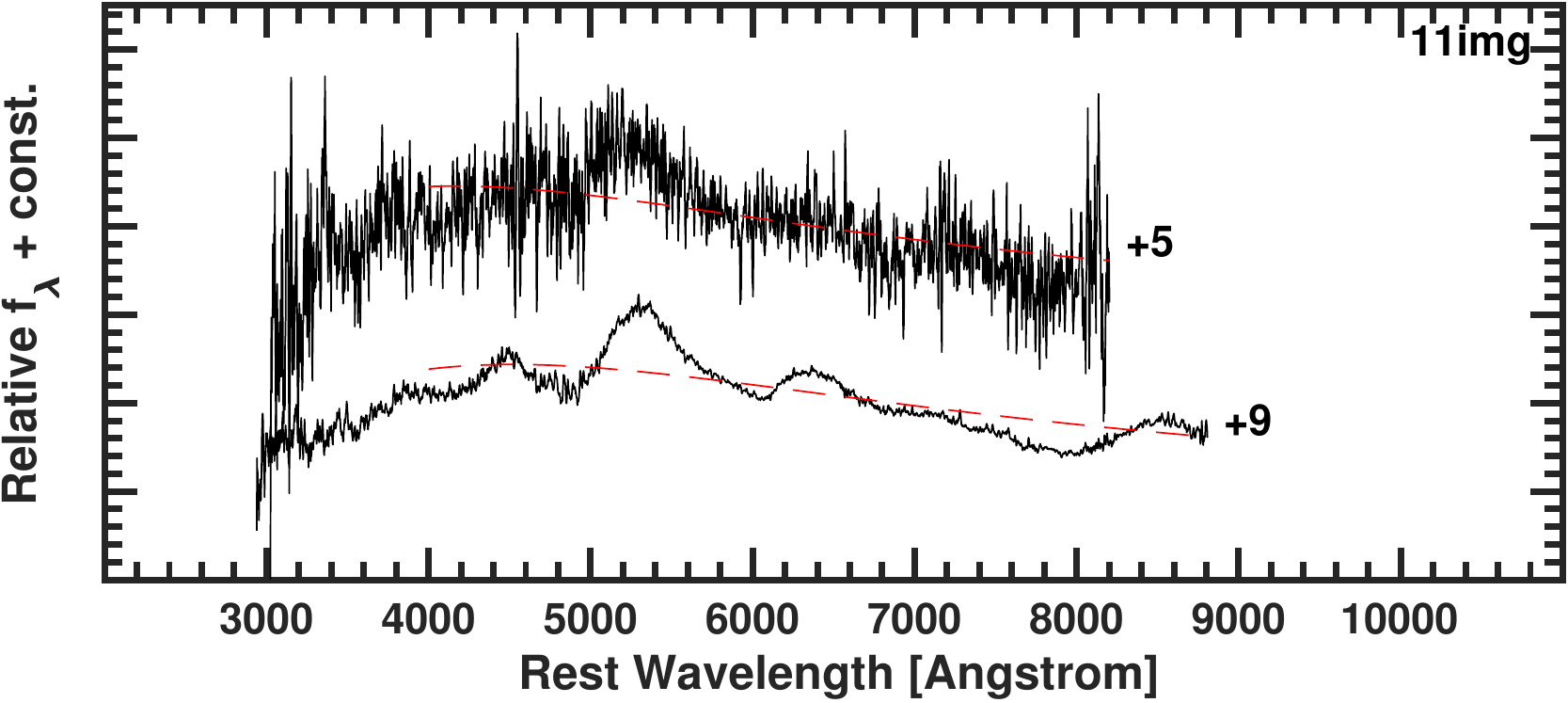} \\
  \includegraphics[width=9cm]{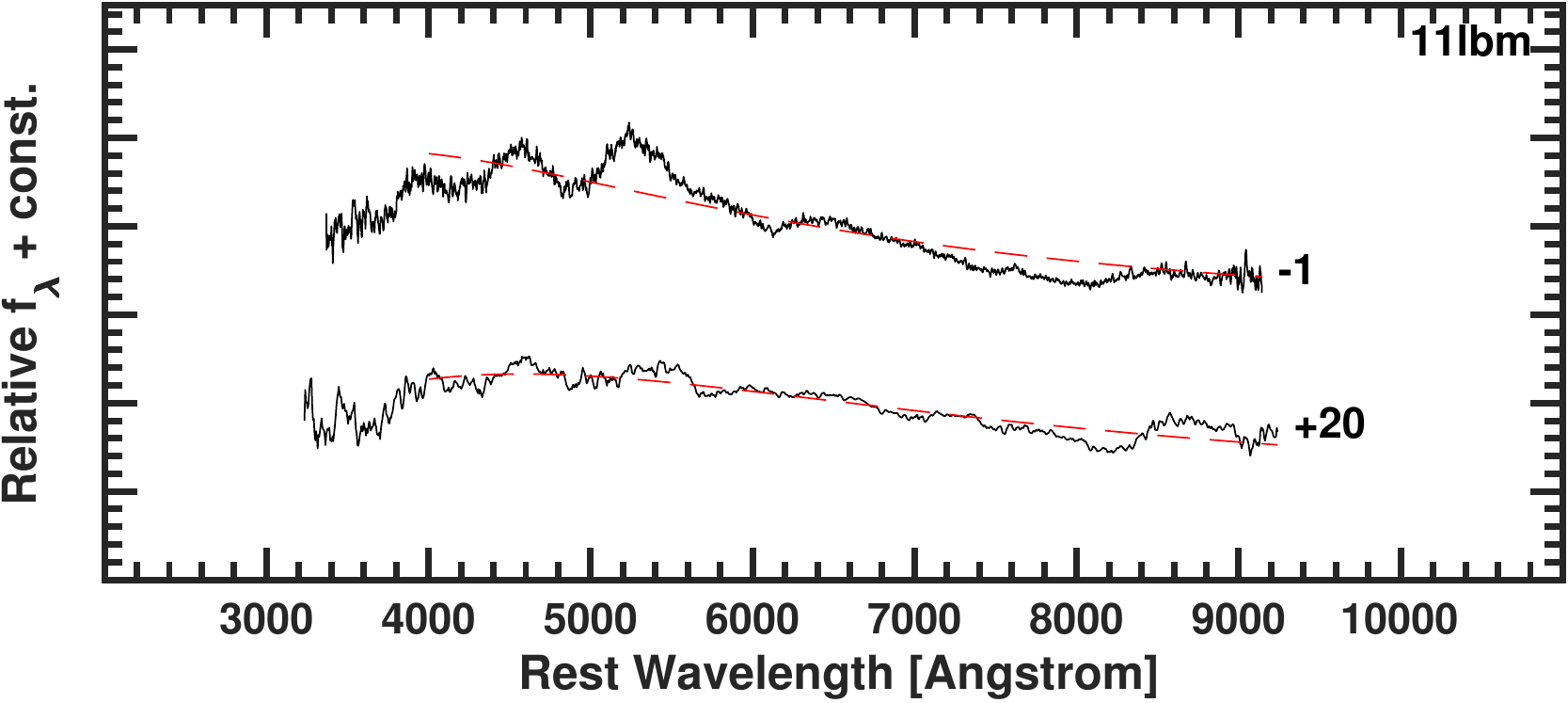} \\
  \includegraphics[width=9cm]{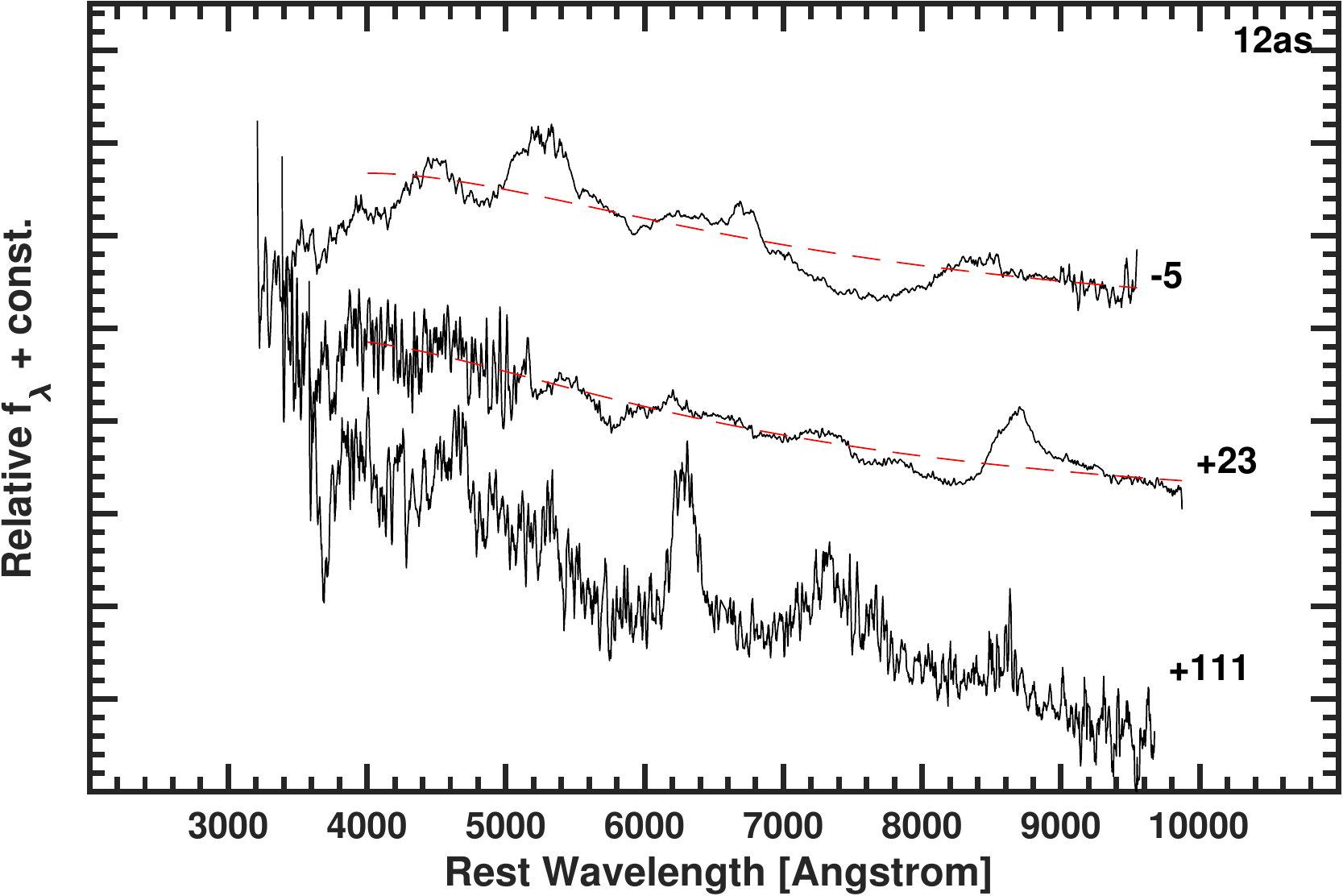}  \\
    \end{array}$
  \caption{\label{spec_seq5}Spectral sequences of PTF11cmh, PTF11img, PTF11lbm, and PTF12as.}
 \end{figure}

 \clearpage 
 
  \begin{figure}
 \centering
 $\begin{array}{c}
  \includegraphics[width=9cm]{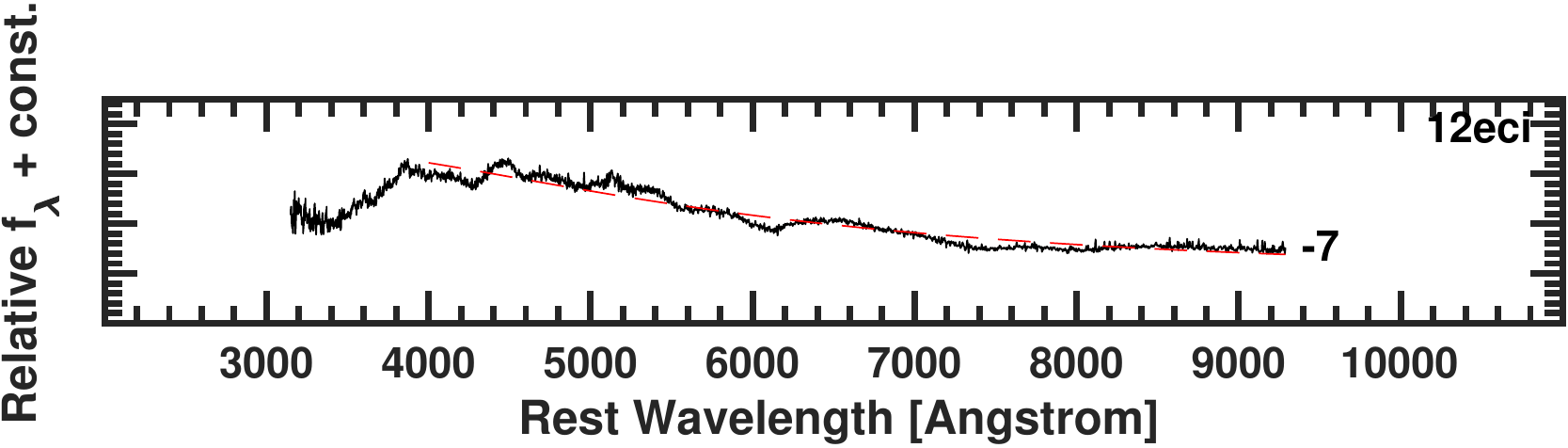} \\
  \includegraphics[width=9cm]{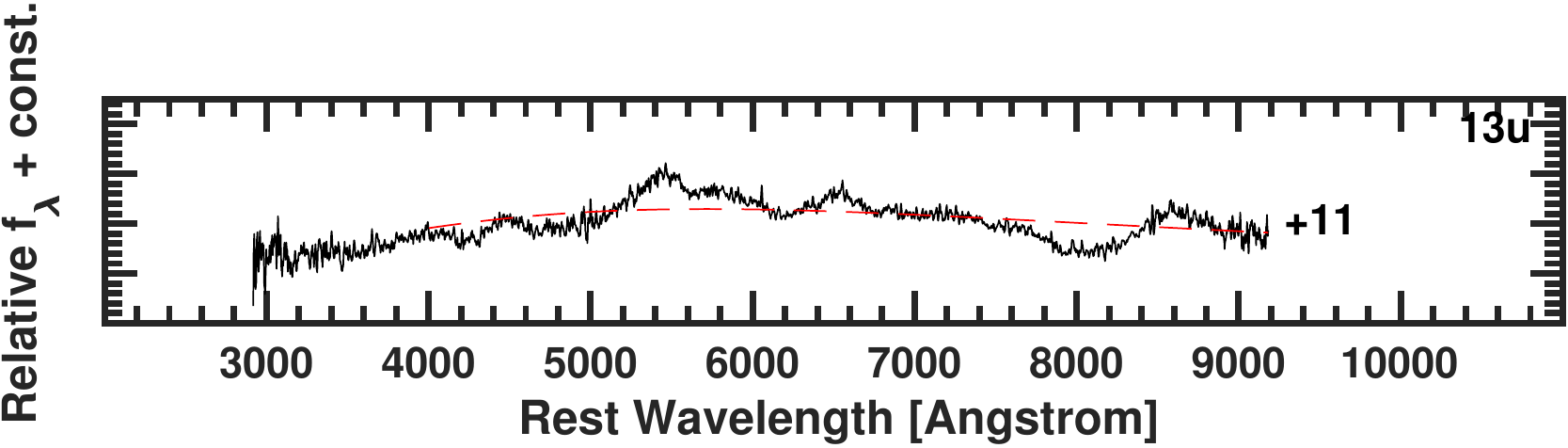}   \\
  \includegraphics[width=9cm]{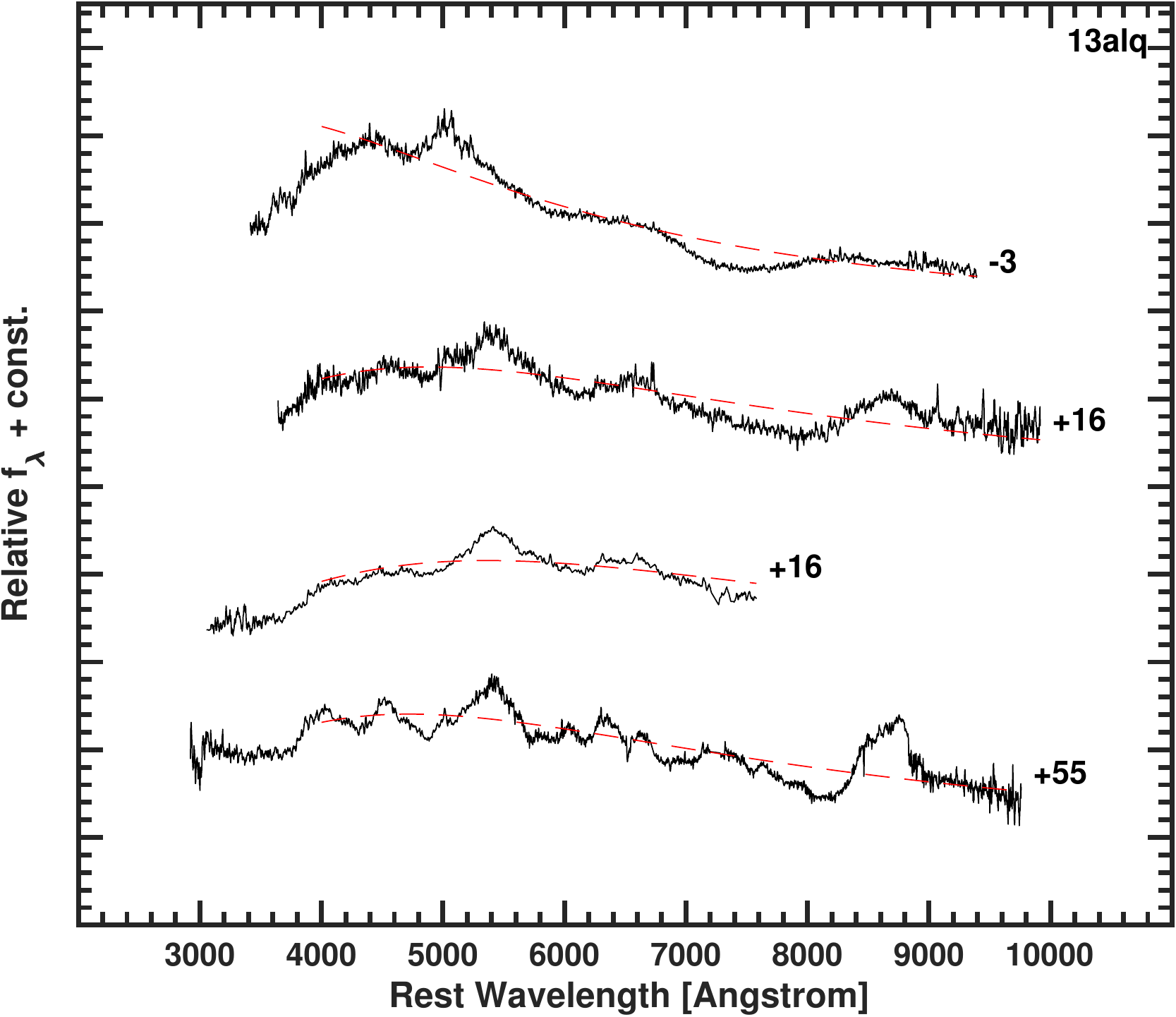} \\
    \end{array}$
  \caption{\label{spec_seq6}Spectral sequences of PTF12eci, iPTF13u, and iPTF13alq.}
 \end{figure}
 
   \begin{figure}
  \centering
 $\begin{array}{c}
\includegraphics[width=9cm]{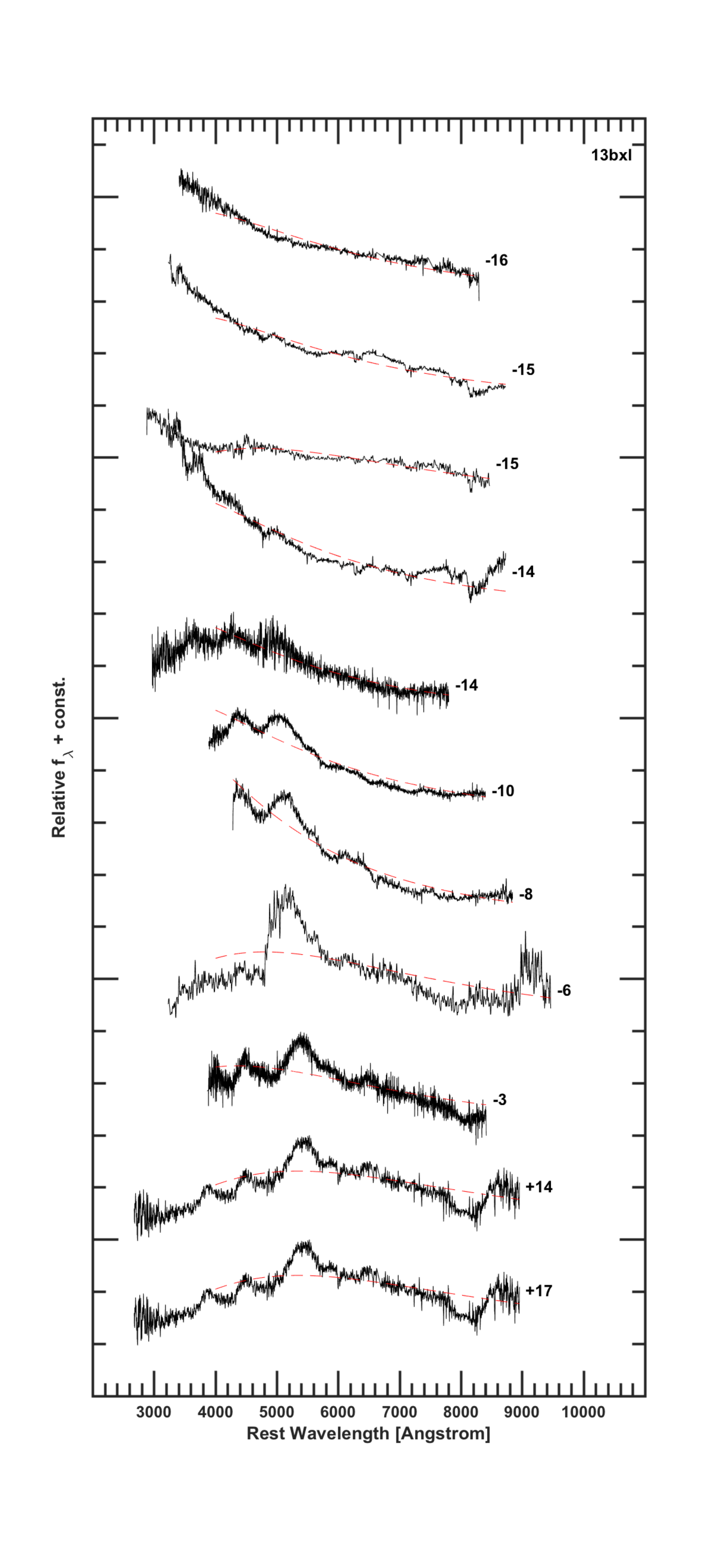} \\
  \includegraphics[width=9cm]{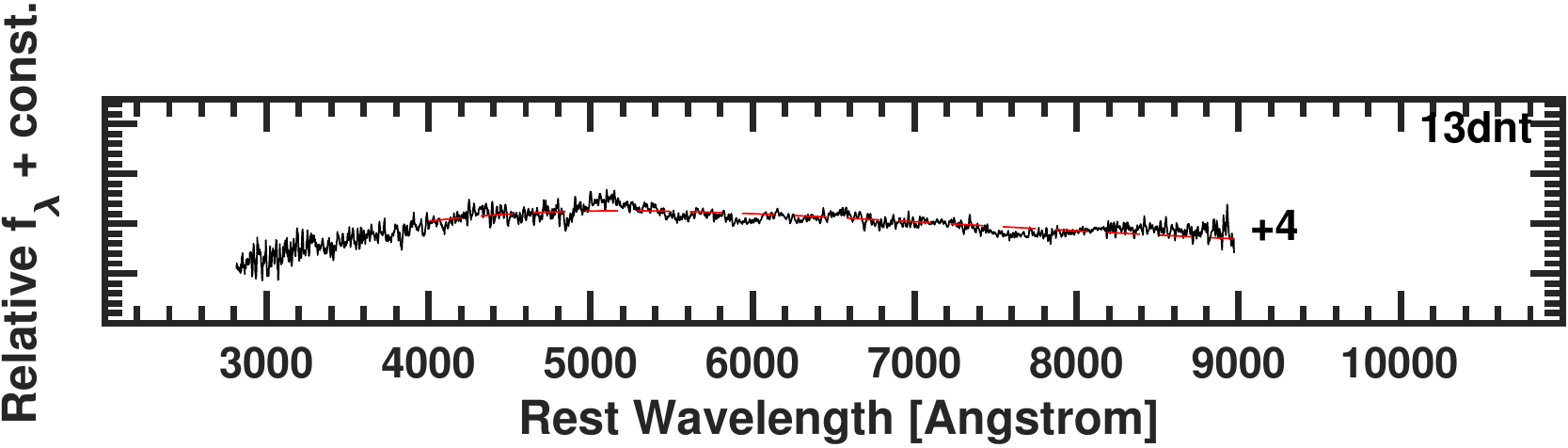} \\
    \end{array}$
  \caption{\label{spec_seq7}Spectral sequences of iPTF13bxl and iPTF13dnt.}
 \end{figure}
 
  \clearpage

   \begin{figure}
  \centering
 $\begin{array}{c}
   \includegraphics[width=9.0cm,height=7.0cm]{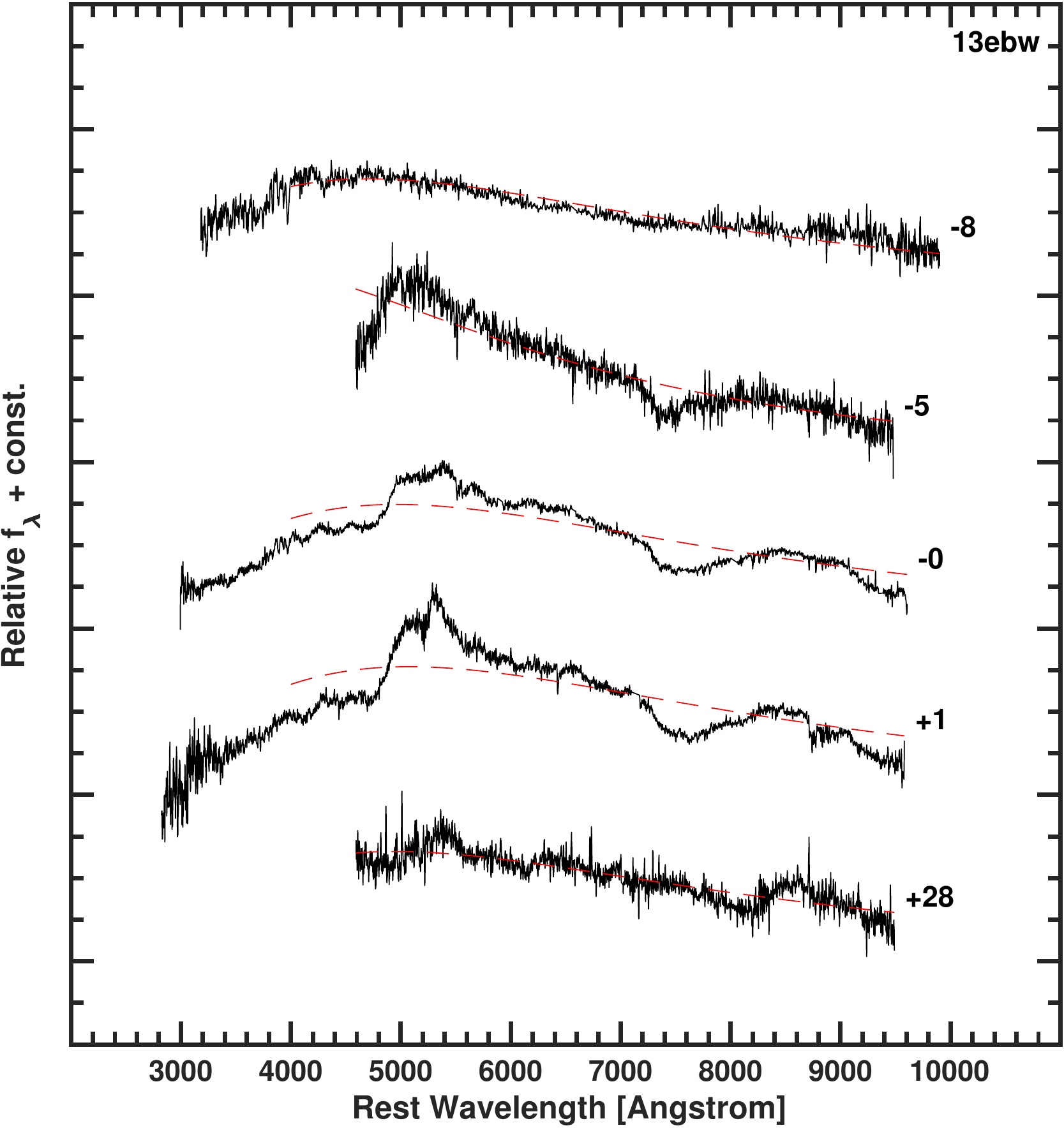} \\
\includegraphics[width=9.0cm,height=6.0cm]{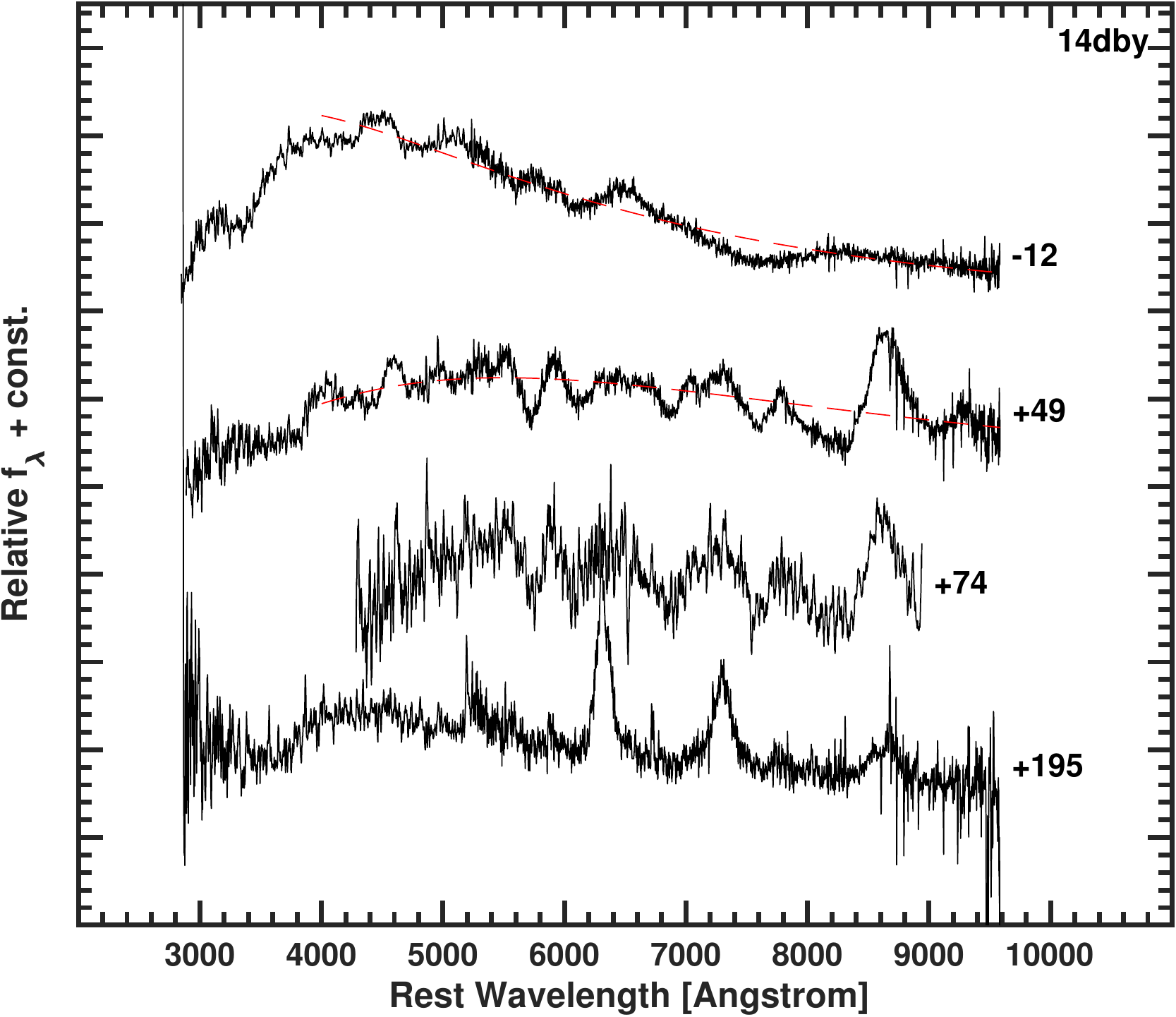} \\
  \includegraphics[width=9.0cm,height=4.5cm]{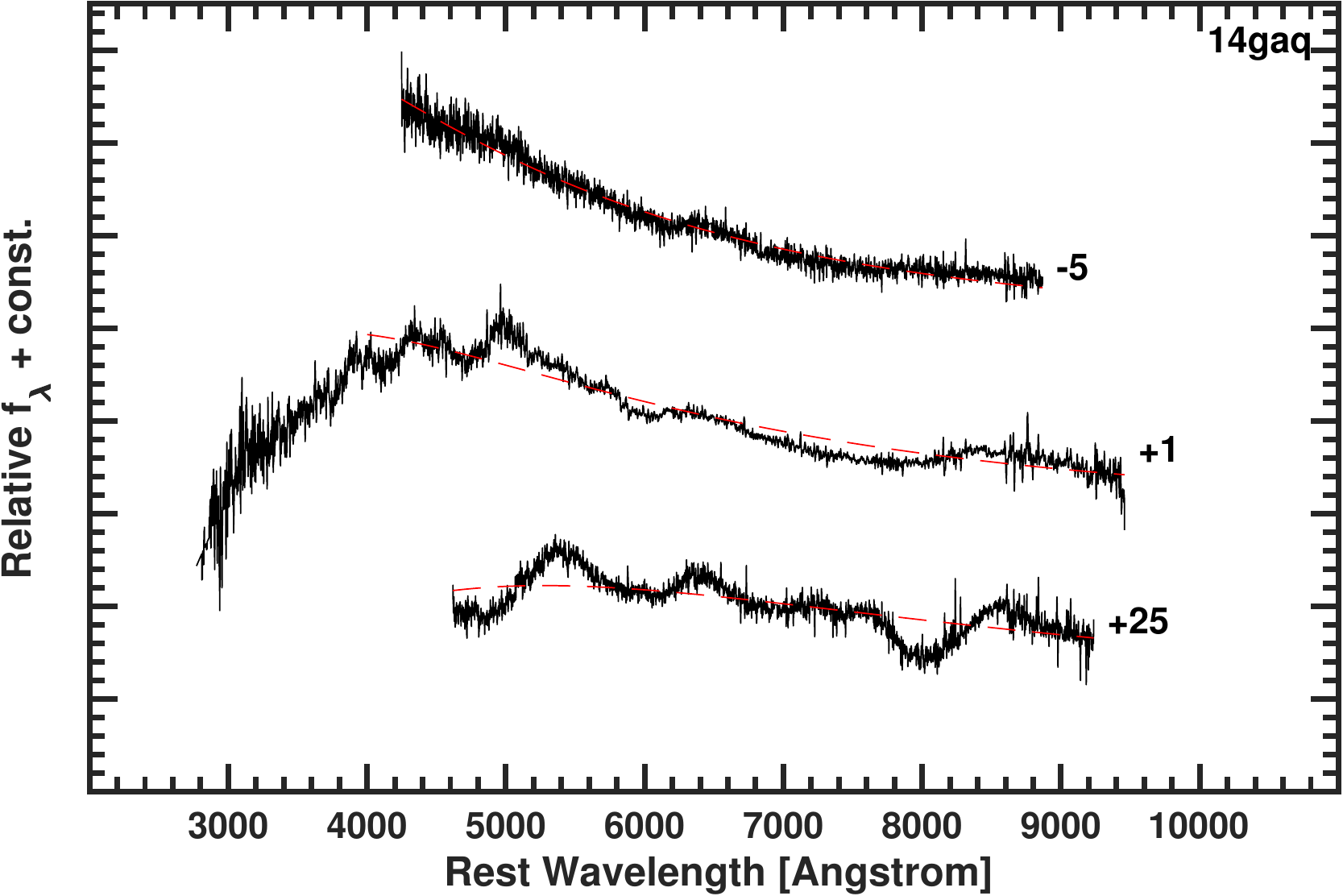} \\
  \includegraphics[width=9.0cm,height=4.5cm]{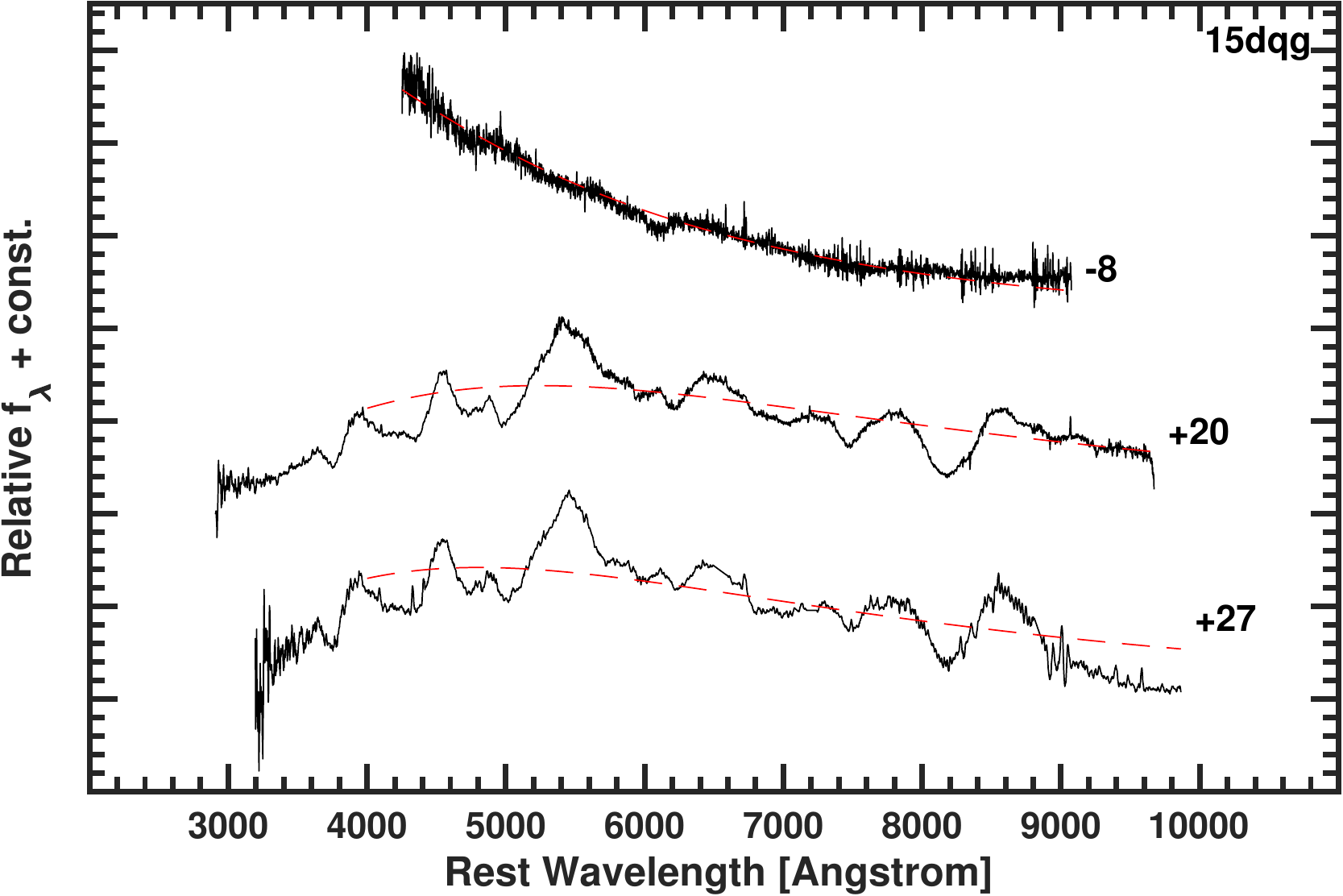} \\
    \end{array}$
  \caption{\label{spec_seq8}Spectral sequences of iPTF13ebw, iPTF14dby, iPTF14gaq, and iPTF15dqg.}
 \end{figure} 

   \begin{figure}
  \centering
 $\begin{array}{c}
\includegraphics[width=9.0cm,height=23cm]{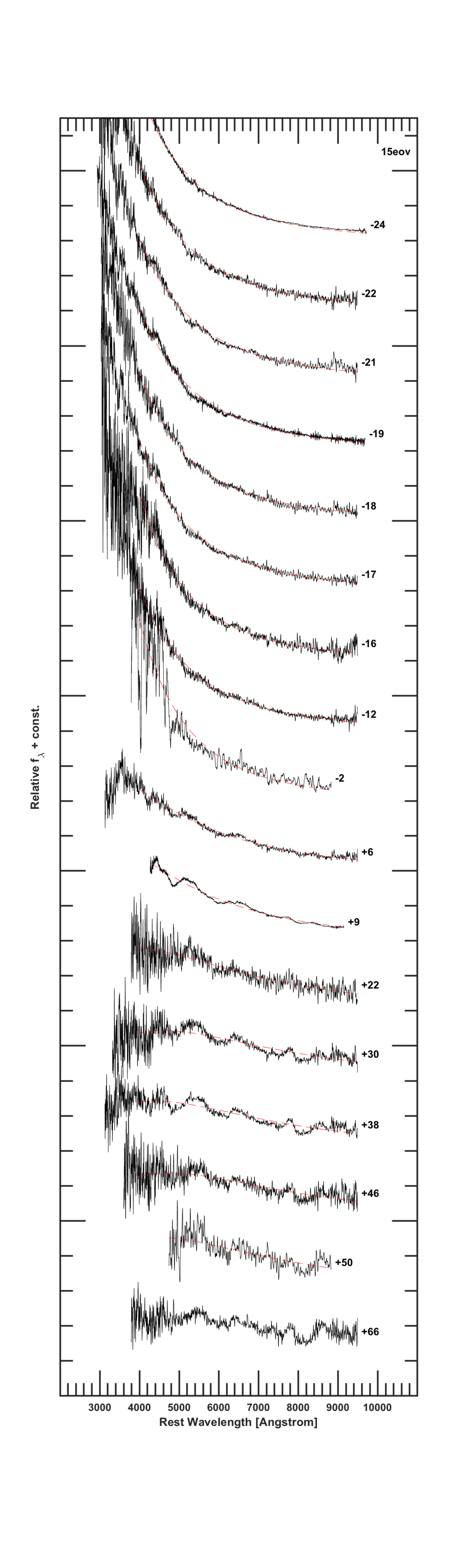} \\
    \end{array}$
  \caption{\label{spec_seq9}Spectral sequence of iPTF15eov.}
 \end{figure} 
 
  \clearpage 
 
    \begin{figure}
  \centering
 $\begin{array}{c}
   \includegraphics[width=9cm]{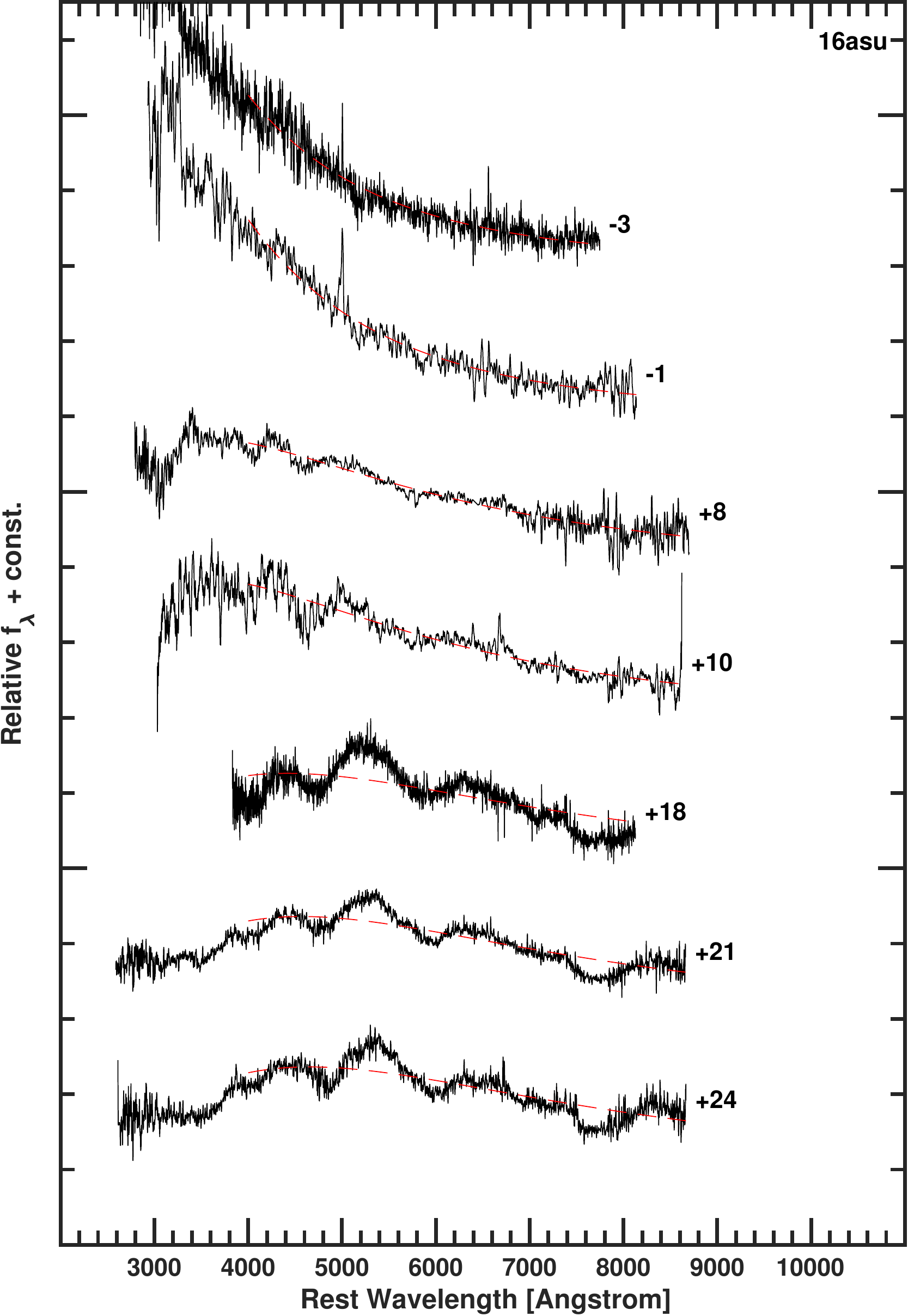} \\
  \includegraphics[width=9cm]{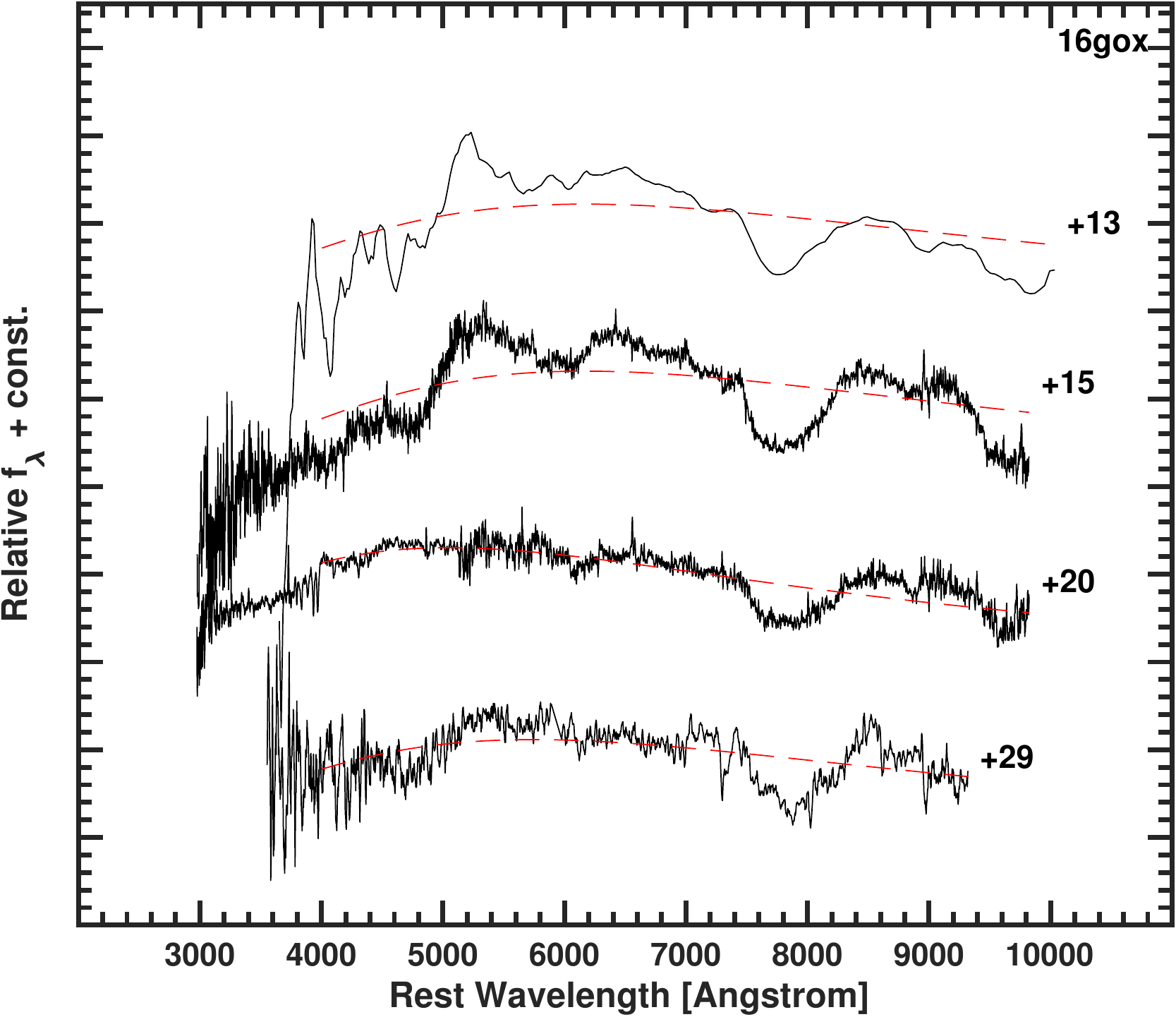} \\
    \end{array}$
  \caption{\label{spec_seq10}Spectral sequences of iPTF16asu and iPTF16gox.}
 \end{figure} 
 
     \begin{figure}
  \centering
 $\begin{array}{c}
   \includegraphics[width=9cm]{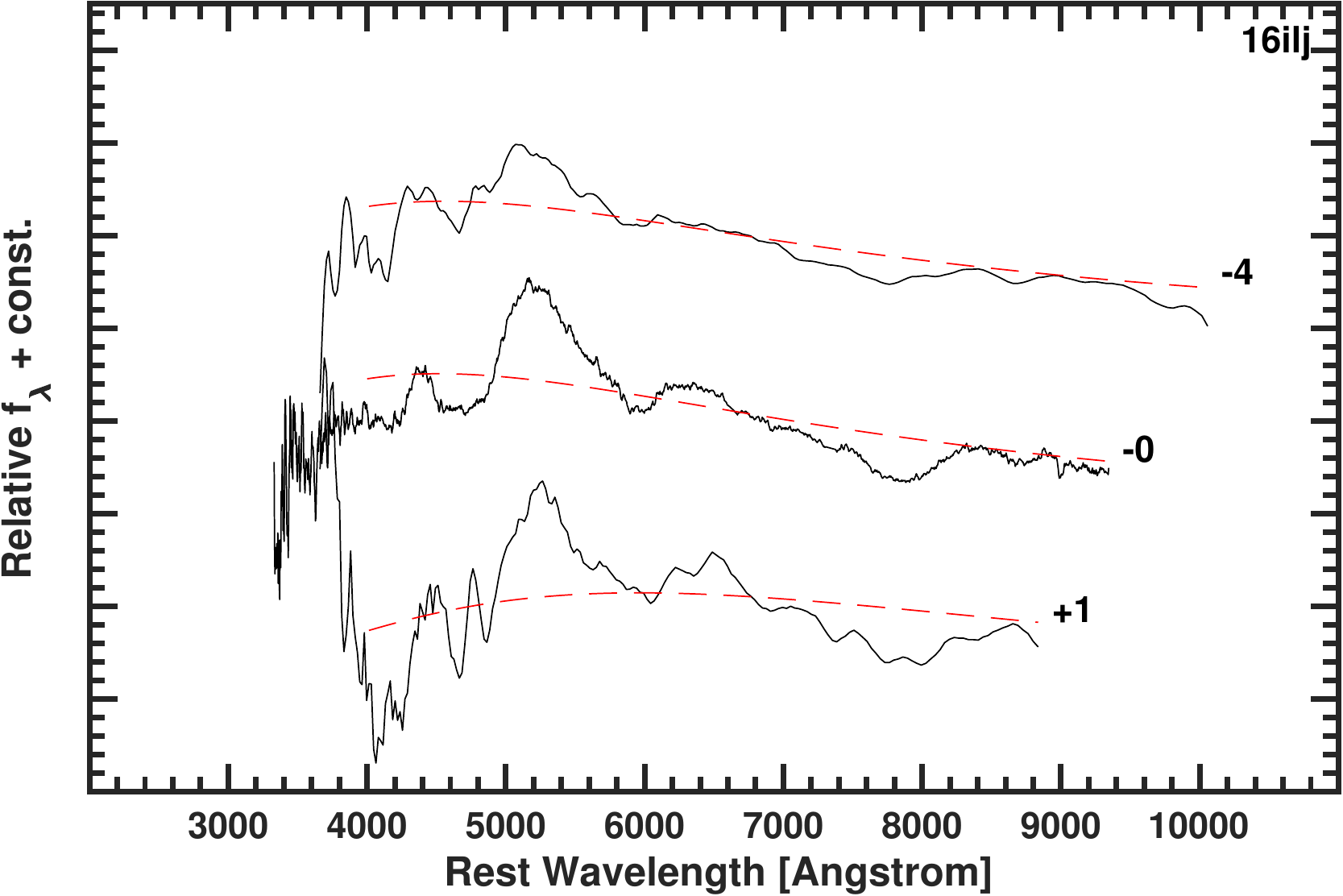} \\
  \includegraphics[width=9cm]{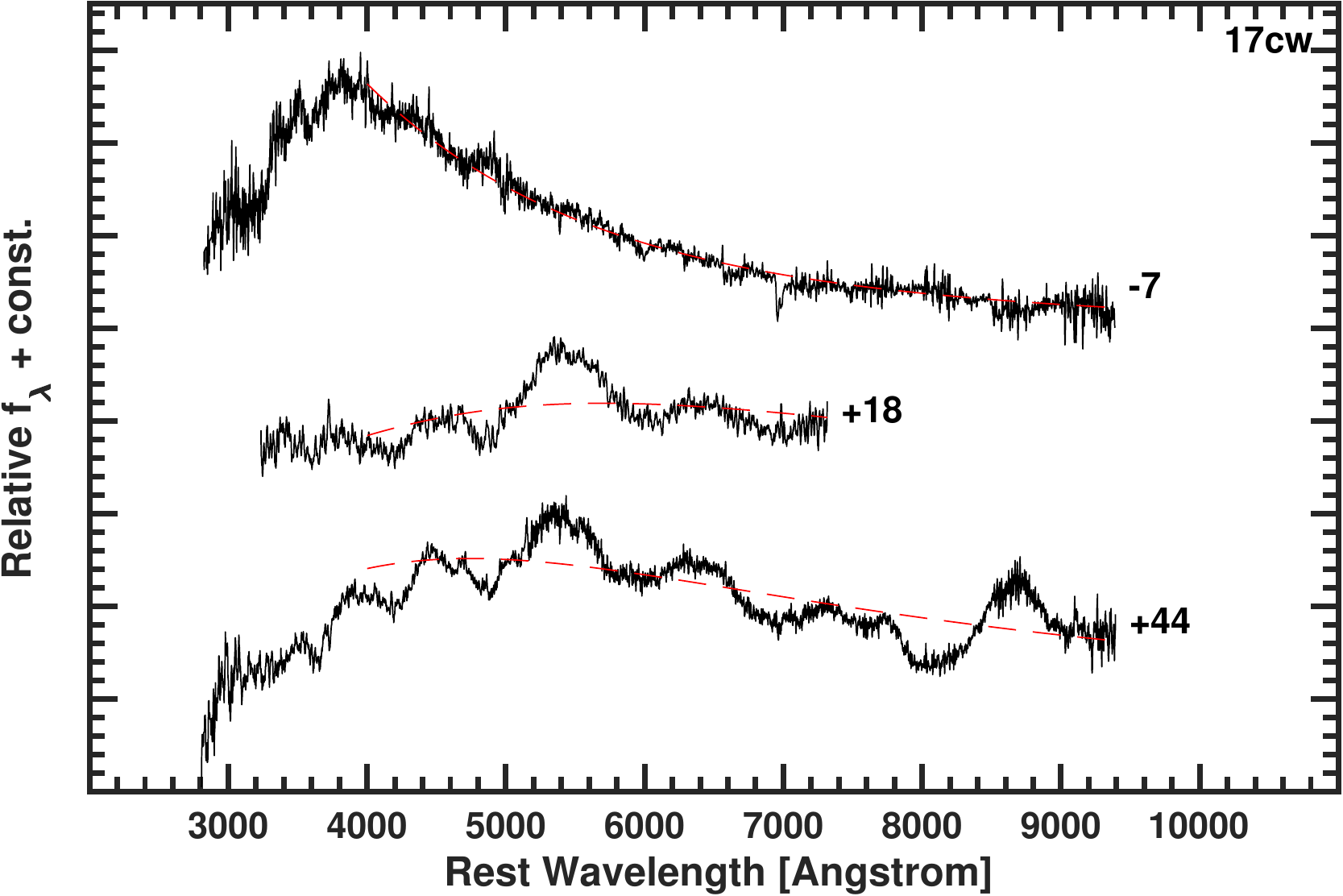}  \\
    \end{array}$
  \caption{\label{spec_seq11}Spectral sequences of iPTF16ilj and iPTF17cw.}
 \end{figure} 
 
      \begin{figure}
  \centering
 $\begin{array}{c}
   \includegraphics[width=9cm]{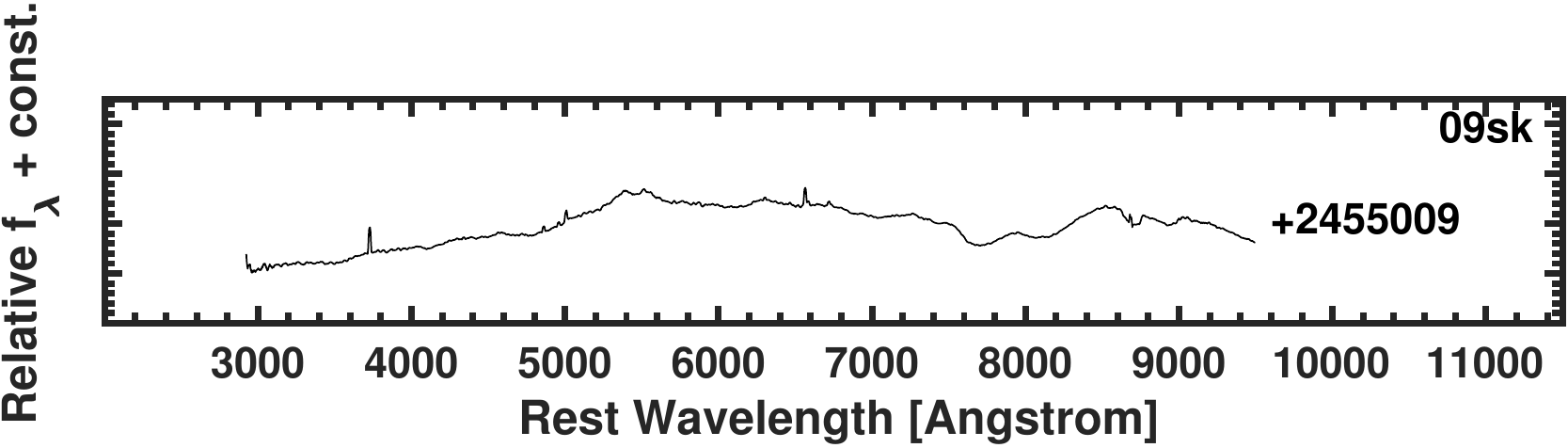} \\
  \includegraphics[width=9cm]{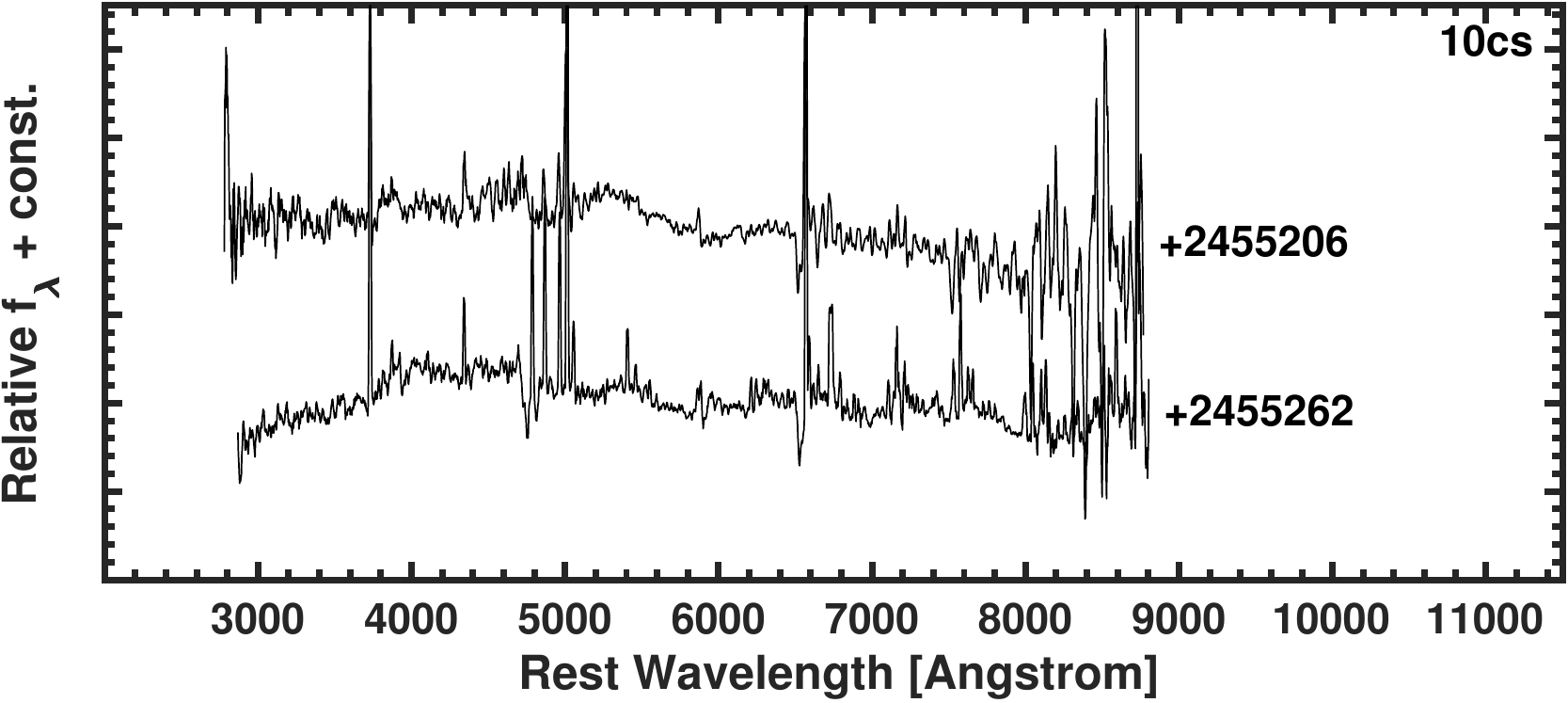}  \\
    \includegraphics[width=9cm]{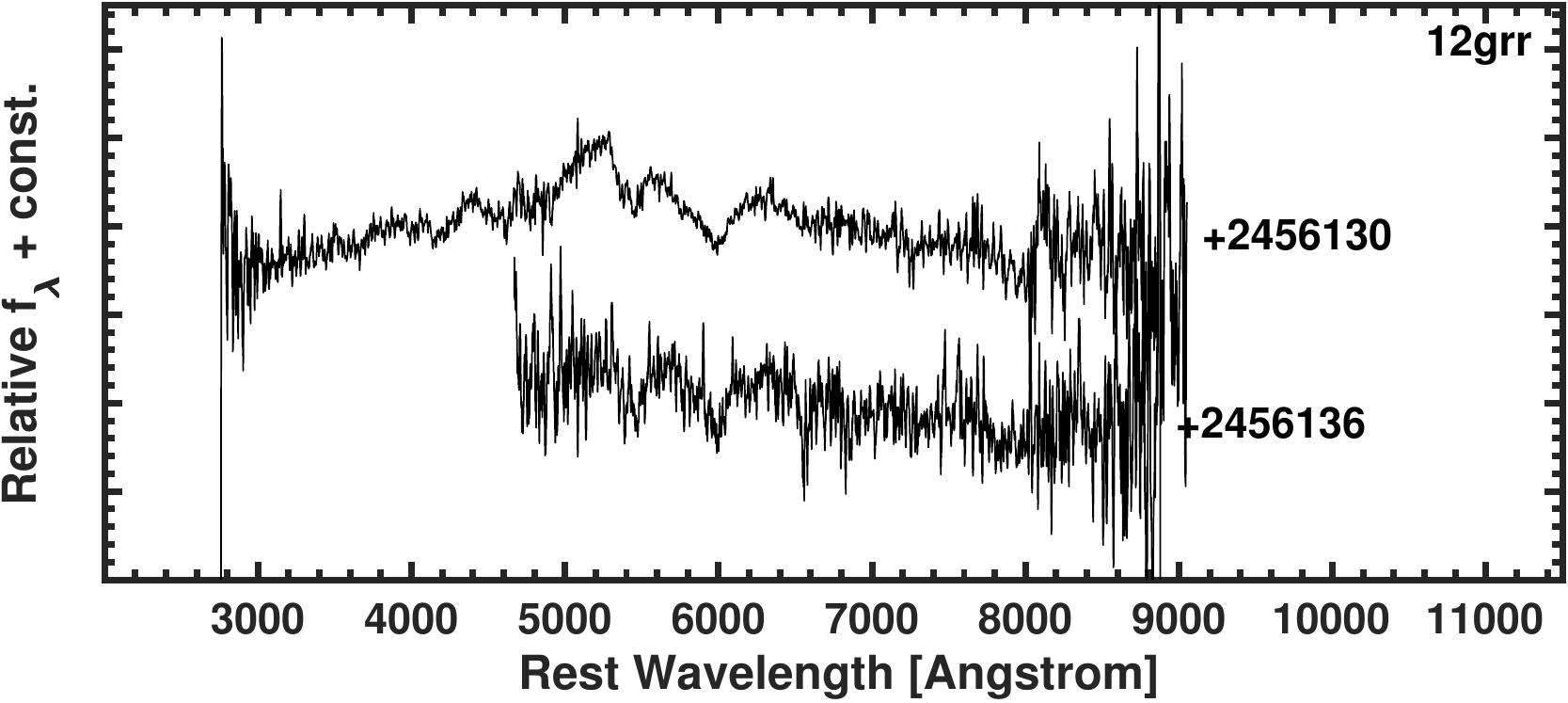}  \\
    \end{array}$
  \caption{\label{spec_seq12}Spectral sequences of the SNe observed after $r$-band peak.}
 \end{figure}

       \begin{figure}
  \centering
 $\begin{array}{c}
 \includegraphics[width=9cm]{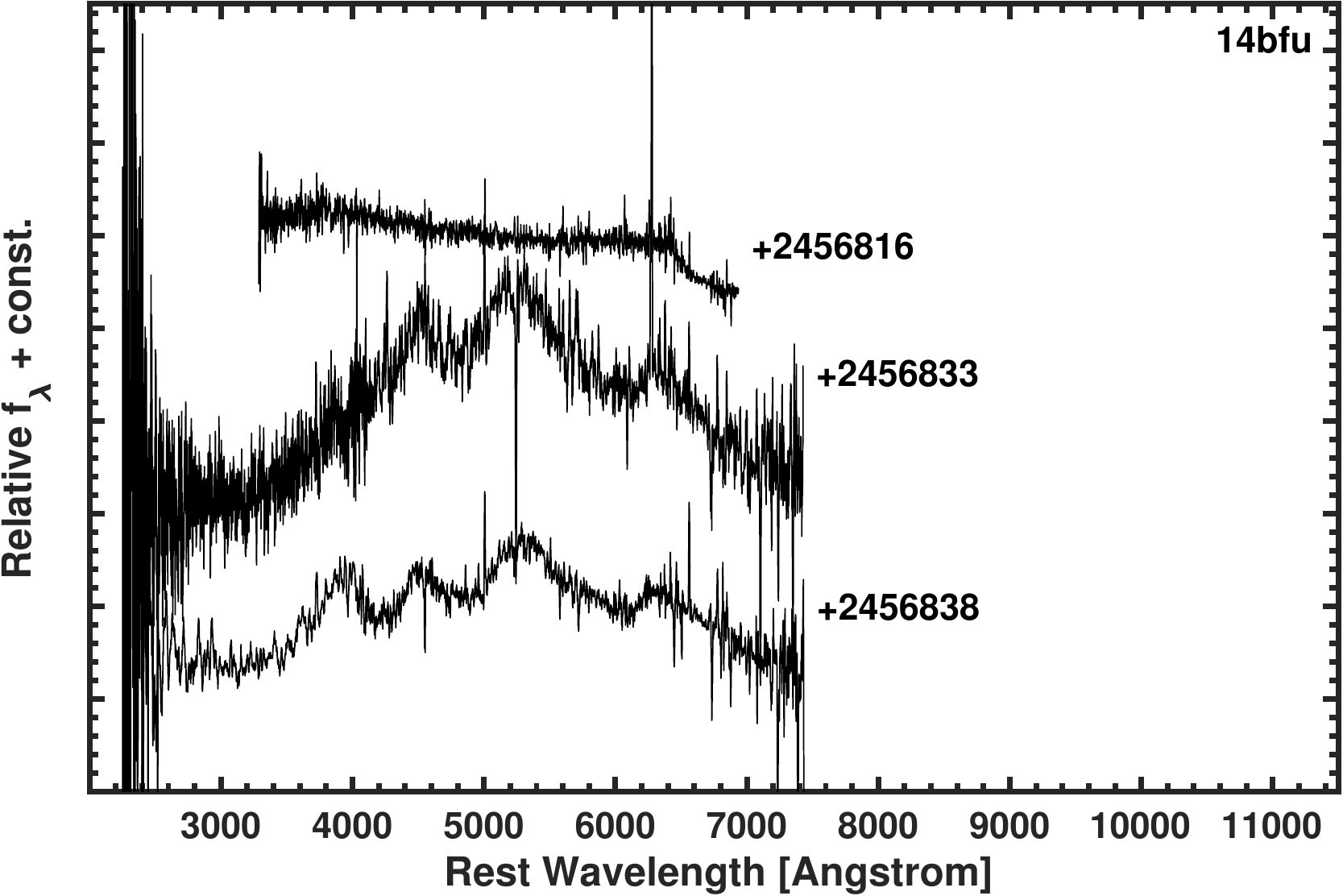}  \\
 \includegraphics[width=9cm]{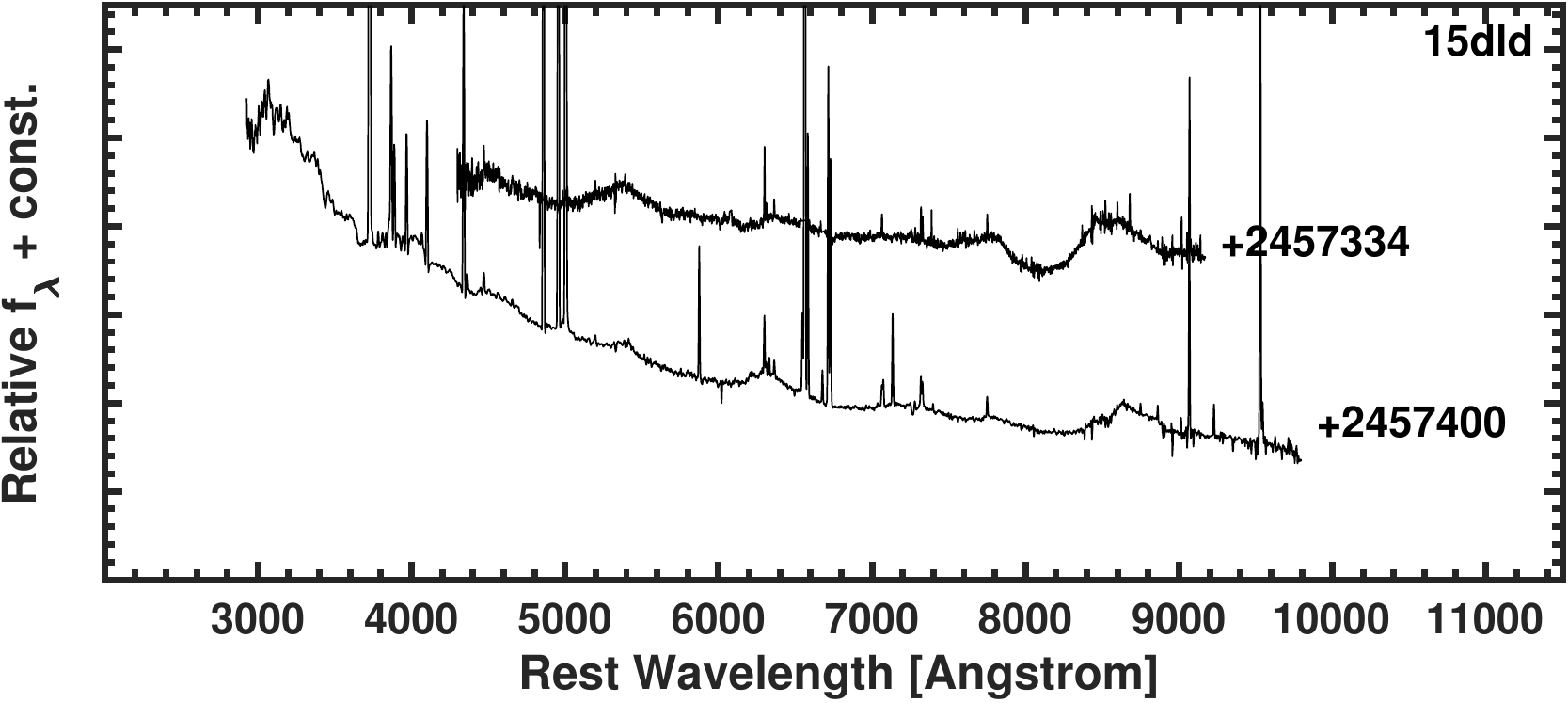}  \\
     \includegraphics[width=9cm]{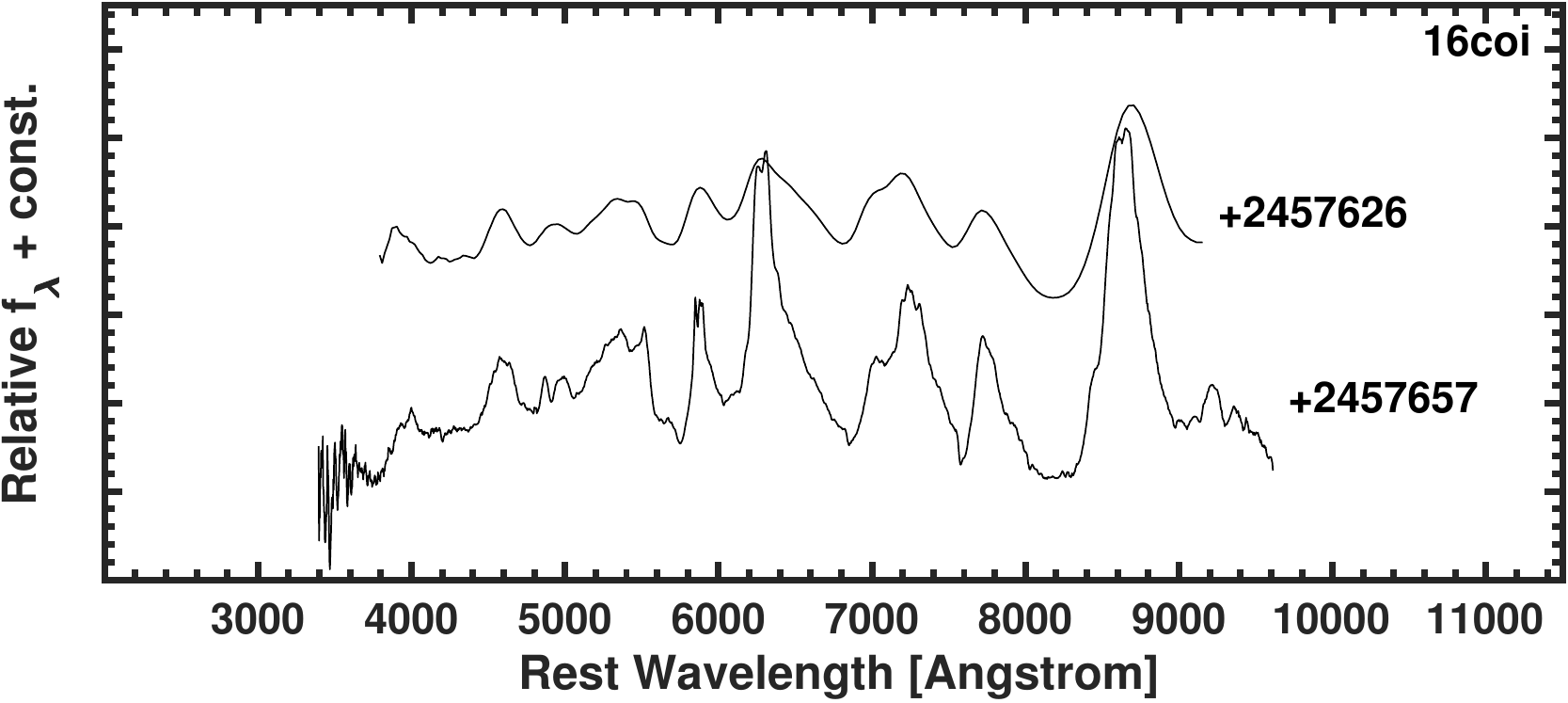}  \\
    \end{array}$
  \caption{\label{spec_seq13}Spectral sequences of the SNe observed after $r$-band peak.}
 \end{figure}

\end{appendix}

\end{document}